\documentclass[sigconf,screen]{acmart}

\usepackage{enumitem}

\usepackage{color}
\usepackage{xcolor}
\usepackage{subcaption}
\usepackage{multirow}
\usepackage{makecell}
\usepackage{stfloats}
\usepackage{array}
\usepackage{siunitx}

\definecolor{diffneg}{RGB}{220,50,47}   % red
\definecolor{diffpos}{RGB}{38,139,210}  % blue

\usepackage{colortbl}

\usepackage{booktabs}

\usepackage{array}

\usepackage{colortbl}
\usepackage{acmart-taps}

\usepackage{array}

\usepackage{soul}

\usepackage{graphicx}
\usepackage{array}
\usepackage{multirow}
\usepackage{fp}

\NewDocumentEnvironment{revtable}{ O{revisionColor} }{%
  \begingroup
  \color{#1}%
}{%
  \endgroup
}

\newcommand*\quotes[1]{``#1''}

\newcolumntype{L}[1]{>{\raggedright\arraybackslash}p{#1}}
\newcolumntype{C}[1]{>{\centering\arraybackslash}p{#1}}
\newcolumntype{R}[1]{>{\raggedleft\arraybackslash}p{#1}}

\definecolor{minColor}{HTML}{67001F}  % low end (dark red)
\definecolor{zeroColor}{HTML}{FFFFFF}  % center (white)
\definecolor{maxColor}{HTML}{053061}  % high end (dark blue)

\definecolor{kfbg}{HTML}{E3E3E3} 
\sethlcolor{kfbg}

\definecolor{mygray}{rgb}{0.89,0.89,0.89}
\aptLtoXcmd{\def\keyfinding#1{\colorbox{mygray}{#1}}}{
\newcommand{\keyfinding}[1]{\hl{#1}}}

\newcommand{\gradientcell}[2][blue]{%
  \begingroup
    \FPset\TMP{#2}%
    \FPmul\TMP\TMP{100}%
    \FPround\TMP\TMP{0}%
    \edef\pct{\TMP}%

    \FPset\D{\pct}%
    \FPsub\D{45}{\D}% D = 45 - pct
    \FPabs\D\D%
    \FPmul\D\D{2}%
    \FPround\D\D{0}%
    \edef\mix{\D}%

    \FPmin\D{\mix}{100}\edef\mix{\D}%
    \FPmax\D{\mix}{0}\edef\mix{\D}%

    \def\ColorSpec{}%
    \FPiflt\pct{50}%
      \edef\ColorSpec{minColor!\mix!zeroColor}%
    \else
      \edef\ColorSpec{#1!\mix!zeroColor}%
    \fi

    \expandafter\cellcolor\expandafter{\ColorSpec}%

    \FPifgt\pct{99}%
      \textcolor{white}{#2}%
    \else
      #2%
    \fi
  \endgroup
}

\definecolor{myOrange}{rgb}{0.84, 0.37, 0.00}  % #D55E00
\definecolor{myGreen}{rgb}{0.00, 0.62, 0.45}   % #009E73
\definecolor{myBlue}{rgb}{0.00, 0.45, 0.70}    % #0072B2
\definecolor{myPink}{rgb}{0.80, 0.47, 0.65}    % #CC79A7

\newcommand{\DivNegColor}{red}     % color for negatives
\newcommand{\DivPosColor}{blue}    % color for positives
\newcommand{\DivZeroColor}{white}  % center color
\newcommand{\DivTextSwitch}{65}    % % at which text turns white (0..100)

\newcommand{\DivGlobalMin}{-1}
\newcommand{\DivGlobalMax}{ 1}

\NewDocumentCommand{\divcell}{ O{\DivGlobalMin} O{\DivGlobalMax} m }{%
  \begingroup
    \pgfmathsetmacro{\val}{#3}
    \pgfmathsetmacro{\vmin}{#1}
    \pgfmathsetmacro{\vmax}{#2}
    \pgfmathsetmacro{\range}{\vmax-\vmin}
    \pgfmathsetmacro{\rangeOK}{ifthenelse(abs(\range)<1e-9,1,\range)}
    \pgfmathtruncatemacro{\straddle}{(\vmin<0) * (\vmax>0)}

    \ifnum\straddle=1
      \pgfmathsetmacro{\den}{max(abs(\vmin),abs(\vmax))}
      \pgfmathsetmacro{\mag}{min(1,max(0,abs(\val)/\den))}
      \pgfmathtruncatemacro{\pct}{round(100*\mag)}
      \ifdim \val pt < 0pt
        \edef\bgcmd{\noexpand\cellcolor{\DivNegColor!\pct!\DivZeroColor}}%
      \else
        \edef\bgcmd{\noexpand\cellcolor{\DivPosColor!\pct!\DivZeroColor}}%
      \fi
    \else
      \pgfmathsetmacro{\t}{(\val-\vmin)/\rangeOK}
      \pgfmathsetmacro{\tclamp}{min(1,max(0,\t))}
      \pgfmathtruncatemacro{\pct}{round(100*\tclamp)}
      \ifdim \vmax pt > 0pt
        \edef\bgcmd{\noexpand\cellcolor{\DivPosColor!\pct!\DivZeroColor}}%
      \else
        \pgfmathtruncatemacro{\pctneg}{round(100*(1-\tclamp))}
        \edef\bgcmd{\noexpand\cellcolor{\DivNegColor!\pctneg!\DivZeroColor}}%
      \fi
    \fi

    \bgcmd
    \pgfmathtruncatemacro{\usewhite}{(\pct > \DivTextSwitch)}
    \ifnum\usewhite=1 \color{white}\fi
    #3%
  \endgroup
}

\definecolor{sigcol}{RGB}{33,113,181}   % blue (color-blind friendly)
\definecolor{nonsiggray}{gray}{0}        % dark gray

\definecolor{sigcol}{RGB}{33,113,181}   % blue (color-blind friendly)
\definecolor{nonsig}{gray}{0.35}        % dark gray

\definecolor{sigblue}{HTML}{1F77B4}   % noticeable but accessible blue
\definecolor{nonsigColor}{gray}{0}  % neutral gray for non-sig

\definecolor{revisionColor}{RGB}{0, 0, 255} % blue
\newlength{\CIwidth}
\newcolumntype{Y}{>{\centering\arraybackslash}m{\CIwidth}}

\AtBeginDocument{%
  }

\definecolor{diffneg}{RGB}{220,50,47}   % red
\definecolor{diffpos}{RGB}{38,139,210}  % blue

\newcommand{\setdiffscale}[2]{%
  \def\DiffMin{#1}%
  \def\DiffMax{#2}%
}
\setdiffscale{-0.7}{0.7}

\newcommand{\DiffBox}[2]{%
  \begingroup
    \setlength{\fboxsep}{0pt}%
    \colorbox{#1}{\strut\small\,(#2)}%
  \endgroup
}

\newcommand{\diffpercent}[1]{%
  \FPset\RAW{#1}%
  \FPabs\RAWABS\RAW%
  \FPabs\MINABS{\DiffMin}%
  \FPabs\MAXABS{\DiffMax}%
  \FPmax\DENOM\MINABS\MAXABS%
  \FPifgt\DENOM{0}%
    \FPdiv\TMP\RAWABS\DENOM%
    \FPmul\TMP\TMP{100}%
  \else
    \FPset\TMP{0}%
  \fi
  \FPmin\TMP\TMP{100}%
  \FPmax\TMP\TMP{0}%
  \FPround\TMP\TMP{0}%
  \edef\Percent{\TMP}%
}

\newcommand{\diffcell}[1]{%
  \begingroup
    \diffpercent{#1}%
    \FPifgt\RAW{0}%
      \DiffBox{diffpos!\Percent!white}{#1}%
    \else
      \DiffBox{diffneg!\Percent!white}{#1}%
    \fi
  \endgroup
}

\newcommand{\diffcellCL}[1]{%
  \begingroup
    \diffpercent{#1}%
    \FPmul\GOOD\RAW{-1}%
    \FPifgt\GOOD{0}%
      \DiffBox{diffpos!\Percent!white}{#1}%
    \else
      \DiffBox{diffneg!\Percent!white}{#1}%
    \fi
  \endgroup
}

\copyrightyear{2026}
\acmYear{2026}
\setcopyright{cc}
\setcctype{by}
\acmConference[CHI '26]{Proceedings of the 2026 CHI Conference on Human Factors in Computing Systems}{April 13--17, 2026}{Barcelona, Spain}
\acmBooktitle{Proceedings of the 2026 CHI Conference on Human Factors in Computing Systems (CHI '26), April 13--17, 2026, Barcelona, Spain}
\acmPrice{}
\acmDOI{10.1145/3772318.3790704}
\acmISBN{979-8-4007-2278-3/2026/04}

\begin{document}

\title{Scrollytelling as an Alternative Format for Privacy Policies}

\author{Gonzalo Gabriel M\'{e}ndez}
\orcid{0000-0002-3440-1115}
\affiliation{%
  \institution{Universitat Polit\`{e}cnica de Val\`{e}ncia}
  \city{Valencia}  
  \country{Spain}}  
\affiliation{%
  \institution{Inria}
  \city{Rennes}  
  \country{France}}
\email{ggmenco1@upv.es}

\author{Jose Such}
\orcid{0000-0002-6041-178X}
\affiliation{%
  \institution{INGENIO (CSIC-Universitat Polit\`{e}cnica de Val\`{e}ncia)}
  \city{Valencia}  
  \country{Spain}}  
\email{jose.such@csic.es}

\renewcommand{\shortauthors}{M\'{e}ndez and Such}

\begin{abstract}

%%%% 010-Abstract.tex starts here %%%%

Privacy policies are long, complex, and rarely read, which limits their effectiveness in informed consent. We investigate scrollytelling, a scroll-driven narrative approach, as a privacy policy presentation format. We built a prototype that interleaves the full policy text with animated visuals to create a dynamic reading experience. In an online study ($N=454$), we compared our tool against text, two nutrition-label variants, and a standalone interactive visualization. Scrollytelling improved user experience over text, yielding higher engagement, lower cognitive load, greater willingness to adopt the format, and increased perceived clarity. It also matched other formats on comprehension accuracy and confidence, with only one nutrition-label variant performing slightly better. Changes in perceived understanding, transparency, and trust were small and statistically inconclusive. These findings suggest that scrollytelling can preserve comprehension while enhancing the experience of policy reading. We discuss design implications for accessible policy communication and identify directions for increasing transparency and user trust.

%%%% Sections/010-Abstract.tex ends here %%%%

\end{abstract}




\maketitle

%%%% 020-Intro.tex starts here %%%%

\section{Introduction}

Privacy policies are meant to inform users about how their personal data is collected, used, and shared. In practice, however, they often fail to achieve this goal. These documents are usually long, written in complex legal or technical language, and presented in formats that discourage reading~\cite{McDonald2008Cost,Obar2016Biggest,Reidenberg2015Disagreeable,abu-salma2025grand}. As a result, users seldom review privacy policies, and those who do often struggle to extract the information necessary to make informed decisions~\cite{Solove2013Nothing}. This gap between the intended and actual function of privacy policies has been widely recognized~\cite{McDonald2008Cost, Reidenberg2015Disagreeable, Obar2016Biggest, Solove2013Nothing, acquisti2015privacy, Staff_2011, Schaub2017Designing}. Prior work has proposed a range of alternatives to traditional policy text, aiming to highlight key practices, reduce cognitive burden, and increase user engagement. These include at-a-glance summaries and icon sets~\cite{Habib2021Icons}, interactive Q\&A systems~\cite{Harkous2018Polisis}, and privacy \quotes{nutrition labels} in both static~\cite{Kelley2009Nutrition,Kelley2010Standardizing} and interactive forms~\cite{ReinhardtInteractivePP}. While such approaches show promise, adoption has been limited, and evidence on their effectiveness remains mixed, which leaves open the question of which other formats could help in practice.

Improving privacy policy communication is hard for several reasons. First, policies must satisfy legal completeness while remaining approachable to lay readers. Second, comprehension is multi-faceted (recall of facts, interpretation, confidence) and not straightforward to measure outside tightly controlled settings. Third, interface changes that aid comprehension can raise other costs (e.g., cognitive load) or be dismissed as marketing, undermining trust. These challenges motivate designs that preserve fidelity to the underlying policy while scaffolding understanding and engagement.

In this paper, we investigate the potential of scrollytelling as a candidate format for presenting privacy policies. Scrollytelling has been effective in other information-dense domains for sustaining attention and conveying structure through narrative sequencing and visual cues~\cite{Segel2010NarrativeVisualization,Lee2015MoreThanTelling}. We hypothesize that it can similarly help people grasp the \textit{what}, \textit{why}, and \textit{who} of data practices without the friction of long, static text. To investigate this, we developed a scrollytelling interface that transforms policy text into a scroll-driven narrative, progressively revealing content alongside visual explanations and light interaction. We conducted a between-subjects online study ($N=454$) comparing our scrollytelling tool against four alternative formats: traditional text, an interactive nutrition label, a nutrition label with a link to the full text, and a standalone interactive visualization. Each participant explored one real policy in one format and completed comprehension, experience, and pre- and post-perception questionnaires.

Our results indicate that scrollytelling performed competitively on comprehension and clearly better on experience. Comprehension accuracy under scrollytelling was statistically indistinguishable from most alternatives; only one nutrition-label variant showed a modest advantage. Comprehension confidence followed a similar pattern. By contrast, scrollytelling consistently increased engagement, enjoyment, and perceived clarity, reduced cognitive load relative to text, and yielded equal or higher willingness to adopt the format. Pre–post perception shifts were small across formats. Together, these results suggest scrollytelling as a practical alternative that preserves comprehension while improving how people experience privacy policies.

Our contributions are threefold: (1) We introduce scrollytelling as a novel format for privacy policy communication and design an interface that combines narrative guidance with policy-faithful exploration; (2) We conduct a head-to-head empirical comparison of scrollytelling against textual, tabular, and visualization-only formats across comprehension, experience, and perception outcomes; and (3) We show that scrollytelling preserves comprehension while substantially enhancing user experience, offering a promising direction for more engaging and accessible privacy communication.

%%%% Sections/020-Intro.tex ends here %%%%

%%%% 030_RelatedWork.tex starts here %%%%

\section{Background and Related Work}

Efforts to improve privacy communication have long highlighted the inadequacy of traditional privacy policy formats and proposed alternatives to make them more accessible, engaging, and informative. In what follows, we first review presentation formats aimed at overcoming these challenges.
We then situate our work within broader literature on narrative techniques in visual communication, with a focus on scrollytelling.

\subsection{Privacy Policy Presentation Formats}

Decades of research document the limitations of privacy policies as long, dense legal texts~\cite{McDonald2008Cost,Obar2016Biggest,Reidenberg2015Disagreeable,Solove2013Nothing}. Numerous formats have emerged in response. Some emphasize condensed overviews, such as layered or standardized short notices that present key points with optional drill-down~\cite{Kelley2010Standardizing,Cranor2012Necessary,Schaub2015DesignSpace}, while others introduce glanceable cues via icons or pictograms~\cite{Schaub2017Designing,Soumelidou2019Visualization,Mazza2023Pictograms,Habib2021Icons}. Interactive and AI-assisted designs further aid navigation and comprehension by surfacing relevant policy content~\cite{Guo2020Polisee,Bui2021PIExtract,Adhikari2023PolicyPulse}. While these approaches often improve usability and reduce effort compared to plain text, each involves trade-offs. Summaries may omit nuance~\cite{Kelley2010Standardizing}; icons risk ambiguity if not validated~\cite{Habib2021Icons}; and interactive tools can overload users if poorly tuned~\cite{Guo2020Polisee,Bui2021PIExtract}.

Among the alternative formats for policy communication, the privacy \textit{nutrition label} is one of the most prominent and studied. Inspired by food labeling, it presents key practices in a familiar, tabular layout~\cite{Kelley2009Nutrition}. Across lab and large-scale studies, privacy nutrition labels consistently outperform long-form text on lookup speed, perceived usability, and satisfaction~\cite{Kelley2009Nutrition,Kelley2010Standardizing,Cranor2012Necessary}. Variants adapted for mobile and IoT contexts similarly show comprehension gains~\cite{EmamiNaeini2020IoTLabels}. More recently, Reinhardt et al.~\cite{ReinhardtInteractivePP} extended this format with interactive elements, enabling users to expand sections and access additional explanations. The interactive label increased perceived attractiveness and encouraged exploration, though effects on trust and acceptance were limited.

Another line of research has explored narrative-based formats to make policies more approachable. \emph{Textured Agreements}, for instance, embedded short vignettes into agreements to draw readers in and improve comprehension relative to plain text~\cite{Kay2010Textured}. Other work has used full comic treatments to explain notice and choice elements~\cite{Knijnenburg2016Comics}. Empirical findings, however, are mixed: some studies report increased engagement~\cite{Tabassum2018ComicPolicy}, but others find no improvements (or even declines) in comprehension and self-efficacy~\cite{Anaraky2019TestingAComic}. In permission dialogs, comics can increase user caution~\cite{Watson2021InvestigationComic,Watson2023ComicToPermission}, but their effects on understanding remain variable. Narrative formats show stronger results in educational contexts. Comic- and game-based privacy lessons improve retention and engagement in children~\cite{ZhangKennedy2017ChildrenComics,Kumar2018CoDesign,ZHANGKENNEDY201710}, and narrative-driven training is also gaining traction for usable security~\cite{Dincelli2019ChooseYourHackingAdventure,Dincelli2025CoDesigningCybersecurity}. Tools such as PrivacyToon~\cite{Suh2022PrivacyToon}, which let users create privacy-themed comics, highlight growing interest in storytelling workflows for privacy communication.

% Other work explores narrative formats to make policies more approachable. In that regard, our work is closest to that of Anaraky et al.~\cite{Anaraky2019TestingAComic}, who tested a comic-style privacy policy against plain text and found mixed results: while the comic increased engagement, it sometimes led to lower comprehension. Narrative techniques, however, have shown stronger effects in privacy \textit{education}, particularly with children. For instance, PrivacyToon~~\cite{Suh2022PrivacyToon} allows learners to create privacy-related comics, fostering engagement and understanding.

The empirical evidence behind these approaches suggests that structured overviews and narrative framing can help users engage with privacy content. Yet few designs combine the benefits of narrative pacing with support for structured exploration and fidelity to the full policy text. We fill this gap by investigating \emph{scrollytelling} as one such format, asking whether a narrative presentation supports readers in extracting factual policy information and whether it complements prior findings focused primarily on attention and engagement outcomes~\cite{Tabassum2018ComicPolicy,Anaraky2019TestingAComic}.

\subsection{Scrollytelling and Narrative Visualization}

Scrollytelling is a form of narrative presentation that couples the act of scrolling with the unfolding of content. As the reader scrolls down a page, text and graphics transition in a choreographed way to tell a story or explain a concept step by step~\cite{Segel2010NarrativeVisualization,Tjarnhage2023Scrollytelling,Mendez2021Scrollytelling}. This technique emerged in online journalism (often for long-form features and data-driven articles) to create an immersive, continuous narrative experience without requiring clicks or page loads. In a typical scrollytelling piece, scrolling triggers animations or reveals new visuals at key points, while narrative text provides context~\cite{Heer2007AnimatedTransitions,Segel2010NarrativeVisualization}. The narrative can be linear or with slight branching, but generally follows a predetermined path (sometimes called an elastic narrative that can stretch with the user's pace)~\cite{Segel2010NarrativeVisualization,Hullman2011Rhetoric}. Though visuals and animations are central, the narrative text remains the backbone: it sets the pace, anchors interpretation, and ties transitions together~\cite{Hullman2011Rhetoric,Heer2007AnimatedTransitions}.

In the context of information visualization, scrollytelling is understood as a form of narrative visualization that blends author-driven sequencing with user-driven pace~\cite{Segel2010NarrativeVisualization}. It evolved from formats like magazine-style or slideshow visualizations, but instead offers a single, continuous canvas where scenes unfold progressively. This allows for tighter integration between text and visual explanation while maintaining structural cohesion.

From a usability standpoint, scrollytelling offers affordances that could benefit the presentation of privacy policies. First, it naturally segments information into digestible chunks, only revealing the next segment when the user has scrolled enough, implementing a form of progressive disclosure~\cite{Shneiderman1996Eyes,Nielsen2006ProgressiveDisclosure,Springer2018ProgressiveDisclosure}. This prevents overwhelming the reader with a wall of text. Second, the intermixing of visuals with text can aid comprehension via established multimedia and dual-coding effects~\cite{Mayer2009MultimediaLearning,Paivio1991DualCoding}. Third, the continuous narrative encourages readers to follow along to the end in a way that static documents often fail to do. Unlike static documents, scroll-based pacing can sustain engagement by rewarding each action with a new layer of the story~\cite{Tjarnhage2023Scrollytelling}.

% Because scrollytelling often feels like \textit{story consumption} rather than tedious reading, users may be more engaged and less likely to quit early~\cite{Tjarnhage2023Scrollytelling}. 

All this makes scrollytelling a promising but as-yet unexplored approach for privacy policy presentation. We investigate whether this format can support readers in extracting factual policy content while maintaining fidelity to the full text.

%%%% Sections/030_RelatedWork.tex ends here %%%%

%%%% 040-Tool.tex starts here %%%%

\section{A Scrollytelling Tool for Privacy Policies}
\label{sec:tool}

In this section, we first describe the design goals and principles that guided the implementation of our scrollytelling tool to present and explore privacy policies. We then describe the narrative and interactive components of our interface, provide a detailed account of our design process, and conclude with key technical details.

\subsection{Design Goals and Principles}
\label{sec:tool:goals_principles}

The final design of our tool follows four goals that address long-standing critiques of privacy policies and practical constraints of alternative formats. These goals are informed by: (i) prior work on privacy notices~\cite{Jensen2004,McDonald2008Cost,Schaub2015DesignSpace} and narrative visualization~\cite{Segel2010NarrativeVisualization}; and (ii) our own design process, as described in Section~\ref{app:tool:design_process}. The design goals of our tool are:\\

\textbf{G1:} \textit{Preserve legal fidelity while enhancing accessibility.} Simplified notices can omit nuance or overstate guarantees, raising questions about legal validity and user trust~\cite{Kelley2009Nutrition,Schaub2015DesignSpace}. This goal ensures that any policy simplification remains tied to the official policy text, allowing users to always access the original wording.

\textbf{G2:} \textit{Scaffold comprehension through progressive but accountable disclosure.} Chunked, staged presentation can reduce cognitive load \cite{spanjers2012explaining,Liu2024EffectsOfSegmentation,Rey2019Metaanalysis} and underpins layered and just-in-time notices~\cite{Schaub2015DesignSpace,feng_design_2021} and narrative visualization~\cite{Segel2010NarrativeVisualization}. At the same time, progressive disclosure can also enable dark patterns that bury key details~\cite{Mathur2019DarkPatterns}. This goal ensures that progressive disclosure is used to structure visual explanations, without omitting or downplaying any data collection or sharing practices described in the policy.

\textbf{G3:} \textit{Support orientation and maximizing the value of limited attention.} Reading policies in full is costly \cite{Jensen2004,McDonald2008Cost}, so it is unrealistic to expect users to engage with every detail. To help them navigate efficiently, the tool should show what has been covered, where users are, and what remains. Narrative scaffolds and progress cues~\cite{Segel2010NarrativeVisualization} should provide this structure, enabling users to extract meaningful insights even from brief interactions with the policy.

\textbf{G4:} \textit{Enable verification and critical reflection.} We not only anchor collection or sharing claims to policy passages but also explicitly represent any vague or conditional wording, rather than smoothing it away~\cite{Eiband2018TransparencyDesign}. This supports verification and critical reflection on whether the policy aligns with users' expectations and risk tolerance~\cite{Sengers2005ReflectiveDesign}.\\

\noindent{From these goals, we derived a set of design principles (DPs). Each principle provides a general guideline that informed the design of our tool, and together they articulate trade-offs in balancing fidelity, comprehension, and engagement:}

\textbf{DP1:} \textit{Anchor simplifications in the source.} In the tool, every summary, label, and visual element must link to the exact policy passage it derives from, so users can trace any simplification back to the original text (G1).

\textbf{DP2:} \textit{Unfold complexity step by step.} The tool should reveal policies incrementally, using staged visuals that allow readers to build understanding progressively rather than confronting all details at once, while keeping collection and sharing practices accessible (G2).

\textbf{DP3:} \textit{Scaffold orientation throughout the narrative.} Users should be able to see where they are in the narrative, which steps they have completed, and what remains to be seen, so that the attention they choose to invest is guided rather than aimless scrolling (G3).

\textbf{DP4:} \textit{Visualize uncertainty rather than conceal it.} When a privacy policy uses vague or conditional wording (e.g., conditional data flows, qualifiers, ranges), the tool should encode that uncertainty explicitly so users can judge the strength of the claims (G4).

\textbf{DP5:} \textit{Ensure bidirectional traceability.} The tool should support navigation in both directions between visuals and full text, preserving mappings from actors and data to their definitions in the full policy (G1, G4).

\textbf{DP6:} \textit{Balance overview with selective inquiry.} The tool should provide a clear big-picture view of the policy's owner's data practices while also supporting flexible exploration of specific actors or data types when users choose to investigate further (G2, G4).

\subsection{The Interface}
\label{sec:system:scrollytelling}

\begin{figure*}[htbp]
    \centering
    \includegraphics[width=\linewidth]{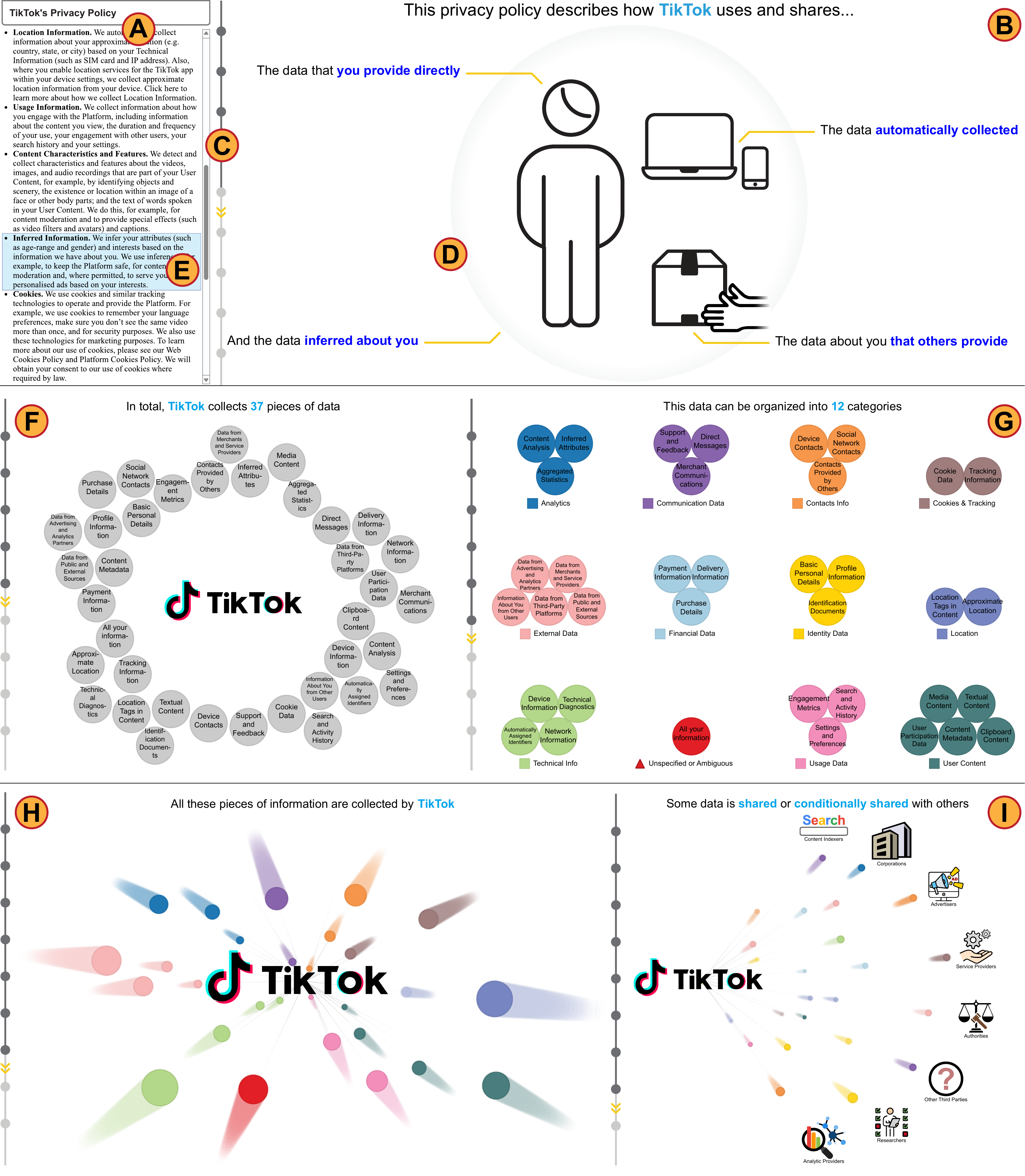}
    \caption{Storyboard of our scrollytelling interface, showing representative moments of the narrative: (A) full-text policy, (B) animation stage, (C) vertical progress bar signaling that scrolling can continue, (D) introduction of inferred personal attributes or interests, (E) synchronized highlighting of the corresponding clause in the textual policy, (F \& G) enumeration and clustering of data items, (H) visualization of data ingestion, and (I) depiction of sharing flows to third-party actors. See Section~\ref{sec:system:scrollytelling} for a full annotated description of the interface.}
    \label{fig:scrollytellingTool}
    \Description{Storyboard of the scrollytelling interface, shown as a multi-part annotated illustration. The figure presents key interface elements and narrative stages. (A) The full privacy policy appears on the left pane in text format. (B) The right pane contains animated visual explanations driven by scroll gestures. (C) A vertical progress bar separates the panes, filling in segments as users scroll. (D) An icon for inferred attributes is introduced, depicted by a glowing halo. (E) A matching clause in the policy text is highlighted, showing synchronization. (F) Data items are shown as individual circles labeled with terms like "IP address" or "Engagement metrics." (G) These items cluster into 12 categories, each labeled (e.g., “Technical Info,” “Contacts Info,” “User Content”). (H) Arrows show data being collected into the platform. (I) Arrows show selected data being shared with third parties like advertisers, researchers, and authorities. The figure illustrates how the interface supports traceability, progressive disclosure, and narrative engagement.}
\end{figure*}

We now describe our scrollytelling interface for presenting privacy policies, highlighting how its features implement the design principles introduced above. References (e.g., \textbf{DP1}) indicate the motivation behind each feature. The description is complemented by the annotated Figures~\ref{fig:scrollytellingTool} and~\ref{fig:interactiveVis}, which depict TikTok's policy, and the demonstration video included in the supplementary materials.

\subsubsection{Tool-wide Features}

Our interface is a scroll-driven, two-pane web page that integrates animated narrative with full-text policy content to support user comprehension. When users first land on the page, they see the full privacy policy rendered in a single-column layout, taking the entire screen real estate. Pressing a prominent \quotes{Start} button reconfigures the interface into a persistent split screen: the left pane displays the complete textual policy (Figure~\ref{fig:scrollytellingTool}.A), while the right pane functions as an animation stage driven by scroll gestures (Figure~\ref{fig:scrollytellingTool}.B). Each scroll gesture triggers a new explanatory step in the right pane. When a step directly corresponds to a specific clause in the policy, the left pane automatically scrolls to that section and applies a highlight to draw the user's attention. If no direct mapping is needed for a given step, the textual policy view remains static. This conditional synchronization addresses \textbf{DP1} and \textbf{DP5}, keeping explanations anchored in the source text and ensuring bidirectional traceability.

To give users a constant sense of orientation in the unfolding sequence, the tool features a vertical progress bar that runs between the text and the animation stage (Figure~\ref{fig:scrollytellingTool}.C). At the beginning, the circles and connecting lines of the progress bar are unfilled and shown in light gray \raisebox{-0.2ex}{\includegraphics[height=0.68em]{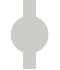}}. As the user scrolls, each step progressively fills \raisebox{-0.2ex}{\includegraphics[height=0.68em]{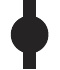}} the corresponding segment, providing a visual marker of advancement. Whenever further content is available, animated yellow arrows \raisebox{-0.2ex}{\includegraphics[height=0.68em]{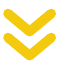}} appear onto the bar, signaling that scrolling can continue, helping users track their position in the narrative and encouraging engagement. These elements implement \textbf{DP3}, scaffolding orientation and sustaining engagement over the course of the explanation.

\subsubsection{The Explanatory Narrative}
\label{sec:ExplanatoryNarrative}

Our scrollytelling narrative is organized into three main segments: data sources, data types, and third-party sharing. In the opening segment, the platform's data sources are introduced through iconography and transitions. A user avatar \raisebox{-0.2ex}{\includegraphics[height=1.4em]{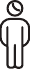}} appears first, labeled \quotes{\textit{Information you provide}}, followed by a scroll to the relevant textual policy section. As the user continues scrolling, a laptop \raisebox{-0.2ex}{\includegraphics[height=0.68em]{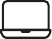}} and smartphone \raisebox{-0.2ex}{\includegraphics[height=0.68em]{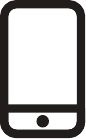}} appear to depict automatically collected data. This is followed by a package icon \raisebox{-0.2ex}{\includegraphics[height=0.68em]{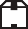}} handed by a pair of hands \raisebox{-0.2ex}{\includegraphics[height=0.68em]{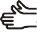}} to signify data received from others (i.e., external sources). When appropriate, an additional step introduces a glowing halo \raisebox{-0.2ex}{\includegraphics[height=0.68em]{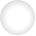}} labeled \quotes{\textit{data inferred about you}} (Figure~\ref{fig:scrollytellingTool}.D), which is accompanied by the corresponding scroll and highlight in the textual policy (Figure~\ref{fig:scrollytellingTool}.E)\footnote{This step is included only for policies that explicitly mention inferred data. For example, TikTok's policy states: \quotes{\textit{We \textbf{infer} your attributes (such as age-range and gender) and interests based on the information we have about you.}} By contrast, OpenAI’s policy does not mention inferred data, so the corresponding visuals and explanation are skipped in that case, and the sequence advances directly to the next stage.}. Along the entire narrative, each icon is animated smoothly into position. This staged progression exemplifies \textbf{DP2}, revealing complexity incrementally rather than all at once.

\begin{figure*} [b]
    \centering
    \includegraphics[width=1\linewidth]{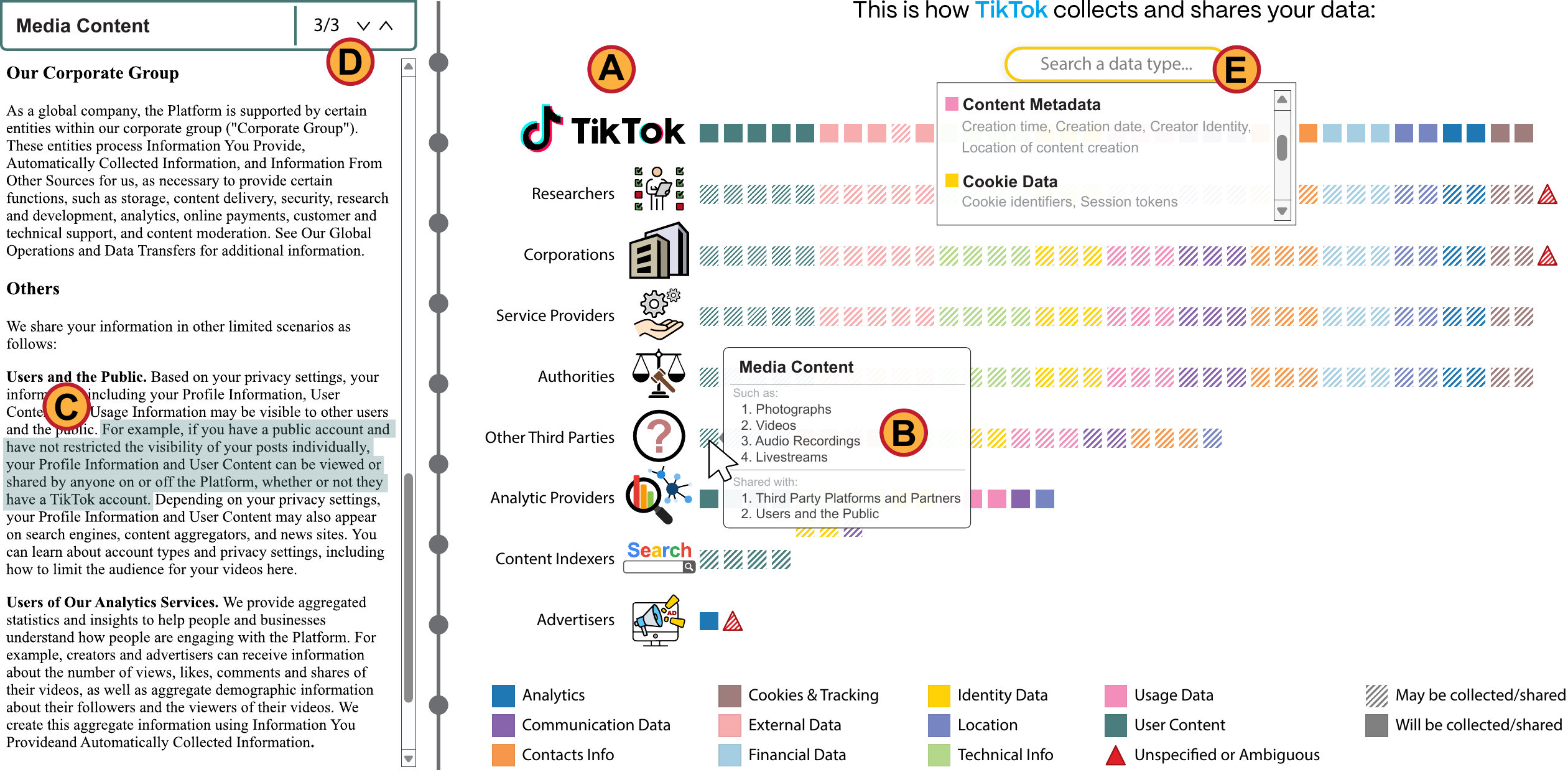}
    \caption{Annotated view of the interactive summary visualization. (A) Actors are displayed in rows, sorted by the volume of data they receive. (B) Hovering over a rectangle reveals the associated data type and its sharing destinations. (C) Clicking a rectangle scrolls and highlights the corresponding passages in the full-text policy. (D) When multiple references exist, a navigation aid allows cycling through them. (E) A search bar enables filtering of data items by keyword. Rectangles use color and pattern to encode both the category of data and the certainty of collection or sharing.}
    \Description{Detailed screenshot of the interactive summary visualization shown at the end of the scrollytelling sequence. (A) Rows represent data recipients (e.g., Service Providers, Advertisers, Authorities), ordered by how much data they receive. (B) Rectangles represent specific data types (e.g., Cookie Data, Media Content), and vary in color and pattern. Solid rectangles indicate definite collection or sharing; striped rectangles indicate conditional phrasing; shaded or ambiguous patterns mark vague practices. (C) Clicking a rectangle highlights corresponding passages in the textual policy. (D) A navigation aid appears for rectangles linked to multiple sections. (E) A search bar allows users to filter data types by keyword. This figure demonstrates how the summary view supports exploration and maintains linkages with the full policy text.}
    \label{fig:interactiveVis}
\end{figure*}

The second narrative segment details what data is collected by the policy owner. The stage fills with circles (Figure~\ref{fig:scrollytellingTool}.F), each labeled with a specific data item extracted from the policy (e.g., \textit{Approximate Location}, \textit{IP Address}, \textit{Engagement Metrics}). A central header dynamically displays the message \quotes{\textit{In total, [Platform] collects n pieces of data},} where $n$ reflects the number of individual items parsed from the policy, and \quotes{\textit{[Platform]}} is replaced with the name of the corresponding service (TikTok in the example shown in Figure~\ref{fig:scrollytellingTool}). These circles then animate into color-coded clusters (Figure~\ref{fig:scrollytellingTool}.G) representing high-level data categories (e.g., \textit{Location}, \textit{Technical Info}, \textit{Usage Data}), visually organizing the policy's contents into a structure that supports both overview and detail. This design reflects \textbf{DP2} and \textbf{DP6}, combining incremental disclosure with the ability to shift between overview and targeted exploration.

The third narrative segment visualizes the dynamics of data collection and sharing with third parties. After the full set of data items is shown in clusters, they collapse inward, visually \quotes{sucked} toward the platform's logo---signifying data ingestion (Figure~\ref{fig:scrollytellingTool}.H). The next scroll gesture triggers a caption that reads \quotes{\textit{Some data is shared or conditionally shared with others}}. This is immediately followed by the appearance of recipient icons aligned on the right side of the stage, representing actors such as \textit{Content Indexers}, \textit{Advertisers}, \textit{Analytics Providers}, \textit{Public Authorities}, and other third parties. Pieces of data then animate outward from the platform's logo toward these actors (Figure~\ref{fig:scrollytellingTool}.I). The directionality and color of these animations communicate both the flow and category of shared data, highlighting, for example, that \textit{Analytics Providers} often receive \textit{technical and usage data}, while \textit{Public Authorities} could receive identity-related items. By differentiating definite from conditional flows, this stage applies \textbf{DP4}, making uncertainty in policy wording explicit.

\subsubsection{Interactive Visualization}
\label{sect:InteractiveVisualization}

Our tool's narrative concludes in an interactive summary that consolidates the information flows introduced earlier (Figure~\ref{fig:interactiveVis}). Actors are displayed in rows, sorted in descending order by the volume of data they receive. The policy owner always appears in the first row (Figure~\ref{fig:interactiveVis}.A), since it is typically the entity that collects the largest share. The colored circles that were previously animated outward from the platform logo now morph into rectangles, each representing a specific data item. These rectangles retain the color palette from the narrative stage, visually connecting the earlier animation to the summary view.

Hovering over a rectangle reveals a tooltip with additional detail about the corresponding data item and its sharing destinations. Rectangles also vary in visual style to encode the certainty of the policy's wording. Solid colors denote data items that the policy explicitly states will be collected or shared (e.g., \quotes{\textit{We collect the content you create}} [TikTok]; \quotes{\textit{If you communicate with us, we collect your name}} [OpenAI]). Striped patterns indicate conditional phrasing (e.g., \quotes{\textit{We \textbf{may} collect or receive information about you from organisations}} [TikTok]; \quotes{\textit{the companies that host our social media pages \textbf{may} provide us with aggregate information and analytics about our social media activity}} [OpenAI]). A third style highlights ambiguous or unspecified instances where the scope or conditions of collection remain unclear (e.g., \quotes{\textit{We share \textbf{your information} directly with advertisers}} [TikTok]; \quotes{\textit{When you use our Services, we collect \textbf{Personal Data that is included in the input}}} [OpenAI]). By distinguishing between definite, conditional, and ambiguous practices, the visualization supports \textbf{DP4}.

Clicking a rectangle of the visualization scrolls the left pane to the corresponding section within the textual privacy policy and highlights it, ensuring direct traceability between the summary view and the source text (Figure~\ref{fig:interactiveVis}.C). When multiple references exist for the same data-actor pair, a navigation aid appears at the top of the policy view (Figure~\ref{fig:interactiveVis}.D), allowing users to cycle through all relevant mentions. A search bar above the visualization (Figure~\ref{fig:interactiveVis}.E) supports targeted exploration: entering a keyword (e.g., \quotes{\textit{IP address}}) highlights matching items while dimming others. These interactive features together realize \textbf{DP5} by ensuring bidirectional traceability between visualization and text, and \textbf{DP6} by balancing overview with targeted open exploration.

% Design cycles

% Participants

% Design team composition: a privacy researcher and an HCI/Vis researcher

% Workshops: 

%     two workshops with lay people who say different versions of the tool. 
%     Participants first sketched their vision on how a narrative-based tool would present a privacy policy in a more engaging way. 
%     The first group saw an initial version 

\subsection{Design Process}
\label{app:tool:design_process}

Our tool was developed through a nine-month iterative design process that combined expertise in HCI/Information Visualization and Privacy/Security. The design team met weekly to sketch, review, and refine ideas until a working prototype emerged. We began by surveying existing literature (e.g.,~\cite{Perez2018Review,Wagner2023,adhikari2023evolution,Zaeem2020,Winkler2016}) and closely reading several privacy policies (TikTok, OpenAI, Amazon, Google, Samsung) to identify structural elements that recur across providers and are suitable for narration. This comparative analysis led to an initial narrative scaffold organized around three recurring themes: \textit{where data comes from}, \textit{what data is collected}, and \textit{with whom it is shared}. Once this scaffold stabilized, we began to implement exploratory prototypes that allowed us to test candidate encodings, transitions, and see how our sketches translated to functioning elements on the screen.

{
\setlength{\fboxsep}{0pt}    % no gap between image and frame
\setlength{\fboxrule}{0.3pt} % thin black border
\begin{figure}[b]
  \centering
  % Row 1
  \begin{subfigure}[t]{0.49\columnwidth}
    \centering
    \fbox{\includegraphics[width=0.99\columnwidth]{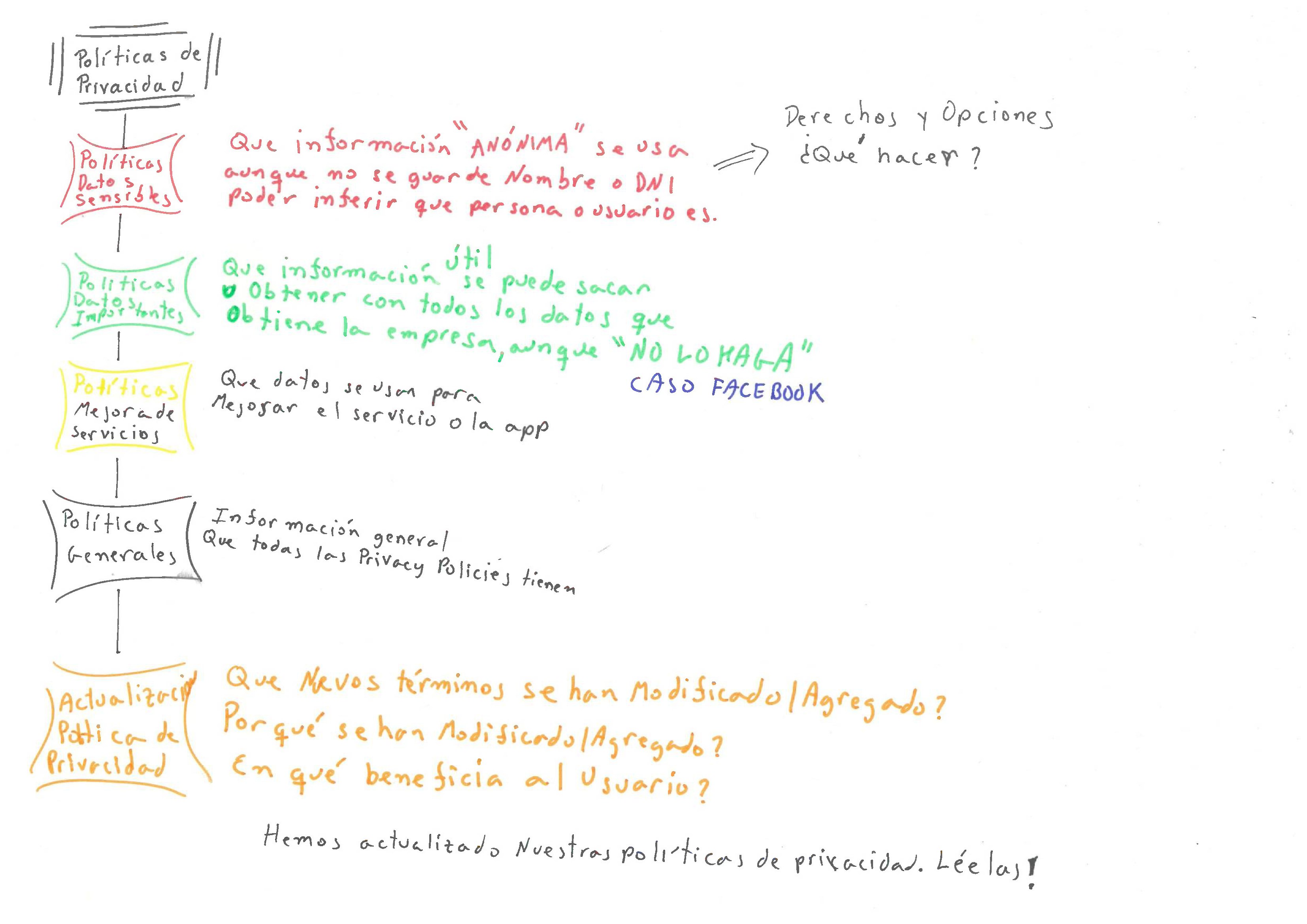}}
    \caption{Participant 1}
     \label{figure:workshop1:sketch01}
  \end{subfigure} \hfill
  \begin{subfigure}[t]{0.49\columnwidth}
    \centering
    \fbox{\includegraphics[width=0.99\columnwidth]{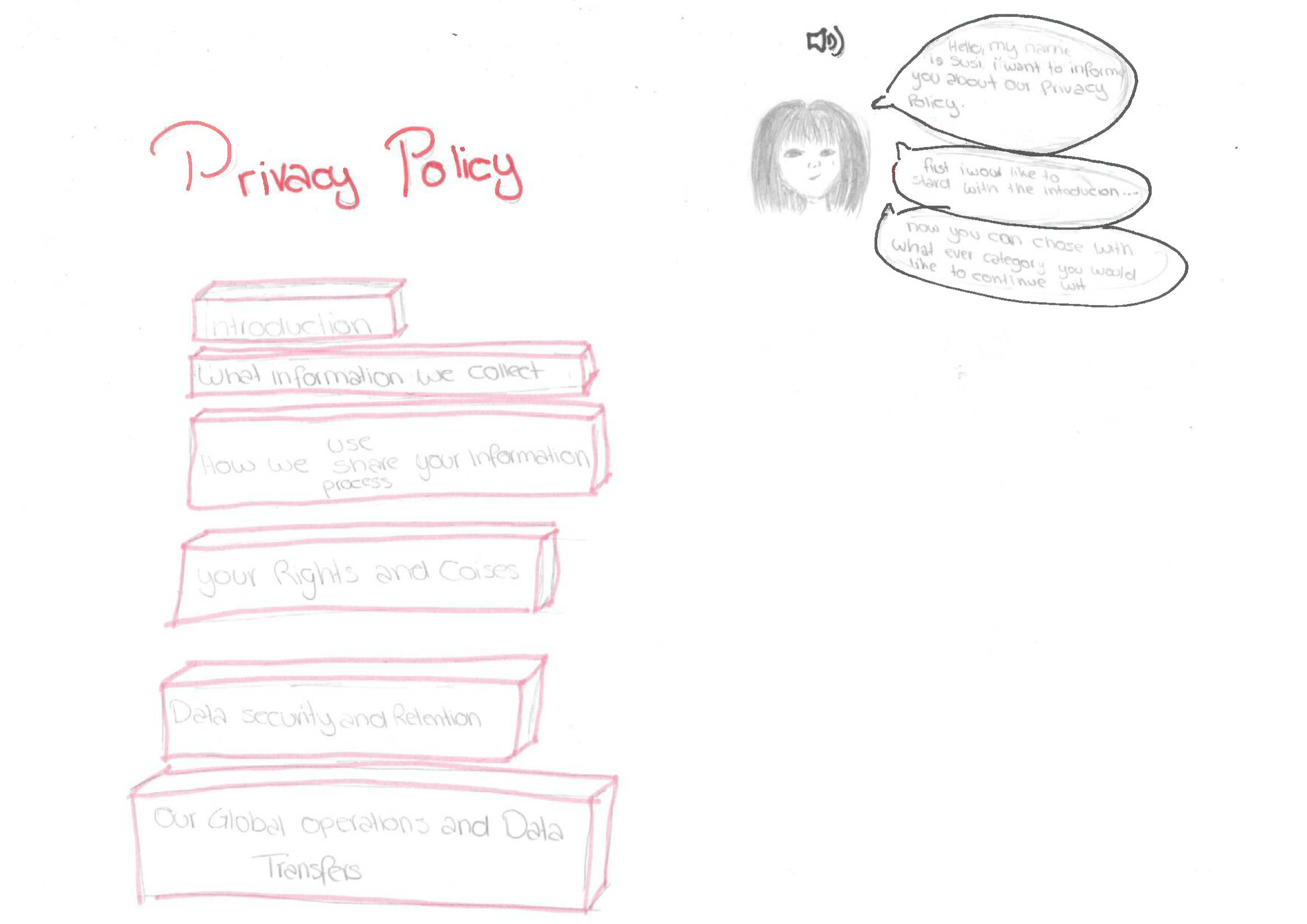}}
    \caption{Participant 2}
     \label{figure:workshop1:sketch02}
  \end{subfigure}

  \vspace{0.1in}

  % Row 3
  \begin{subfigure}[t]{0.49\columnwidth}
    \centering
    \fbox{\includegraphics[width=0.99\columnwidth]{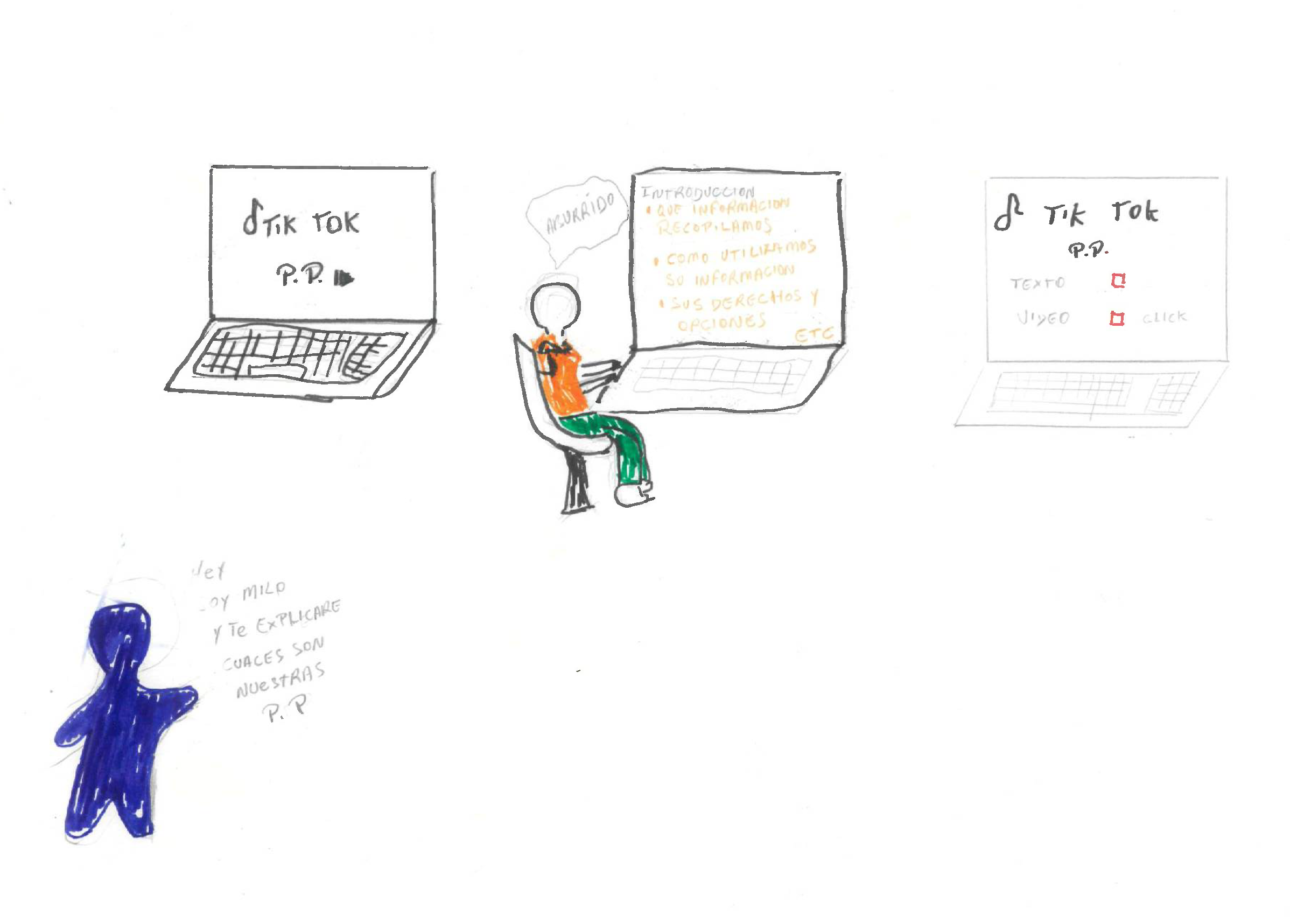}}
    \caption{Participant 3}
     \label{figure:workshop1:sketch03}
  \end{subfigure} \hfill
  \begin{subfigure}[t]{0.49\columnwidth}
    \centering
    \fbox{\includegraphics[width=0.99\columnwidth]{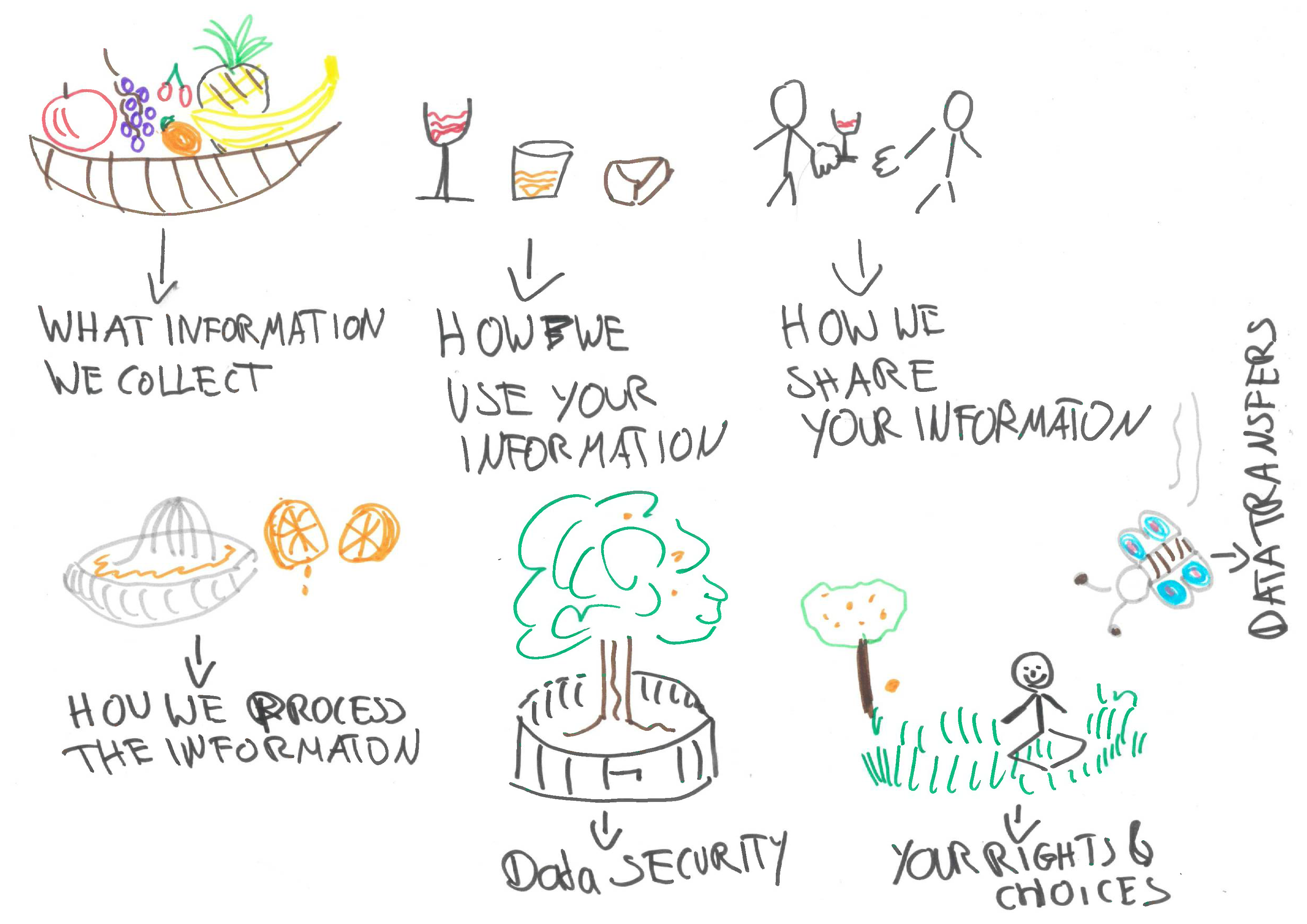}}
    \caption{Participant 4}
     \label{figure:workshop1:sketch04}
  \end{subfigure}

  \caption{Sketches from the first participatory design workshop. Without being primed on narrative structures, participants explored diverse ways of visually communicating privacy policy content. Ideas included hierarchical outlines, metaphors such as food and gardening, conversational formats, and user prompts. Some focused on summarizing key sections (a, d), while others proposed interaction scenarios or analogies for how data is used and shared (b, c). These early sketches helped us surface user expectations, preferences, and concerns, which informed the overall framing and scope of our tool.}
    \Description{This figure displays four hand-drawn sketches produced by participants during the first participatory design workshop, each offering a different visual approach to communicating privacy policy content. Participant 1 (a) created a color-coded list in Spanish with thematic sections and reflective questions about data use, rights, and user benefit. Participant 2 (b) illustrated a hierarchical stack of labeled policy sections, paired with a drawn character expressing doubts through thought bubbles. Participant 3 (c) depicted a user interacting with two laptops labeled TikTok, highlighting a scenario of information access or comparison, with an additional character expressing uncertainty. Participant 4 (d) used food and social metaphors (e.g., a fruit bowl, drinks, a tree) to represent stages of data handling—from collection to processing and sharing—along with references to user rights and security. These varied sketches surfaced user expectations, metaphors, and concerns, informing the design and framing of the final tool.}
  \label{fig:workshop1:sketches}
\end{figure}
}

\subsubsection{Co-Design Workshops}

Our tool design was also shaped by two participatory design workshops we conducted to assess two different early prototypes of the tool. In both workshops, participants first read TikTok's full privacy policy to ground their understanding before proceeding to sketch possible visual or narrative representations.

\begin{figure}[b!]
  \centering

  % Row 1
  \begin{subfigure}[t]{0.49\columnwidth}
    \centering
    \includegraphics[width=0.99\columnwidth]{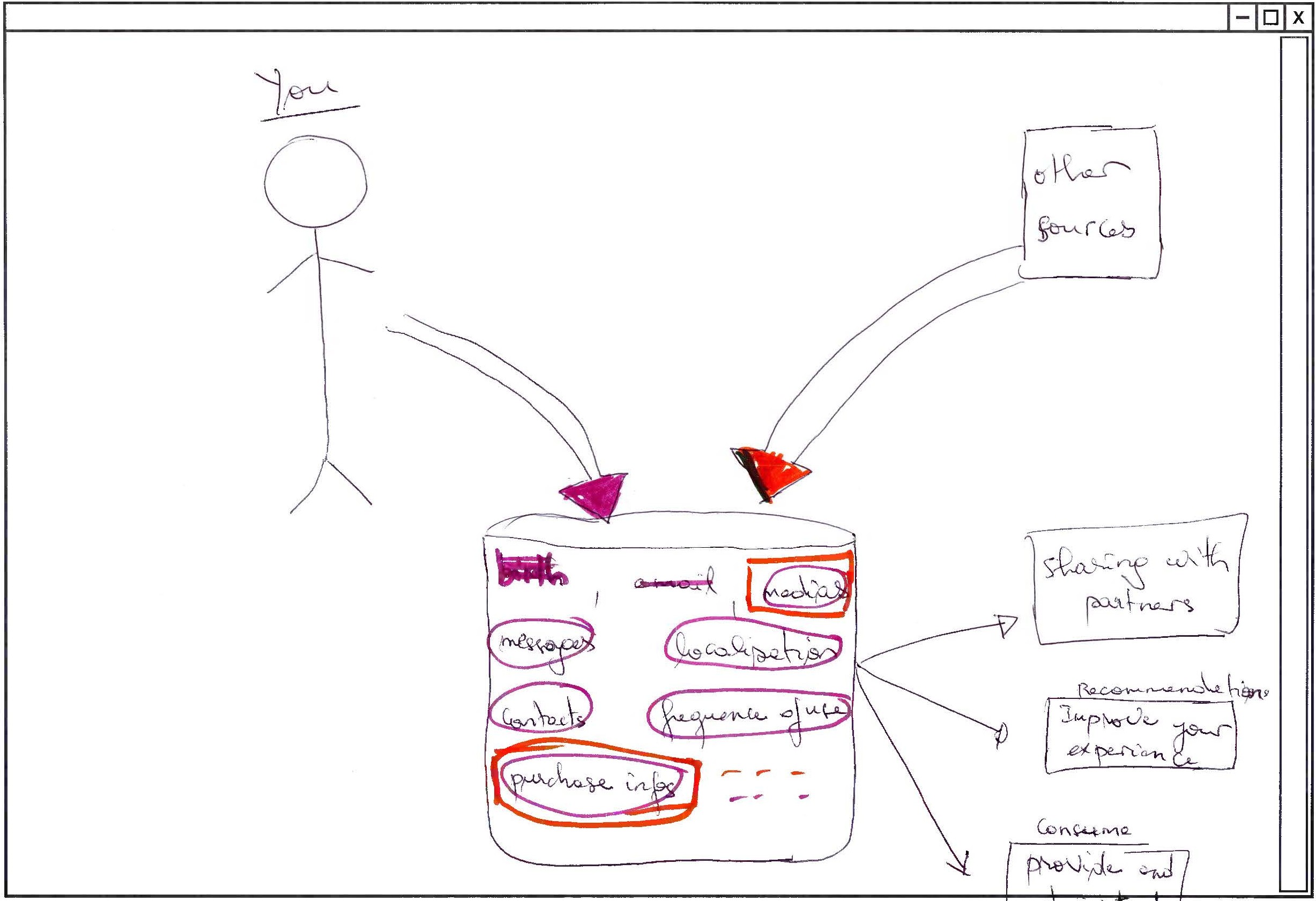}
    \caption{Participant 1}
     \label{figure:workshop2:sketch01}
  \end{subfigure} \hfill
  \begin{subfigure}[t]{0.49\columnwidth}
    \centering
    \includegraphics[width=0.99\columnwidth]{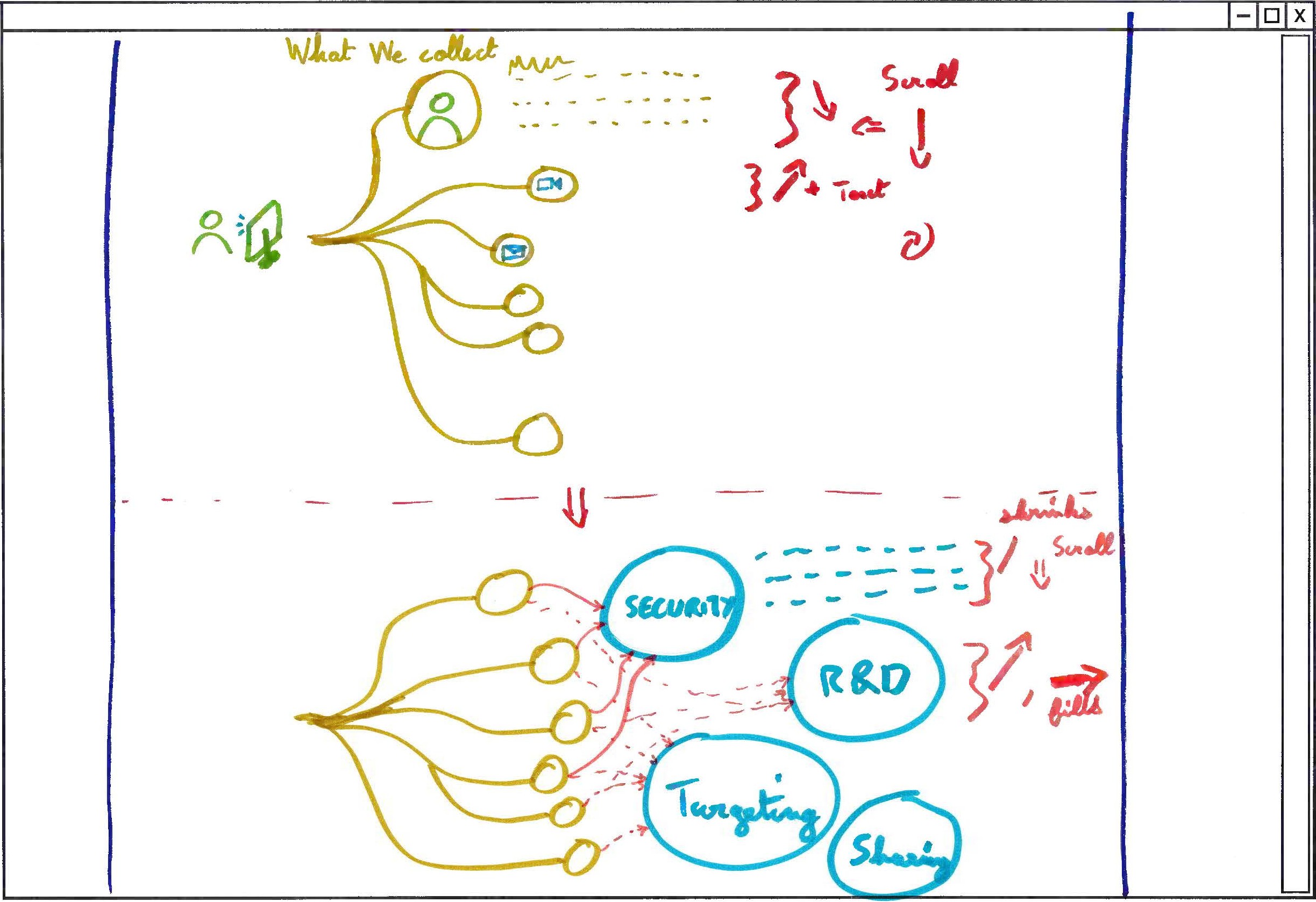}
    \caption{Participant 2}
     \label{figure:workshop2:sketch02}
  \end{subfigure}

  \vspace{0.1in}

  % Row 3
  \begin{subfigure}[t]{0.49\columnwidth}
    \centering
    \includegraphics[width=0.99\columnwidth]{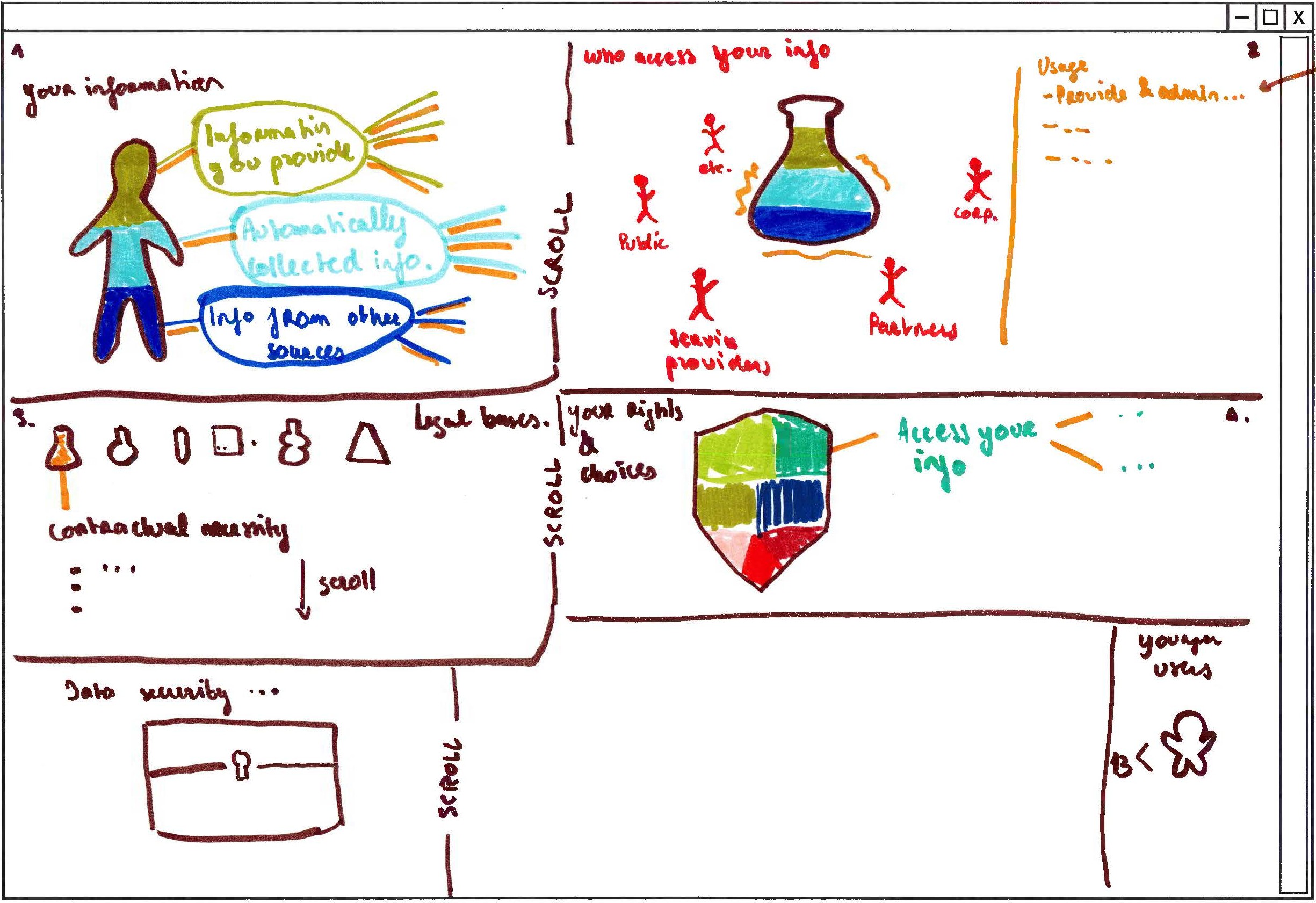}
    \caption{Participant 3}
     \label{figure:workshop2:sketch03}
  \end{subfigure} \hfill
  \begin{subfigure}[t]{0.49\columnwidth}
    \centering
    \includegraphics[width=0.99\columnwidth]{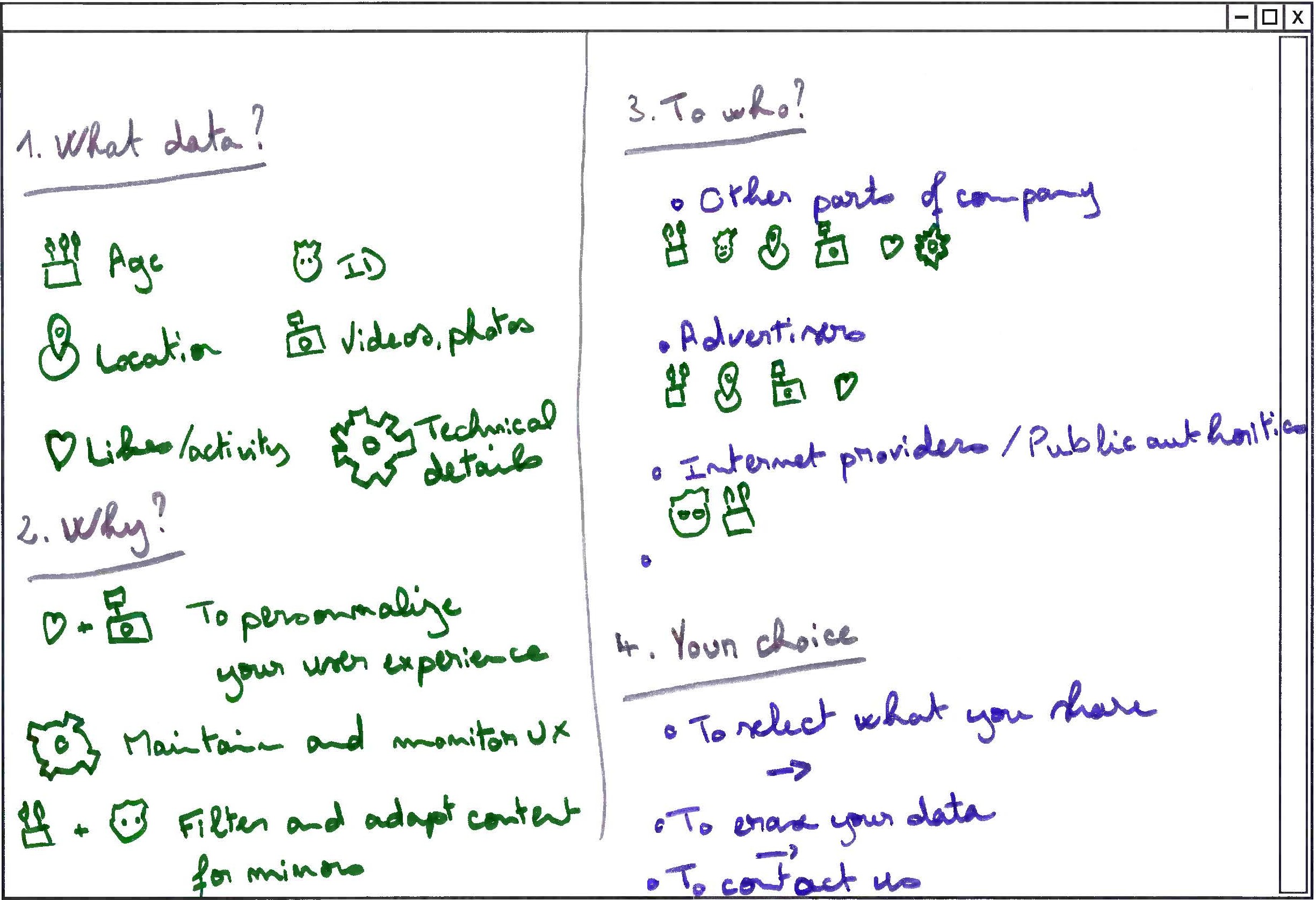}
    \caption{Participant 4}
     \label{figure:workshop2:sketch04}
  \end{subfigure}  

  \caption{Sketches from the second participatory design workshop. Participants were primed with examples of scrollytelling narratives and asked to sketch how privacy policies might be explained using similar formats. Three of the four sketches are centered around a user avatar.}
\Description{This figure presents four participant-generated sketches from the second participatory design workshop, where participants were primed with examples of scroll-driven storytelling and asked to envision how privacy policies could be communicated in similar narrative formats. Participant 1 (a) drew a stick-figure user linked via arrows to a layered container diagram representing different categories of data use, connected to external parties through labeled flows. Participant 2 (b) illustrated branching tree-like structures showing stages of data collection and use, with outputs labeled as Targeting, R&D, and Sharing, visually conveying how data flows through decision paths. Participant 3 (c) combined a user avatar with vivid icons like potion flasks, pie charts, and sliders to illustrate the transformation and analysis of user data, set within a structured layout that suggests narrative progression. Participant 4 (d) used a list-based layout organized into sections answering What data?, Why?, and To whom?, using colored icons and bullet points to simulate scrollytelling beats. Three of the four sketches are centered around a user figure, highlighting narrative framing as a core design strategy.}
  \label{fig:workshop2:sketches}
\end{figure}

\paragraph{First Workshop}

It included two male and two female participants, aged 20, 28, 36, and 40 years. Two participants reported postgraduate degrees, one had completed secondary education, and one had vocational training. At the time of the workshop, one participant was employed full-time, one was unemployed, and two were students. Their areas of work or study varied: two participants were in information technology (one of whom also indicated education), one in health sciences, and one in jewelry making.

Three participants had previously read privacy policies, while one had not. Self-rated understanding of privacy policies was measured on a 5-point Likert scale from \textit{very low} to \textit{very high}; two participants selected \textit{low}, one \textit{medium}, and one \textit{high}. Reported time typically spent reading a privacy policy varied: one participant indicated 1--5 minutes, one 5--10 minutes, one more than 10 minutes, and one reported usually not reading them.

Participants were asked to sketch how privacy policies might be communicated visually. The presented ideas included a compact visual summary with a traffic-light scheme to signal the sensitivity of different data items (Figure~\ref{figure:workshop1:sketch01}), two conversational agents (Figures~\ref{figure:workshop1:sketch02} and~\ref{figure:workshop1:sketch03}), and a \quotes{data garden} metaphor (Figure~\ref{figure:workshop1:sketch03}), where data pieces were depicted as seeds and fruits. This last sketch also addressed broader topics (e.g., data transformation, user rights) that we had decided not to include in our narrative. After presenting and discussing their sketches, participants interacted with an early prototype instantiated with TikTok's policy. The subsequent group discussion revealed two issues: some participants believed the narrative ended before the final visualization, prompting stronger orientation cues with a progress indicator and animated scroll prompts (\textbf{DP3}); and no one used the search function, which at the time was too inconspicuous, leading us to redesign its placement to better support targeted exploration (\textbf{DP6}).

\paragraph{Second Workshop} It followed the same sketch-present-discuss format, but this time participants were first introduced to the concept of scrollytelling and shown examples of well-known implementations (\cite{Bloomberg2015USAutoSales, FlaggCraigBruno2014_CaliforniasGettingFracked, ChowBestCollegeMajors}). This priming encouraged them to think directly in scrollytelling terms. 

The group included two male and two female participants, aged 22, 24, 25, and 26 years. All four reported postgraduate education in information technology. All participants were PhD students in Artificial Intelligence or Machine Learning, and they were members of a research lab at another university. Three had previously read privacy policies on websites or apps. On the same 5-point self-report scale used in the first workshop, all four rated their understanding of privacy policies as \textit{medium}. Two participants reported spending less than one minute on average reading a privacy policy, while the other two reported 1--5 minutes.

Three of the four sketches proposed in this workshop centered on a user avatar (Figures~\ref{figure:workshop2:sketch01}---\ref{figure:workshop2:sketch03}), a direction we had begun to explore between workshops. The recurrence of this motif validated our decision to keep the user as a focal anchor in the narrative, aligning with incremental introduction of concepts (\textbf{DP2}) while supporting later, focused inquiry into user-specific data practices (\textbf{DP6}). This workshop's fourth sketch (Figure~\ref{figure:workshop2:sketch04}) reinforced our decisions on data categorization and visual encodings.

\subsubsection{Expert Critique Sessions}

Additional design feedback came from two critique sessions with external experts. A male usability \& UX specialist from industry emphasized the importance of deferring interactive elements until after the narrative, to avoid drawing attention away from the narrative's unfolding logic. This insight reinforced our decision to reserve most interaction (search, filtering, and inspection of policy-linked data flows) for the final stage of the tool. There, users can freely explore connections to the full policy text without fragmenting the narrative sequence, aligning with our goals of maintaining clarity (\textbf{DP2}) and promoting traceable exploration (\textbf{DP5}, \textbf{DP6}).

The second critique session was with a female communication expert from academia, who helped us shape the accessibility of the language used throughout the tool. Their suggestions led us to refine textual phrasing, tighten connections between visual elements and their explanatory text, and ensure that the narrative style remained approachable without oversimplifying. Together, these critiques reinforced our design decisions to stage the narrative in discrete, scroll-driven steps while keeping the full policy visible for verification, directly supporting \textbf{DP1}, \textbf{DP2}, and \textbf{DP5}.

\subsection{Technical Choices and Implementation Details}
\label{sec:Implementation}

We now outline key technical choices underlying our tool, including decisions about the granularity of narrative elements, the rendering pipeline and graph construction process, and the technologies used to implement the interface.

\subsubsection{Granularity of Narrative Elements}

% Early prototypes showed that representing each data field and recipient separately overloaded the display, so we aggregated them into higher-level data categories and actor classes to balance progressive disclosure with overview (\textbf{DP2}, \textbf{DP6}) while preserving traceability to the full text (\textbf{DP1}, \textbf{DP5}). 

A central technical decision concerned the granularity of policy elements, both for data and for actors. Our early prototypes revealed that representing the items mentioned in a privacy policy (e.g., \textit{email}, \textit{age}, \textit{address}) in an atomic way, or listing every individual recipient entity (e.g., individual advertising firms, analytics providers, or affiliated brands), quickly overloaded the display. To address this, we decided to group related elements into higher-level data categories (e.g., \textit{Identity data}) and more abstract actor classes (e.g., \textit{Authorities}). This approach reduced visual clutter while maintaining conceptual fidelity and structuring detail so that it could be unpacked when needed. It also balanced incremental complexity (\textbf{DP2}) with the need for an overview users could explore more deeply (\textbf{DP6}), while maintaining traceability to the original policy text (\textbf{DP1}, \textbf{DP5}).

\subsubsection{Rendering Pipeline}

Our tool generates the scrollytelling narrative algorithmically from two structured inputs. In our current implementation, these inputs were manually crafted to ensure fidelity to the source policy, although they could, in principle, be produced through automated or semi-automated means (see also Section~\ref{sec:TowardsAdoption} in the Discussion).

The first input is a lightweight configuration file that specifies the policy's data-source facets (e.g., user-provided, automatically collected, third-party, inferred), their display order, and scroll anchors into the full policy text. This file also specifies user-facing labels, icons, and the groupings that organize individual data items and third-party actors into higher-level categories. The elements specified in the configuration file drive the three segments of the explanatory narrative described in Section~\ref{sec:ExplanatoryNarrative}.

The second input encodes the policy's data collection and sharing practices as a directed graph. In this file, each node represents a data item or actor, and each directed edge corresponds to statements about collection or sharing (i.e., instances where a specific piece of data is either collected by the policy owner or shared with a third party). Each edge of the graph includes metadata specifying whether the practice is conditional or ambiguous, as well as a list of verbatim quotes from the policy text where the collection or sharing is described. These quotes are used to synchronize the scrolling and highlighting in the policy pane (e.g., when a rectangle representing a data piece is clicked), and to support iteration when the same data-actor pair appears multiple times in the policy (as depicted in Figure~\ref{fig:interactiveVis}.D). The data collection and sharing relationships specified in the graph file are used to produce the interactive visualization described in Section~\ref{sect:InteractiveVisualization}.

\subsubsection{Collection \& Sharing Graph Construction}

During the initial stages of our design process, we explored how much of the process of extracting and representing actors and data items from the policies could be automated. To this end, we experimented with PolyGraph~\cite{Cui2023PoliGraph}, which uses natural language processing (NLP) to generate graph representations of what a policy collects, shares, and with whom. PolyGraph, however, treated section headers (e.g., \textit{Information you provide}) as atomic nodes, so when the text later stated something along the lines of \quotes{\textit{We may share information you provide with third parties},} the graph referenced only the header rather than the many concrete data items defined inside that section (e.g., \textit{name}, \textit{email}, \textit{phone number}). Moreover, PolyGraph's output provided no way of representing such containment relationships, meaning that references to a section could not be expanded to the set of data items it encompassed. This systematically underestimated what was actually being shared. 

Thus, to preserve fidelity and traceability (\textbf{DP1}, \textbf{DP5}), we abandoned full automation and adopted a hybrid approach: for a given privacy policy, we manually identified and encoded all necessary elements---data-source facets, actors, data types, groupings, and data collection and sharing relationships---based on close reading of the source text. The tool then renders the narrative algorithmically from the corresponding inputs, ensuring consistent execution while grounding the representation in a qualitative, carefully curated analysis similar in spirit to a thematic analysis of qualitative data. We instantiated this hybrid pipeline for the two privacy policies evaluated in our study.

% We followed this approach with the two policies used in our study 

% we decided to adopt a hybrid pipeline: we manually coded actors and data for two real privacy policies (TikTok and OpenAI) and built the tool to render the narrative algorithmically.

% For a given privacy policy, we manually identified all relevant elements for both files, ensuring that every visual or interactive element in the tool is grounded in the source text. 

\subsubsection{Implementation}

The tool itself was implemented using standard web technologies. We used D3.js\footnote{\url{https://d3js.org}} for rendering data-driven visual elements, and GSAP\footnote{\url{https://gsap.com}} (GreenSock Animation Platform) to choreograph the animated transitions throughout the narrative. All other functionality was built using vanilla JavaScript and HTML.

% For TikTok, this yielded four data sources, each tied to a corresponding section; for OpenAI, which does not mention inference, only three are defined.

% This grounded approach ensures the narrative reflects what is explicitly stated, without introducing unsupported categories. While facet detection could be assisted by NLP in future versions, we treat human verification as essential for transparency and fidelity.

%%%% Sections/040-Tool.tex ends here %%%%

%%%% 050-Study.tex starts here %%%%

\section{Study}
\label{sec:study}

\begin{figure}[b]
    \centering
    \includegraphics[width=\columnwidth]{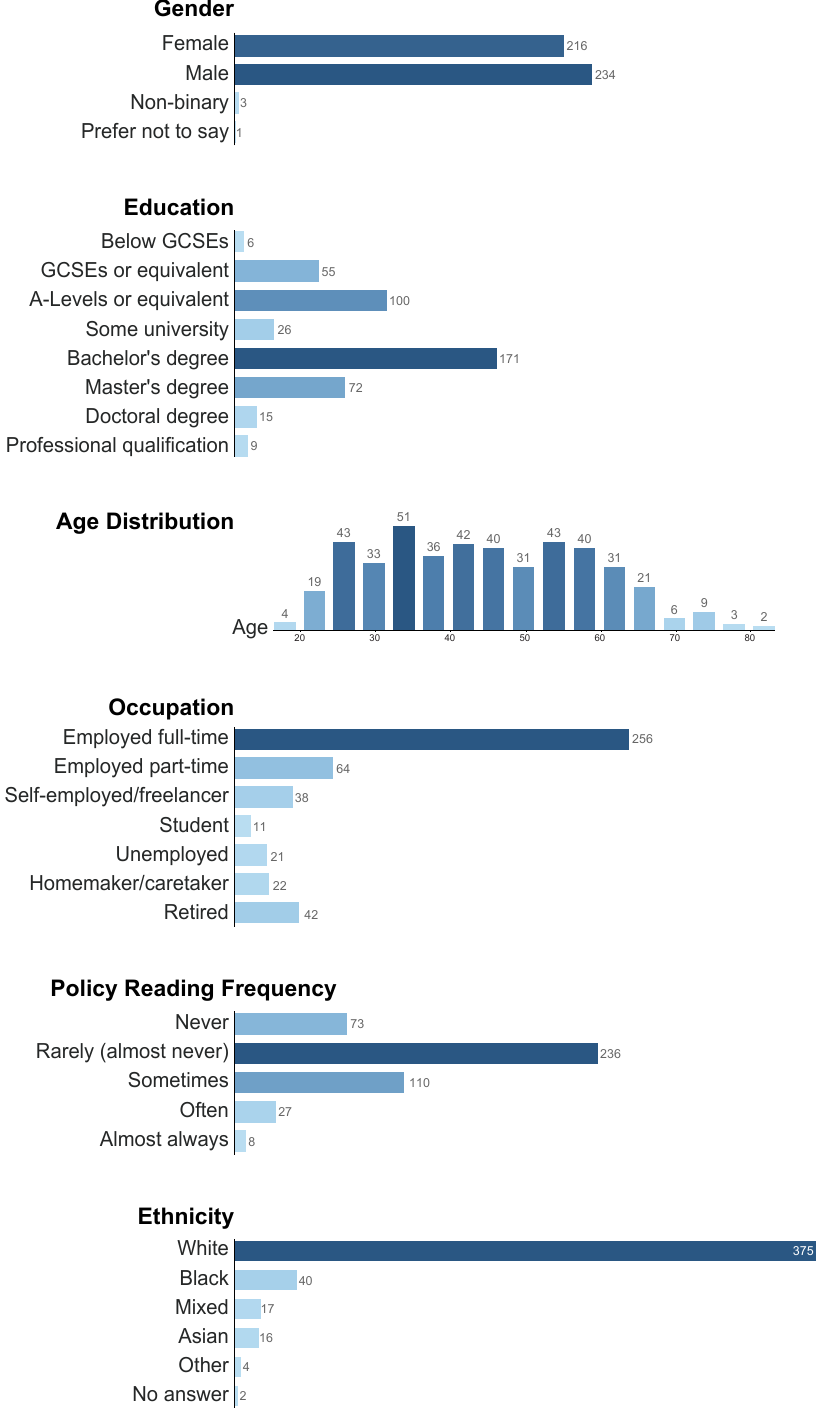}
    \caption{Demographic makeup of our study participants.}
    \label{fig:demographics}
    \Description{Demographic summary of participants in bar chart form. Multiple subplots show: Gender (with majority female and male, few non-binary or prefer-not-to-say); Education (ranging from below GCSE to doctoral degree); Occupation (e.g., full-time employed, student, unemployed); Frequency of reading privacy policies (ranging from never to almost always); Ethnicity (majority White, followed by Asian, Black, Mixed, and Other); Age distribution (shown as a histogram with most participants between 20 and 50). Each category is clearly labeled, with counts shown on a horizontal axis.}
\end{figure}

We conducted an online experiment to compare the effectiveness of \textit{scrollytelling} against four privacy policy formats in terms of user comprehension, perceptions, and subjective experience. Each participant reviewed a single privacy policy presented in one format and completed a series of questionnaires.

% We examined how format influenced participants' responses and, in some cases, whether effects varied by policy owner.

We powered the study for the \textit{Format} main effect in a $5{\times}2$ between-subjects design, including \textit{Owner} as a fixed effect (additive adjustment). Monte Carlo simulations that reproduced the fixed-effect structure of the planned analyses, with two-sided $\alpha=.05$, indicated that \emph{approximately $n\approx45$ per \textit{Format}$\times$\textit{Owner} cell} (analyzable $N\approx450$) provides $\ge .95$ power for the confirmatory \textit{Format} main effects (Appendix~\ref{app:samplesize}). We do not include interactions in all analyses: smaller per-cell $n$ leaves \textit{Format}$\times$\textit{Owner} interactions underpowered. As such, we treat them as exploratory and, where relevant, report them separately. To accommodate anticipated exclusions of $15$--$30\%$ due to inattentiveness or failed attention checks~\cite{Peer2017,Palan2018,Chmielewski2020}, we set a recruitment target of $N{=}600$ (60 people per cell).

\subsection{Recruitment}

We recruited 600 participants through Prolific, restricting eligibility to adults residing in the UK, fluent in English, and with a 95--100\% approval rating to support data quality. UK-based recruitment was chosen to minimize language-related comprehension confounds, as all study materials---including the platform-specific privacy policies---were presented in English and reflected the UK legal context. In addition, Prolific’s UK participant pool is widely regarded as stable and reliable in terms of engagement and data quality~\cite{Peer2017,Palan2018}.

To maintain experimental integrity, we created ten separate Prolific studies---one for each cell (format/policy combination). We used Prolific prescreeners to ensure participants could enroll in only one condition and to automatically exclude those who participated in any prior pilot studies (see Appendix~\ref{app:pilots}). Participants received £2.25 for completing the 18-minute study (approximately £7.50/hour), a rate consistent with Prolific's compensation guidelines. The demographic makeup of our participants and self-assessed familiarity with privacy policies are summarized in Figure~\ref{fig:demographics}.

\subsection{Procedure}
\label{sec:Procedure}

Participants completed the study via a custom web-based wizard hosted on our institutional servers. The interface first presented an introductory page explaining the study's purpose, data collection procedures, and compensation terms.\\

After providing explicit informed consent, the interface guided them through the following sequential steps:

\aptLtoX{\begin{itemize}
    \item[\textbf{Step 1:}] \textbf{Background Questionnaire.} Participants provided demographic information and reported their prior experience with privacy policies, including frequency of reading and confidence in explaining them.
    
    \item[\textbf{Step 2:}] \textbf{Pre-Perception Questionnaire.} This section assessed baseline perceptions of the assigned platform's data practices. Items included frequency of platform use, familiarity with data practices, perceived transparency, trust, appropriateness of data collection amounts, and third-party data sharing.
    
    \item[\textbf{Step 3:}] \textbf{Privacy Policy Format Exposure.} Participants then reviewed their assigned policy, shown in a new browser window. To discourage superficial engagement, our custom web interface enforced a 90-second minimum review period. During this time, a countdown timer and instructional message prompted participants to engage with the policy content. The study wizard allowed advancing to the next step only once the timer had elapsed.

    \item[\textbf{Step 4:}] \textbf{Comprehension Questionnaire.} Participants then answered eight true/false statements designed to assess their understanding of the policy content. The questionnaire included four \textit{factual} and four \textit{interpretive} items. Factual items could be answered by referring to a specific, clearly stated part of the policy (e.g., whether the platform collects location data). Interpretive items required reasoning about implications or synthesizing information across multiple sections (e.g., whether law enforcement agencies can access all of a user's data). Each true/false statement was followed by a confidence rating on a 5-point Likert scale ranging from \quotes{\textit{Not at all confident}} to \quotes{\textit{Extremely confident}}, capturing participants' assessment of their own comprehension accuracy. Instructions explicitly directed participants to use the assigned policy format to answer each statement, not to rely on their memory. Although items were identically worded for both privacy policies, correct answers and difficulty could vary by policy owner (e.g., the same statement could be true for \textit{OpenAI} and false for \textit{TikTok}).
    
    \item[\textbf{Step 5:}] \textbf{Experience Questionnaire.} Participants evaluated the usability and overall subjective experience of the policy format. This questionnaire assessed six constructs: perceived clarity, cognitive load, engagement, enjoyment, behavioral intentions, and likelihood of adopting the format in future interactions. An open-ended prompt also invited participants to reflect on their experience in their own words, requiring a minimum of 50 characters.
    
    \item[\textbf{Step 6:}] \textbf{Post-Perception Questionnaire.} Participants then reassessed their perceptions of the platform's data practices, using the same constructs as in the pre-perception questionnaire of Step~2.    
\end{itemize}}{\begin{enumerate}[label=\textbf{Step~\arabic*:}, leftmargin=*, itemsep=0em, wide]

    \item \textbf{Background Questionnaire.} Participants provided demographic information and reported their prior experience with privacy policies, including frequency of reading and confidence in explaining them.
    
    \item \textbf{Pre-Perception Questionnaire.} This section assessed baseline perceptions of the assigned platform's data practices. Items included frequency of platform use, familiarity with data practices, perceived transparency, trust, appropriateness of data collection amounts, and third-party data sharing.
    
    \item \textbf{Privacy Policy Format Exposure.} Participants then reviewed their assigned policy, shown in a new browser window. To discourage superficial engagement, our custom web interface enforced a 90-second minimum review period. During this time, a countdown timer and instructional message prompted participants to engage with the policy content. The study wizard allowed advancing to the next step only once the timer had elapsed.

    \item \textbf{Comprehension Questionnaire.} Participants then answered eight true/false statements designed to assess their understanding of the policy content. The questionnaire included four \textit{factual} and four \textit{interpretive} items. Factual items could be answered by referring to a specific, clearly stated part of the policy (e.g., whether the platform collects location data). Interpretive items required reasoning about implications or synthesizing information across multiple sections (e.g., whether law enforcement agencies can access all of a user's data). Each true/false statement was followed by a confidence rating on a 5-point Likert scale ranging from \quotes{\textit{Not at all confident}} to \quotes{\textit{Extremely confident}}, capturing participants' assessment of their own comprehension accuracy. Instructions explicitly directed participants to use the assigned policy format to answer each statement, not to rely on their memory. Although items were identically worded for both privacy policies, correct answers and difficulty could vary by policy owner (e.g., the same statement could be true for \textit{OpenAI} and false for \textit{TikTok}).
    
    \item \textbf{Experience Questionnaire.} Participants evaluated the usability and overall subjective experience of the policy format. This questionnaire assessed six constructs: perceived clarity, cognitive load, engagement, enjoyment, behavioral intentions, and likelihood of adopting the format in future interactions. An open-ended prompt also invited participants to reflect on their experience in their own words, requiring a minimum of 50 characters.
    
    \item \textbf{Post-Perception Questionnaire.} Participants then reassessed their perceptions of the platform's data practices, using the same constructs as in the pre-perception questionnaire of Step~2.    
\end{enumerate}}

All questionnaire items were adapted from established, validated instruments and, when appropriate, reworded to fit the context of privacy policy reading while preserving the intended constructs. Engagement items were inspired by the short form of the User Engagement Scale~\cite{obrien2018ues}; enjoyment from the interest/enjoyment subscale of the Intrinsic Motivation Inventory~\cite{mcauley1989imi}; cognitive load from a multi-item cognitive load scale~\cite{leppink2013cls}; and perceived clarity from clarity and understandability in the information quality and transparency literature~\cite{wixom2005integration,schnackenberg2021transparency}. Items for format adoption and behavioral intention were based on intention-to-use measures from TAM and UTAUT~\cite{venkatesh2003utaut,venkatesh2012utaut2}, while trust and perceived transparency drew on scales in trust and transparency research~\cite{mcknight2002trust,schnackenberg2021transparency}. All questionnaires are available in the Supplementary Materials.

% Full item wordings are provided in Appendix~\ref{app:questionnaires}.

To ensure reliable measurement, the questionnaires in Steps 4--6 included both direct and reverse-coded Likert-scale items, with item order randomized for each participant to mitigate ordering effects. Each of these questionnaires also embedded one attention check to help identify low-effort or disengaged responses. Additionally, in all three steps, the interface provided a link to reopen the policy format (in case participants had closed it before answering). 

Before the main study, we conducted multiple pilot runs to refine the interface, adjust task timing, and improve item wording. Full details about our pilot studies are reported in Appendix~\ref{app:pilots}.

% The complete list of constructs and questionnaire items used in the study is provided in Section~\ref{app:questionnaires} of the Appendix.

\subsection{Privacy Policies and Policy Formats}

Our study focused on the European privacy policy of OpenAI (effective February 15, 2024) and TikTok's EEA/UK/CH privacy policy (effective November 19, 2023). We selected these documents for three key reasons: a) both platforms enjoy broad recognition across age groups and socioeconomic backgrounds; b) they represent fundamentally different service types with contrasting data practices; and c) their substantial length disparity: OpenAI's concise policy ($\approx2894$ words) compared to TikTok's longer, more comprehensive document ($\approx6544$ words). Selecting these policies allowed us to assess format effectiveness across widely used services that differ in privacy context, complexity, and informational density, factors that are central to the challenges of effective privacy communication.\\

\noindent{We evaluated five distinct privacy policy formats, each providing a different approach to privacy communication:}\\

\begin{figure*}[hb!]
    \centering
    \includegraphics[width=0.999\linewidth]{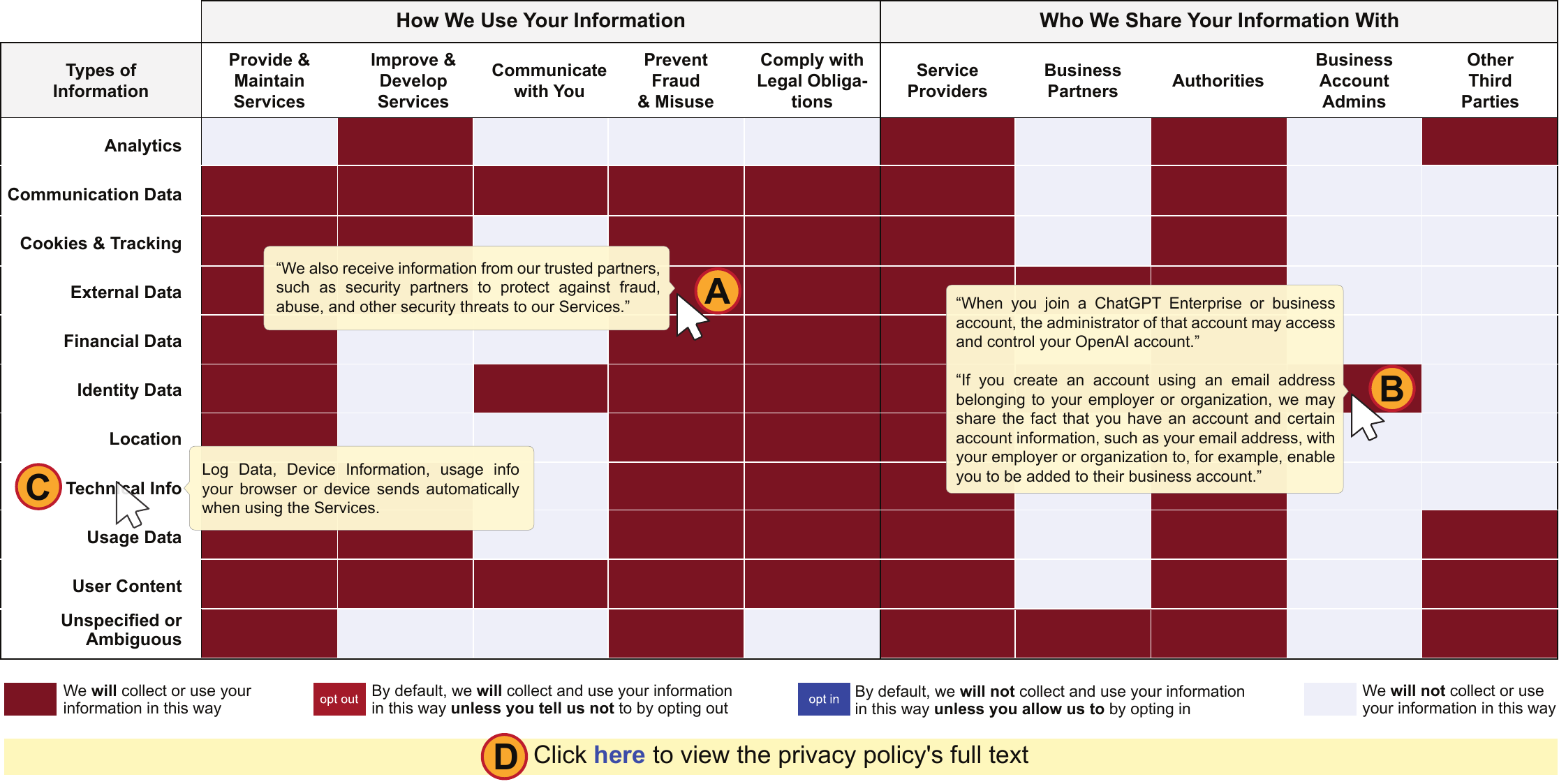}
    \caption{Interactive Nutrition Label we developed. The table presents policy information in a matrix of data types (rows) and purposes or recipients (columns). Cell colors indicate whether collection or disclosure occurs. Hovering over a cell (A \& B) reveals verbatim policy excerpts that explain the specific practice, while hovering over a data type (C) displays the individual data elements included in that category. In the \quotes{nutrition + text} condition, a direct link to the full privacy policy (D) was provided below the table, allowing participants to consult the complete document.}
    \Description{Interactive nutrition label format we implemented. The label is structured as a matrix, with rows representing data categories (e.g., Identity Data, Technical Info) and columns representing purposes or recipients (e.g., Analytics, Fraud Prevention, Legal Obligations). (A and B) Hovering over a cell displays a tooltip with an excerpt from the policy explaining that specific data-use pairing. (C) Hovering over a row reveals the individual data elements included in that category. (D) A link to the full policy text is placed below the table. Cell background color indicates whether data is always collected, optionally collected, or never collected.}
    \label{fig:nutritionlabel}
\end{figure*}

\textbf{Textual Format:} This condition presented the privacy policy as it appears on the official website of the respective platform, with no substantive alterations to content, structure, or language. It closely matched real-world exposure, serving as a control condition for comparison with the other formats.

\textbf{Interactive Nutrition Label:} This format presented the policy as a matrix-style table, with rows representing data types/categories and columns detailing usage or disclosure (see Figure~\ref{fig:nutritionlabel}). The nutrition label structure was inspired by prior work on privacy nutrition labels~\cite{Kelley2009Nutrition, Kelley2010Standardizing, KelleyAppDecision}, while the decision to incorporate interactivity was based on findings by Reinhardt et al.~\cite{ReinhardtInteractivePP}. In our implementation of this format, users could hover over each cell to reveal a tooltip containing verbatim policy excerpts that explained the relationship between the intersecting row and column (Figure~\ref{fig:nutritionlabel}.A \& B). Hovering over a data type displays the individual data elements included in that category (Figure~\ref{fig:nutritionlabel}.C).

% This design supported at-a-glance exploration of what data is collected and for what purposes, while enhancing accessibility and comprehension through interactivity.

\textbf{Interactive Nutrition Label + Link to Textual Policy:} This condition was identical to the nutrition label format, but also included a link below the table to the full textual privacy policy (Figure~\ref{fig:nutritionlabel}.D). Clicking the link opened the official policy in a new tab, providing participants with access to the complete document if needed. The policy content was identical to that used in the textual condition. This format supported tabular-based exploration while allowing optional access to the full policy, without embedding it within the main interface.

\textbf{Scrollytelling:} In this condition, the privacy policy was presented using the tool described in Section~\ref{sec:system:scrollytelling}, which progressively revealed content through a scroll-driven narrative. The narrative concluded with the interactive visualization depicted in Figure~\ref{fig:interactiveVis} persistently displayed alongside the text, allowing participants to explore data types and access corresponding policy excerpts by interacting with visual elements. Thus, this format combined guided narration with opportunities for user-driven exploration.

\textbf{Interactive Visualization Only:} This condition presented the privacy policy only through the interactive visualization shown in Figure~\ref{fig:interactiveVis} (omitting the preceding scrollytelling sequence). That is, participants directly interacted with the visual grid of data types/categories (right side), while the associated policy references appeared in a panel on the left. We decided to introduce this format to isolate the effects of visual exploration without narrative guidance or progressive disclosure.\\

All participant responses were collected through our custom web-based study wizard, which automatically saved data upon completion of each section. In addition to questionnaire responses, the system logged detailed metadata per participant, including item-level timestamps and randomized question order. For participants assigned to the \textit{Nutrition Label + Text} format, the wizard also recorded whether the full privacy policy text was opened.

% Progress was saved locally to allow participants to resume incomplete sessions.

% 

% All data processing and statistical analyses were performed in Python, and all code and anonymized data necessary for replication are available in the Supplemental Materials. 

%%%% Sections/050-Study.tex ends here %%%%

%%%% 060-Results.tex starts here %%%%

\section{Results}
\label{sec:results}

% Original number of participants: 597
% Discarded for not passing attention checks: 49
% Discarded for being suspicious: 100
% Discarded for both reasons: 6

\begin{figure}[t!]
    \centering
    \includegraphics[width=1\columnwidth]{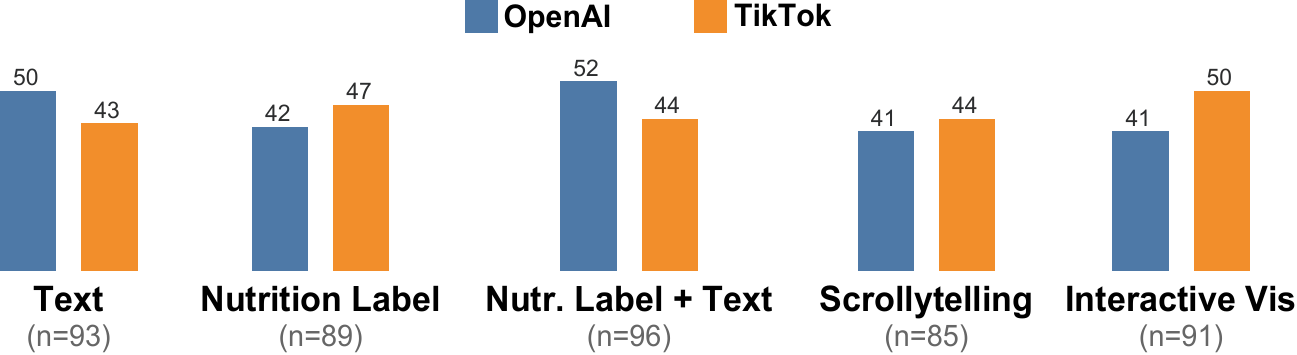}
    \caption{Number of participants assigned to each presentation format, split by policy owner.}
    \Description{Bar charts showing the number of participants in each experimental condition. Five formats are compared (Text, Nutrition Label, Nutrition Label + Text, Interactive Vis, and Scrollytelling), each split by policy owner (OpenAI vs TikTok). For each format-owner pair, a bar indicates the participant count, with OpenAI in blue and TikTok in orange. Bars are labeled with exact counts (ranging from 41 to 52), confirming balanced allocation across the 10 conditions.}
    \label{fig:participantsPerCondition}
\end{figure}

Of the 600 participants recruited, data from 3 could not be fully recorded due to technical issues, leaving 597 complete responses. Among these, 43 were excluded for failing attention checks, 94 for exhibiting high-entropy response patterns indicative of random or unengaged answering, and 6 who met both exclusion criteria. This resulted in a final sample of 454 participants, with 41--52 participants per Format$\times$Owner cell (Figure~\ref{fig:participantsPerCondition}), which preserves high power for \textit{Format} main effects but affords only limited sensitivity for \textit{Format}$\times$\textit{Owner} interactions. All analyses reported below are based on this sample. 

To aid readability, each subsection below begins by introducing the analysis approach before reporting the corresponding results. Key findings are \keyfinding{visually emphasized inline} for quick scanning.

% Results are organized by questionnaire. We treat \textit{Format} main effects for \textit{Comprehension} (accuracy, confidence), \textit{Experience}, and pre--post \textit{Perception} ($\Delta$) as confirmatory, using two-sided tests with $\alpha = .05$. \textit{Owner} is included as a fixed effect to adjust for baseline differences between policies, so \textit{Format} effects are interpreted as marginal (i.e., averaged across owners). For \textit{Comprehension}, we did not model \textit{Format}${\times}$\textit{Owner} interactions; per-cell accuracy is summarized descriptively (Appendix Table~\ref{tab:accuracy:accuracy_by_condition}). For \textit{Experience} and \textit{Perception}, interaction terms are estimated but treated as exploratory; we report 95\% confidence intervals and provide full model outputs in Appendix Table~\ref{app:tab:owner_effects:experience} (Experience) and Table~\ref{tab:appendix:owner_interactions:opinion_shifts} (Perception).

% \input{Sections/Results/051_Comprehension}

%%%% Sections/Results/051_Comprehension_OnlyMainEffects.tex starts here %%%%

% Comprehension results (main-effects only)
\subsection{Policy Comprehension}
\label{sec:results:comprehension}

\begin{table*}[hb!]
\centering
\small
\caption{Fixed effects from the variational Bayes GLMM for comprehension accuracy (main effects only). Estimates are on the log-odds scale; odds ratios (OR) are $=\exp(\hat\beta)$. The rightmost column plots each effect's OR and 95\% confidence interval (CI) on a common log scale axis. The vertical gray line marks OR$=1$ (no difference from the baseline). Filled markers in blue indicate effects reliably different from the \textit{Scrollytelling}/\textit{OpenAI}/\textit{factual} baseline (i.e., CIs that exclude 1); gray open markers indicate CIs including 1.}
\Description{Fixed effects from a Bayesian logistic mixed-effects model predicting comprehension accuracy. Each row reports an effect (e.g., format or owner) with its estimate, standard error, z-score, p-value, odds ratio (OR), and 95\% confidence interval. For example, the Nutrition Label format has a statistically significant positive effect (OR = 1.27, CI [1.04, 1.56]). The TikTok owner shows a strong effect (OR = 4.79), while other format contrasts are not statistically reliable.}
\label{tab:accuracy:vb_fixed_effects}

% Use @{} to remove side padding in the last column to prevent overflow
\begin{tabular}{lcccccc@{}c@{}} 
\toprule
\textbf{Term} & \textbf{Estimate} & \textbf{Post.\ SD} & \textbf{$z$} & \textbf{$p$} & \textbf{OR} & \textbf{OR 95\% CI} & \textbf{CI (OR scale)} \\
\midrule

Intercept (\textit{Scrollytelling}, \textit{OpenAI}, \textit{Factual})
& 0.459 & 0.045 & 10.313 & $<$.001 & 1.583 & [1.45, 1.73]
& \multirow{7}{*}{\includegraphics[height=2.45cm]{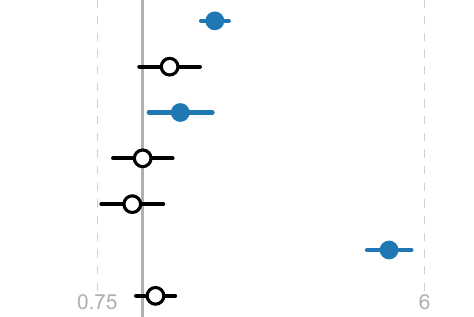}} \\

  % height=7.5\baselineskip, keepaspectratio

Format: \textit{Text}
& 0.172 & 0.099 &  1.740 & 0.082   & 1.188 & [0.98, 1.44] & \\

Format: \textit{Nutrition Label}
& 0.240 & 0.104 &  2.319 & 0.020   & 1.271 & [1.04, 1.56] & \\

Format: \textit{Nutrition Label + Text}
& 0.001 & 0.095 &  0.013 & 0.990   & 1.001 & [0.83, 1.21] & \\

Format: \textit{Interactive Vis}
& -0.065 & 0.099 & -0.660 & 0.509   & 0.937 & [0.77, 1.14] & \\

Owner: \textit{TikTok}
& 1.566 & 0.072 & 21.643 & $<$.001 & 4.789 & [4.16, 5.52] & \\

Item Type: \textit{Interpretative}
& 0.083 & 0.065 &  1.277 & 0.201   & 1.086 & [0.96, 1.23] & \\

\bottomrule
\end{tabular}
\end{table*}

We assessed comprehension using three measures: (i) binary accuracy (0/1), (ii) self-reported confidence on a 5-point Likert scale, and (iii) time spent answering the comprehension items. \textit{Format} and \textit{Owner} were manipulated between participants, while \textit{Item Type} (factual vs.\ interpretive) varied within participants across the Comprehension questionnaire (Step~4 in Section~\ref{sec:Procedure}). 

In the models presented below, the intercept corresponds to the \textit{Scrollytelling} format, and---when included in the model---the \textit{OpenAI} policy owner, and \textit{Factual} question type. This is our reference group: every coefficient we report in this section tells us how much higher or lower a condition is relative to that baseline.

\subsubsection{Comprehension Accuracy}

We modeled accuracy using a Bayesian binomial generalized linear mixed-effects model (GLMM) with logit link, fitted via variational Bayes. The model predicted the log-odds of a correct response based on three fixed effects: \textit{Format}, \textit{Owner}, and \textit{Item type}. We included random intercepts for participant and item to account for individual differences and item-specific variability. Full results appear in Table~\ref{tab:accuracy:vb_fixed_effects}. \keyfinding{Among formats, only \textit{Nutrition Label} showed a reliable accuracy increase compared to \textit{Scrollytelling}} (OR = 1.27, 95\% CI [1.04, 1.56]). Other formats did not differ reliably from the baseline (see the corresponding ORs and CIs in Table~\ref{tab:accuracy:vb_fixed_effects}). 

% To understand the factors influencing comprehension accuracy, we fitted a binomial generalized linear mixed-effects model (GLMM) to the data. The model predicted the log-odds of a correct response based on three fixed effects: policy format, policy owner, and item type. We included random intercepts for participant and item to account for individual differences and item-specific variability. The results, based on Variational Bayes (VB) estimation, are summarized in Table~\ref{tab:accuracy:vb_fixed_effects}. Relative to scrollytelling, only \textit{Nutrition Label} showed a reliable increase in accuracy (OR = 1.27, 95\% CI [1.04, 1.56]). The \textit{Text}, \textit{Nutrition Label + Text}, and \textit{Interactive Vis} formats were not distinguishable from the reference (see the corresponding ORs and CIs in Table~\ref{tab:accuracy:vb_fixed_effects}).

To assess practical impact, we computed predicted marginal accuracies by format, averaging over owner and item type (Table~\ref{tab:accuracy:predicted_accuracy_by_format}). Accuracy was consistently high (72.7--78.1\%). \keyfinding{Compared to \textit{Scrollytelling} (0.755), \textit{Nutrition Label} improved accuracy by +2.5 percentage points}---the only contrast whose \emph{odds-ratio} 95\% CI excludes 1 in Table~\ref{tab:accuracy:vb_fixed_effects}.

\begin{table}[h!]
\centering
\small
\caption{Predicted comprehension accuracy by presentation format, averaging over owner and item type. Baseline format: \textit{Scrollytelling}.}
\label{tab:accuracy:predicted_accuracy_by_format}
\begin{tabular}{lcc}
\toprule
\textbf{Format} & \textbf{$\hat p(\text{correct})$} & \textbf{$\Delta$ vs \textit{Scrollytelling}} \\
\midrule
\textit{Scrollytelling} & 0.755 & +0.000 \\
\textit{Text}           & 0.736 & -0.020 \\
\textit{Nutrition Label}  & 0.781 & +0.025 \\
\textit{Nutrition Label + Text}      & 0.756 & +0.001 \\
\textit{Interactive Vis}            & 0.727 & -0.028 \\
\bottomrule
\end{tabular}
\Description{Predicted comprehension accuracy (proportion correct) for each presentation format, averaged across owners and item types. Accuracy values range from 0.727 (Interactive Vis) to 0.781 (Nutrition Label). Scrollytelling serves as the baseline at 0.755. Each row includes the format name, predicted accuracy, and difference versus Scrollytelling.}
\end{table}

{
\begin{table*}[ht!]
\centering
\small
\caption{Linear model for item-level confidence (main effects only), HC3 standard errors. Coefficients are relative to scrollytelling (Format), OpenAI (Owner), and factual (Item type). The rightmost column plots each coefficient and 95\% confidence interval (CI) on a common \emph{linear} axis. The vertical gray line marks 0 (no difference from the baseline).}
\label{tab:confidence:lmm_fixed_effects}
\Description{Results from a linear model predicting confidence ratings (1–5 scale) on comprehension items. Each row shows a main effect (e.g., format or owner) with estimated coefficient, standard error, z-score, p-value, and confidence interval. Nutrition Label shows a small but statistically significant increase in confidence compared to Scrollytelling; other formats show no reliable differences.}

% Use @{} to remove side padding in the last column to prevent overflow
\begin{tabular}{lccccc@{}c@{}}
\toprule
\textbf{Term} & \textbf{Estimate} & \textbf{SE} & \textbf{$z$} & \textbf{$p$} & \textbf{95\% CI} & \textbf{CI (coefficient scale)} \\
\midrule

Intercept (\textit{Scrollytelling}, \textit{OpenAI}, \textit{factual})
& 3.678 & 0.069 & 53.255 & $<$.001 & [3.54, 3.81]
& \multirow{7}{*}[0em]{%
    \includegraphics[height=2.45cm,keepaspectratio]{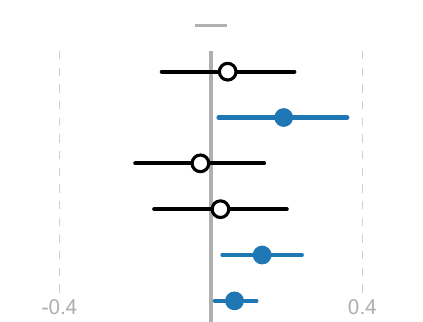}
  } \\

Format: \textit{Text}
& 0.044 & 0.088 & 0.501 & 0.617 & [-0.13, 0.22] & \\

Format: \textit{Nutrition Label}
& 0.192 & 0.087 & 2.193 & 0.028 & [0.02, 0.36] & \\

Format: \textit{Nutrition Label + Text}
& -0.028 & 0.088 & -0.315 & 0.753 & [-0.20, 0.14] & \\

Format: \textit{Interactive Vis}
& 0.025 & 0.088 & 0.285 & 0.776 & [-0.15, 0.20] & \\

Owner: \textit{TikTok}
& 0.135 & 0.055 & 2.433 & 0.015 & [0.03, 0.24] & \\

Item Type: \textit{Interpretative}
& 0.062 & 0.027 & 2.312 & 0.021 & [0.01, 0.12] & \\

\bottomrule
\end{tabular}
\end{table*}
}

For the \textit{Nutrition Label + Text} condition, which included a link to the full policy, log data confirmed that, in total, 90\% of participants assigned to this format accessed the full policy text at least once during the study. Specifically, 49 of 52 participants (94\%) assigned to the \textit{OpenAI} policy and 37 of 44 (84\%) assigned to \textit{TikTok} opened the full text policy link. These rates are consistent with those observed in all our pilot studies and suggest that participants did consult the reference text when available.

% Accuracy was consistently high (72.7–78.1\%). Compared to \textit{Scrollytelling} (0.755), \textit{Nutrition Label} improved accuracy by +2.5 percentage points—the only contrast with a 95\% CI excluding zero. All other differences were small and not statistically reliable.

% To better understand the practical impact of policy format, we calculated the model-based marginal accuracies for each format, averaging over owner and item type. Predicted accuracy was uniformly high across formats (72.7--78.1\%; Table~\ref{tab:accuracy:predicted_accuracy_by_format}). Relative to \textit{Scrollytelling} (0.755), \textit{Nutrition Label} was 2.5 percentage points (pp) higher (0.781), the only contrast with a 95\% CI excluding zero (Table~\ref{tab:accuracy:vb_fixed_effects}). \textit{Nutrition Label + Text} was essentially identical (+0.1 p.p.), while \textit{Text} (-2.0 p.p.) and \textit{Interactive Vis} (-2.8 p.p.) were slightly lower, with confidence intervals overlapping zero. In absolute terms, format differences were small.

% Model-based marginal accuracy by format was: scrollytelling = 0.755, text = 0.736, nutrition = 0.756, nutritionOnly = 0.781, vis = 0.727 (Table~\ref{tab:accuracy:predicted_accuracy_by_format}).

% Independent of format, accuracy was much higher for TikTok than OpenAI (OR = 4.79, [4.16, 5.52]), while item type showed a small, imprecise difference (interpretative vs factual: OR = 1.09, [0.96, 1.23]). .

\keyfinding{Independent of format, \textit{Owner} had a large effect: accuracy was substantially higher for TikTok than for OpenAI} (OR = 4.79, 95\% CI [4.16, 5.52]). \textit{Item type} had no reliable effect: interpretive and factual items yielded comparable accuracy (OR = 1.09, CI [0.96, 1.23]). For descriptive purposes, observed per-cell proportions are reported in Appendix Table~\ref{tab:accuracy:accuracy_by_condition}.

\begin{table}[b!]
\centering
\small
\caption{Predicted item-level confidence by presentation format (1--5 scale), averaging over owner and item type. Baseline format: \textit{Scrollytelling}.}
\begin{tabular}{lcc}
\toprule
\textbf{Format} & \textbf{$\hat{\mu}(\text{confidence})$} & \textbf{$\Delta$ vs \textit{Scrollytelling}} \\
\midrule
\textit{Scrollytelling} & 3.777 & +0.000 \\
\textit{Text}           & 3.821 & -0.044 \\
\textit{Nutrition Label}  & 3.969 & -0.192 \\
\textit{Nutrition Label + Text}      & 3.750 & +0.027 \\
\textit{Interactive Vis}            & 3.802 & -0.025 \\
\bottomrule
\end{tabular}
\Description{Predicted mean confidence ratings for each format, on a 1–5 scale. Nutrition Label has the highest mean (3.969), slightly above Scrollytelling (3.777). Differences are small (max plus minus 0.19).}
\label{tab:confidence:predicted_confidence_by_format}
\end{table}

\subsubsection{Comprehension Confidence} 

Confidence ratings were modeled using a linear mixed-effects model with participant random intercepts and HC3 robust standard errors. Complete estimates are shown in Table~\ref{tab:confidence:lmm_fixed_effects}. \keyfinding{Among formats, only \textit{Nutrition Label} led to a statistically significant increase in confidence over \textit{Scrollytelling}} ($\beta = 0.192$, 95\% CI [0.020, 0.363], $p = .028$). All other formats yielded similar confidence levels (all $p > .05$).

To better illustrate the effect of format, we calculated predicted mean confidence scores from the model, averaging over \textit{Owner} and \textit{Item Type}. Confidence ratings were similar across formats on the 1--5 scale (Table~\ref{tab:confidence:predicted_confidence_by_format}), with the highest mean observed for \textit{Nutrition Label} (3.969).

% . The results are presented in Table~\ref{tab:confidence:predicted_confidence_by_format}, which shows that the \textit{Nutrition Label} format shows the highest predicted confidence.

% Predicted confidence by format was: scrollytelling = 3.777, text = 3.821, nutrition = 3.750, nutritionOnly = 3.969, vis = 3.802 on the 1--5 scale (Table~\ref{tab:confidence:predicted_confidence_by_format}).

We also observed significant main effects for \textit{Owner} and \textit{Item type}. \keyfinding{Participants reported higher confidence when evaluating the \textit{TikTok} policy compared to \textit{OpenAI}} ($\beta = 0.135$, CI [0.03, 0.24], $p = .015$), and slightly higher confidence for interpretive versus factual items ($\beta = 0.062$, CI [0.01, 0.12], $p = .021$).

% \begin{table*}[t]
% \centering
% \caption{Linear model for item-level confidence (main effects only), HC3 standard errors. Coefficients are relative to scrollytelling (Format), OpenAI (Owner), and factual (QType).}
% \label{tab:confidence:lmm_fixed_effects}
% \begin{tabular}{lrrrrr}
% \toprule
% Term & Estimate & SE & z & p & 95\% CI \\
% \midrule
% Intercept (scrollytelling, OpenAI, factual) & 3.678 & 0.069 & 53.255 & $<$.001 & [3.54, 3.81] \\
% Format: nutrition                           & -0.028 & 0.088 & -0.315 & 0.753 & [-0.20, 0.14] \\
% Format: nutritionOnly                       &  0.192 & 0.087 &  2.193 & 0.028 & [ 0.02, 0.36] \\
% Format: text                                &  0.044 & 0.088 &  0.501 & 0.617 & [-0.13, 0.22] \\
% Format: vis                                 &  0.025 & 0.088 &  0.285 & 0.776 & [-0.15, 0.20] \\
% Owner: TikTok                               &  0.135 & 0.055 &  2.433 & 0.015 & [ 0.03, 0.24] \\
% QType: interpretative                       &  0.062 & 0.027 &  2.312 & 0.021 & [ 0.01, 0.12] \\
% \bottomrule
% \end{tabular}
% \end{table*}

\begin{figure}[b!]
\centering
\includegraphics[width=0.99\linewidth]{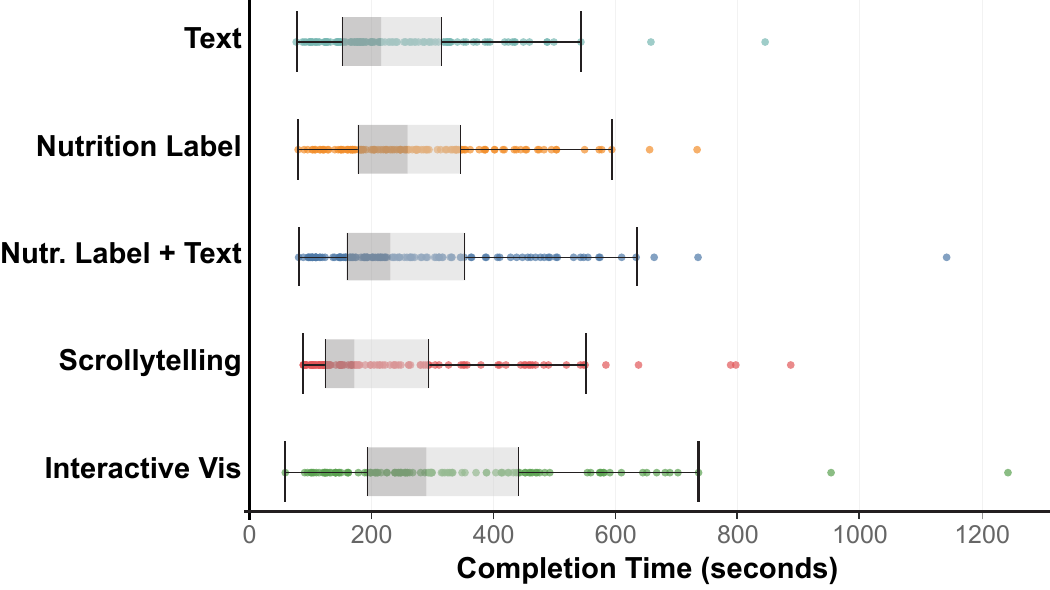}
\caption{Completion times (in seconds) for the Comprehension questionnaire across formats. Each point represents a participant; boxplots show the median and interquartile range (IQR), with whiskers extending to 1.5×IQR. \textit{Scrollytelling} and \textit{Text} formats had the lowest medians, while \textit{Interactive Vis} showed the highest times and greatest variability.}
\Description{This figure presents a boxplot of completion times (in seconds) for the Comprehension questionnaire across five formats: Text, Nutrition Label, Nutrition Label + Text, Scrollytelling, and Interactive Vis. Each data point represents a participant, and the boxplots display the median, interquartile range (IQR), and whiskers extending to 1.5 times the IQR, with outliers shown as individual dots. The y-axis ranges from 0 to approximately 1300 seconds. Participants using Scrollytelling and Text completed the questionnaire fastest, with median times just under 250 seconds. Nutrition Label and Nutrition Label + Text formats resulted in slightly higher times around 280 seconds. The Interactive Vis format led to the longest and most variable completion times, with a median over 325 seconds and several outliers exceeding 800 seconds.}
\label{fig:CompletionTinesComprehension}
\end{figure}

\subsubsection{Comprehension Time}
\label{sec:results:comprehension-time}

We analyzed how long participants took to complete the Comprehension questionnaire. Times were calculated from log timestamps as the number of seconds between opening and submitting the Step~4 form (Section~\ref{sec:Procedure}). Note that this measure excludes the mandatory 90-second exploration period from Step~3. With this data, we also investigated whether more time on the comprehension step predicted higher accuracy, and whether this relationship varied by format.

\paragraph{Format, Owner, and Interaction Effects} 

We analyzed completion time with a two-way ANOVA, using \textit{Format} and \textit{Owner} as fixed effects. Because times were right-skewed, we ran the analysis on raw seconds and, as a robustness check, on $\log(time+1)$. \keyfinding{\textit{Format} significantly affected completion time} (raw: $p = 8.9 \times 10^{-5}$; log: $p = 1.3 \times 10^{-5}$). In contrast, neither the main effect of \textit{Owner} (raw: $p = .768$; log: $p = .591$) nor its interaction with \textit{Format} (raw: $p = .119$; log: $p = .314$) reached significance. Thus, we interpret format differences as robust across policies.\\

For interpretability, we report raw means (in seconds): \textit{Scrollytelling} $\approx$ 238.6 $<$ \textit{Text} $\approx$ 244.7 $<$ \textit{Nutrition Label} $\approx$ 279.4 $\approx$ \textit{Nutrition Label + Text} $\approx$ 281.0 $<$ \textit{Interactive Vis} $\approx$ \textit{328.1} (see also Figure~\ref{fig:CompletionTinesComprehension}). Tukey HSD conducted on $\log(time+1)$ indicated that \keyfinding{participants assigned to the \textit{Interactive Vis} format required significantly more time than those assigned to \textit{Text}} ($\Delta \approx +83$ s, $p < .001$) \keyfinding{and \textit{Scrollytelling}} ($\Delta \approx +90$ s, $p < .001$). No other pairwise differences were significant after correction. Within-owner analyses showed that format differences were significant for \textit{TikTok} ($p = 6.8 \times 10^{-5}$) but not for \textit{OpenAI} ($p = .223$). However, due to the non-significant interaction term, these should be treated as descriptive rather than inferential.

% In short, participants completed the comprehension task fastest with \textit{Scrollytelling} and \textit{Text}, and slowest with \textit{Interactive Vis}, consistently across both policy types.

\paragraph{Time-Accuracy Relationship}

To test whether time spent answering the comprehension items predicted accuracy, and whether this effect varied by format, we fit the model $Accuracy \sim \log(time+1) \times Format,$ using OLS with HC3 robust standard errors. In this model, the dependent variable $Accuracy$ is the participant's percentage of correct responses on the Comprehension questionnaire. \textit{Scrollytelling} served as the reference format. As a descriptive complement, we also report within-format Pearson correlations between log-time and \textit{Accuracy}.

\keyfinding{The association between time and accuracy was weak for most formats}, and although \textit{Interactive Vis} showed a statistically significant positive slope (Table~\ref{tab:time_accuracy_by_format}), \keyfinding{a joint test found no significant \textit{log-time} × \textit{Format} interaction} ($\chi^2(4)=4.51$, $p=.341$), indicating that the slopes are not statistically distinguishable across formats.
Compared to \textit{Scrollytelling}, only \textit{Interactive Vis} showed a marginally larger slope ($\Delta\beta = 0.0782$, SE = 0.0467, $p = .094$), meaning that accuracy increased more sharply with additional time in this format. No other format differed from \textit{Scrollytelling} in a statistically reliable way.

%%%% Sections/Results/051_Comprehension_OnlyMainEffects.tex ends here %%%%

%%%% Sections/Results/052_Experience.tex starts here %%%%

\subsection{Subjective Experience}
\label{sec:results:experience}

\begin{table}[t!]
\centering
\small
\caption{Within-format Pearson correlations ($r$) and OLS slopes ($\beta_{\log\text{ time}}$) of \textit{Accuracy} on $\log(\text{time}+1)$. \textit{Accuracy} refers to the proportion of comprehension questions answered correctly, expressed as a percentage. Slopes come from a single OLS model with HC3 robust SEs and a $\log(\text{time}+1)\times\textit{Format}$ interaction, using \textit{Scrollytelling} as the reference. Here, $r$ is the Pearson correlation and $p$ its two-sided test.}
\label{tab:time_accuracy_by_format}
\begin{tabular}{lcccc}
\toprule
\textbf{Format} & \textbf{$r(p)$} & \textbf{$\beta_{\text{log time}}$} & \textbf{SE} & \textbf{$p$} \\
\midrule
\textit{Scrollytelling} & 0.163 ($.075$) & 0.0591 & 0.0343 & .085 \\
\textit{Text} & 0.129 ($.163$) & 0.0540 & 0.0363 & .137 \\
\textit{Nutrition Label} & 0.138 ($.134$) & 0.0589 & 0.0349 & .091 \\
\textit{Nutrition Label + Text} & 0.205 ($.026$) & 0.0826 & 0.0352 & .019 \\
\textit{Interactive Vis} & \textbf{0.371} ($<.001$) & \textbf{0.1372} & 0.0317 & $<.001$ \\
\bottomrule
\end{tabular}
\Description{This table shows how comprehension accuracy relates to completion time within each format. For each of the five formats—Scrollytelling, Text, Nutrition Label, Nutrition Label + Text, and Interactive Vis—it reports the Pearson correlation between log-transformed completion time and accuracy, along with the slope from a linear regression predicting accuracy from log-transformed time. Accuracy is measured as the percentage of correct answers on the comprehension questionnaire. Most correlations were small and not statistically significant, except for Nutrition Label + Text and Interactive Vis, which showed positive and reliable associations. In particular, Interactive Vis had the strongest correlation (0.371) and the steepest positive slope (0.1372), both significant at p < 0.001, indicating that participants who spent more time tended to achieve higher accuracy in that condition. Slopes were derived from a single model with robust standard errors and a time-by-format interaction, using Scrollytelling as the reference.}
\end{table}

\newlength{\wFmt}\setlength{\wFmt}{0.30\linewidth}
\newlength{\wColA}\setlength{\wColA}{0.24\linewidth}
\newlength{\wCol}\setlength{\wCol}{0.24\linewidth}

\begin{table*}[hb!]
\centering
\small
\setlength{\tabcolsep}{4pt}
\renewcommand{\arraystretch}{1.20}
\caption{Predicted mean ratings by presentation format across subjective experience constructs (fixed-effects predictions, averaged equally over policy owner). Each cell shows the mean, with the difference from \textit{scrollytelling} in parentheses. Parentheses are background-shaded by the magnitude of the difference on a diverging blue-white-red scale (white = 0; blue = better, red = worse). For \textit{Cognitive Load}, colors reflect desirability (higher load is worse): increases are red and decreases are blue.}
\label{tab:xp:pred_means_by_format}

\begin{tabular}{ 
  L{\wFmt} C{\wColA} C{\wCol} C{\wCol} C{\wCol} C{\wCol} C{\wCol} }
\toprule
\textbf{Format} &
\textbf{Behavioral Intentions} &
\textbf{Cognitive Load} &
\textbf{Engagement} &
\textbf{Enjoyment} &
\textbf{Format Adoption} &
\textbf{Perceived Clarity} \\
\midrule
\textit{Scrollytelling} & 2.875 & 2.845 & 2.900 & 2.991 & 3.449 & 3.445 \\
\textit{Text} & 2.656 \diffcell{-0.219} & 3.473 \diffcellCL{+0.629} & 2.254 \diffcell{-0.647} & 2.354 \diffcell{-0.636} & 2.773 \diffcell{-0.676} & 2.941 \diffcell{-0.504} \\
\textit{Nutrition Label} & 2.905 \diffcell{+0.031} & 2.794 \diffcellCL{-0.051} & 3.084 \diffcell{+0.183} & 3.177 \diffcell{+0.186} & 3.390 \diffcell{-0.059} & 3.371 \diffcell{-0.074} \\
\textit{Nutrition Label + Text} & 2.964 \diffcell{+0.089} & 2.823 \diffcellCL{-0.021} & 2.878 \diffcell{-0.022} & 3.080 \diffcell{+0.089} & 3.309 \diffcell{-0.141} & 3.400 \diffcell{-0.045} \\
\textit{Interactive Vis} & 3.046 \diffcell{+0.172} & 2.908 \diffcellCL{+0.064} & 2.994 \diffcell{+0.094} & 3.042 \diffcell{+0.051} & 3.347 \diffcell{-0.102} & 3.430 \diffcell{-0.015} \\
\bottomrule
\end{tabular}

\Description{Predicted means for six subjective experience constructs by format, including: Behavioral Intentions, Cognitive Load, Engagement, Enjoyment, Format Adoption, and Perceived Clarity. Each cell includes the mean score and the difference from Scrollytelling in parentheses. Differences are color-coded by magnitude and direction (positive/negative). Scrollytelling performs better than Text across all constructs. Nutrition Label and Interactive Vis perform similarly to Scrollytelling.}
\end{table*}

To assess how each presentation format shaped participants' policy reviewing experience, we analyzed six subjective constructs (Step~5 in Section~\ref{sec:Procedure}): \textit{Behavioral Intentions}, \textit{Cognitive Load}, \textit{Engagement}, \textit{Enjoyment}, \textit{Format Adoption}, and \textit{Perceived Clarity}. Each participant’s score was calculated as the average of their responses across items in each construct (Likert 1-5). We modeled each construct using ordinary least squares (OLS) with heteroskedasticity-consistent (HC3) standard errors to account for unequal variances. All models included fixed effects for \textit{Format}, \textit{Owner}, and their interaction. The coefficients reported below reflect point differences relative to the \textit{Scrollytelling}/\textit{OpenAI} reference. Before modeling, we assessed internal consistency of the Experience scales using Cronbach's $\alpha$ and McDonald's $\omega_{\text{total}}$. Reliability was adequate to high across all constructs (the specific values are reported in Appendix Table~\ref{app:tab:xp_reliability}).

\subsubsection{Format Effects}

Estimated marginal means (Table~\ref{tab:xp:pred_means_by_format}) highlight clear differences across formats. \keyfinding{Relative to \textit{Scrollytelling}, the \textit{Text} format consistently led to worse experience}: it lowered \textit{Engagement} (-0.65), \textit{Enjoyment} (-0.64), \textit{Format Adoption} (-0.68), and \textit{Perceived Clarity} (-0.50), while increasing \textit{Cognitive Load} by +0.63. \textit{Behavioral Intentions} were also slightly lower for \textit{Text}, though the confidence interval narrowly included zero.

Beyond \textit{Text}, Table~\ref{tab:xp:pred_means_by_format} highlights two broader trends. First, \keyfinding{\textit{Format Adoption} and \textit{Perceived Clarity} are below \textit{Scrollytelling} for every format} (both columns appear shaded red). Second, \textit{Nutrition Label} shows a small advantage in \textit{Engagement} $+0.183$ and \textit{Enjoyment} $+0.186$. \keyfinding{\textit{Behavioral Intentions} are slightly higher for all non-text formats} (\textit{Interactive Vis} $+0.172$, \textit{Nutrition Label + Text} $+0.089$, \textit{Nutrition Label} $+0.031$). \textit{Interactive Vis} also shows potential for higher \textit{Engagement} $+0.094$ and \textit{Enjoyment} $+0.051$ but it also suggests a slightly higher \textit{Cognitive Load} $+0.064$ over \textit{Scrollytelling}.

For inference, Table~\ref{tab:xp:format_main_effects} reports fixed-effects contrasts (OLS with HC3 SEs) vs.\ the \textit{Scrollytelling}/\textit{OpenAI} baseline. In line with the estimated marginal means of Table~\ref{tab:xp:pred_means_by_format}, \textit{Text} reliably reduces \textit{Engagement} ($-0.768$, 95\% CI [$-1.253,\,-0.283$]), \textit{Enjoyment} ($-0.764$, [$-1.239,\,-0.289$]), \textit{Format Adoption} ($-0.886$, [$-1.351,\,-0.421$]), and \textit{Perceived Clarity} ($-0.736$, [$-1.220,\,-0.253$]), and increases \textit{Cognitive Load} ($+0.748$, [$0.300,\,1.195$]); the effect on \textit{Behavioral Intentions} is negative but not reliable ($-0.365$, [$-0.769,\,0.040$]). The \textit{Nutrition Label}, \textit{Nutrition Label + Text}, and \textit{Interactive Vis} formats are close to the baseline, with confidence intervals overlapping zero across constructs (gray open markers in Table~\ref{tab:xp:format_main_effects}).

% ===========================
% Format main effects (rows) × Constructs (columns)
% ===========================
\begin{table*}[ht!]
\centering
\small
\setlength{\tabcolsep}{2pt}
\renewcommand{\arraystretch}{1.20}

\caption{Format main effects on subjective experience constructs (OLS with HC3 SEs). Entries are point differences vs.\ the \textit{scrollytelling}/\textit{OpenAI} baseline (for \textit{Cognitive Load}, higher is worse). Right subcolumns plot 95\% CIs on a common difference axis (vertical line at 0). Filled markers indicate CIs excluding 0.}
\label{tab:xp:format_main_effects}
\Description{Main effects of each format on subjective experience constructs, using OLS with HC3 robust standard errors. Each row shows the format contrast with respect to Scrollytelling on one experience measure, along with its 95\% confidence interval. Filled markers indicate statistically significant differences (e.g., Text format increases Cognitive Load and decreases Engagement, Enjoyment, Clarity, and Adoption). Other formats show no significant deviations.}

% ---------- Panel A ----------
\begin{tabular}{l C{2.9cm} C{2cm} C{2.9cm} C{2cm} C{2.9cm} C{2cm}}
\toprule
& \multicolumn{2}{c}{\textbf{Behavioral Intentions}} &
  \multicolumn{2}{c}{\textbf{Cognitive Load}} &
  \multicolumn{2}{c}{\textbf{Engagement}} \\
\cmidrule(lr){2-3}\cmidrule(lr){4-5}\cmidrule(lr){6-7}
\textbf{Format} & $\Delta$ [95\% CI] & CI & $\Delta$ [95\% CI] & CI & $\Delta$ [95\% CI] & CI\\
\midrule

\textit{Text}
& $-0.365$ [$-0.769,\,+0.040$]
& \multirow{4}{*}[-0.1em]{\includegraphics[height=1.67cm,keepaspectratio]{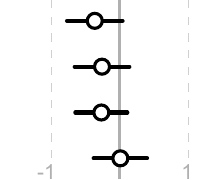}}
& $+0.748$ [$+0.300,\,+1.195$]
& \multirow{4}{*}[-0.1em]{\includegraphics[height=1.67cm,keepaspectratio]{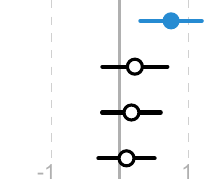}}
& $-0.768$ [$-1.253,\,-0.283$]
& \multirow{4}{*}[-0.1em]{\includegraphics[height=1.67cm,keepaspectratio]{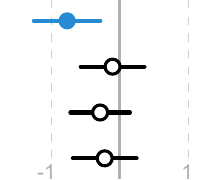}}
\\

\textit{Nutrition Label}
& $-0.259$ [$-0.657,\,+0.139$] & 
& $+0.220$ [$-0.254,\,+0.694$] & 
& $-0.105$ [$-0.569,\,+0.359$] & 
\\

\textit{Nutr.\ Label + Text}
& $-0.269$ [$-0.648,\,+0.110$] & 
& $+0.170$ [$-0.258,\,+0.597$] & 
& $-0.286$ [$-0.720,\,+0.147$] & 
\\

\textit{Interactive Vis}
& $+0.008$ [$-0.381,\,+0.397$] & 
& $+0.098$ [$-0.312,\,+0.508$] & 
& $-0.220$ [$-0.684,\,+0.245$] & 
\\
\end{tabular}

\vspace{0.15em}

% ---------- Panel B ----------
% \begin{tabular}{l C{3cm} Y C{3cm} Y C{3cm} Y}
\begin{tabular}{l C{2.9cm} C{2cm} C{2.9cm} C{2cm} C{2.9cm} C{2cm}}
\toprule
& \multicolumn{2}{c}{\textbf{Enjoyment}} &
  \multicolumn{2}{c}{\textbf{Format Adoption}} &
  \multicolumn{2}{c}{\textbf{Perceived Clarity}} \\
\cmidrule(lr){2-3}\cmidrule(lr){4-5}\cmidrule(lr){6-7}
\textbf{Format} & $\Delta$ [95\% CI] & CI & $\Delta$ [95\% CI] & CI & $\Delta$ [95\% CI] & CI\\
\midrule

\textit{Text}
& $-0.764$ [$-1.239,\,-0.289$]
& \multirow{4}{*}[-0.1em]{\includegraphics[height=1.67cm,keepaspectratio]{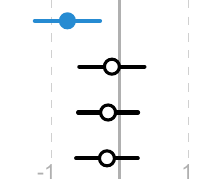}}
& $-0.886$ [$-1.351,\,-0.421$]
& \multirow{4}{*}[-0.1em]{\includegraphics[height=1.67cm,keepaspectratio]{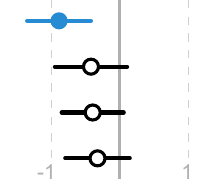}}
& $-0.736$ [$-1.220,\,-0.253$]
& \multirow{4}{*}[-0.1em]{\includegraphics[height=1.67cm,keepaspectratio]{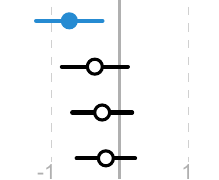}}
\\

\textit{Nutrition Label}
& $-0.115$ [$-0.590,\,+0.359$] &
& $-0.420$ [$-0.947,\,+0.107$] &
& $-0.364$ [$-0.845,\,+0.118$] &
\\

\textit{Nutr.\ Label + Text}
& $-0.171$ [$-0.608,\,+0.266$] &
& $-0.395$ [$-0.847,\,+0.057$] &
& $-0.257$ [$-0.694,\,+0.180$] &
\\

\textit{Interactive Vis}
& $-0.187$ [$-0.636,\,+0.262$] &
& $-0.325$ [$-0.795,\,+0.145$] &
& $-0.203$ [$-0.629,\,+0.223$] &
\\
\bottomrule
\end{tabular}

\end{table*}

% ---- width knobs (tune as needed) ----
\newlength{\wFmtShift} \setlength{\wFmtShift}{0.25\linewidth}
\newlength{\wTxtShift} \setlength{\wTxtShift}{0.30\linewidth}
\newlength{\wCIShift}  \setlength{\wCIShift}{0.12\linewidth}

\begin{table*}[hb!]
\centering
\small
\setlength{\tabcolsep}{2pt}
\renewcommand{\arraystretch}{1.20}

\caption{Format main effects on shifts in perception constructs. Entries are point differences on the 1--5 scale relative to the \textit{scrollytelling}/\textit{OpenAI} baseline; positive values indicate larger post--pre increases. All 95\% CIs include 0, so all markers are hollow.}
\label{tab:shift:format_main_effects}

\begin{tabular}{l C{2.4cm} C{2.4cm} C{2.4cm} C{2.4cm} C{2.4cm} C{2.4cm}}

\toprule
& \multicolumn{2}{c}{\textbf{Confidence in Understanding}} &
  \multicolumn{2}{c}{\textbf{Perceived Transparency}} &
  \multicolumn{2}{c}{\textbf{Trust}} \\
\cmidrule(lr){2-3}\cmidrule(lr){4-5}\cmidrule(lr){6-7}
\textbf{Format} &
$\Delta$ [95\% CI] & CI &
$\Delta$ [95\% CI] & CI &
$\Delta$ [95\% CI] & CI \\
\midrule

\textit{Text}
& $-0.342$ [$-0.81,\,0.13$]
& \multirow{4}{*}{\includegraphics[height=1.67cm,keepaspectratio]{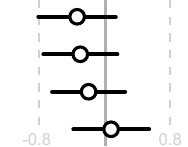}}
& $-0.034$ [$-0.42,\,0.36$]
& \multirow{4}{*}{\includegraphics[height=1.67cm,keepaspectratio]{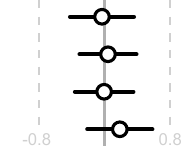}}
& $+0.018$ [$-0.35,\,0.39$]
& \multirow{4}{*}{\includegraphics[height=1.67cm,keepaspectratio]{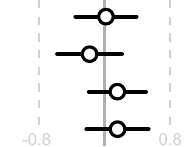}}
\\

\textit{Nutrition Label}
& $-0.303$ [$-0.75,\,0.15$] & 
& $+0.039$ [$-0.31,\,0.39$] &
& $-0.181$ [$-0.58,\,0.21$] &
\\

\textit{Nutr.\ Label + Text}
& $-0.203$ [$-0.65,\,0.24$] &
& $-0.008$ [$-0.36,\,0.35$] &
& $+0.157$ [$-0.19,\,0.50$] &
\\

\textit{Interactive Vis}
& $+0.073$ [$-0.39,\,0.53$] &
& $+0.183$ [$-0.21,\,0.58$] &
& $+0.159$ [$-0.22,\,0.54$] &
\\

\bottomrule
\end{tabular}

\Description{Main effects of format on post--pre shifts in three perception constructs: Confidence in Understanding, Perceived Transparency, and Trust. Each entry shows the difference from Scrollytelling and a 95\% CI. No format yields a statistically significant difference; all confidence intervals include 0.}
\end{table*}

\subsubsection{Owner and Interaction Effects}

For \textit{Scrollytelling}, compared to \textit{OpenAI}, \textit{TikTok}'s policy reduced \textit{Behavioral Intentions} by -0.67 (95\% CI [-1.11, -0.23]). Other effects of policy owner were small and not reliably different from zero. Interactions between format and owner were exploratory due to limited power. One interaction reached significance: \textit{Nutrition Label} increased \textit{Behavioral Intentions} more strongly when paired with the \textit{TikTok} policy (+0.72, 95\% CI [0.15, 1.29]). All other interaction terms were inconclusive (see Appendix Table~\ref{app:tab:owner_effects:experience} for the full estimates).

% Owner effects were weaker and less consistent than format effects. At the \textit{scrollytelling}/\textit{OpenAI} baseline, \textit{TikTok} (vs.\ \textit{OpenAI}) reduced \textit{Behavioral Intentions} by $-0.674$ points (95\% CI [$-1.113,,-0.234$]). Effects on \textit{Cognitive Load} ($+0.144$, [$-0.319,,0.608$]), \textit{Engagement} ($-0.362$, [$-0.830,,0.106$]), \textit{Enjoyment} ($-0.360$, [$-0.862,,0.142$]), \textit{Format Adoption} ($-0.353$, [$-0.861,,0.155$]), and \textit{Perceived Clarity} ($-0.329$, [$-0.792,,0.133$]) were modest and not reliably different from $0$ (Interactions are exploratory given limited power.) All format$\times$owner terms were estimated; aside from \textit{Nutrition Label + Text}$\times$\textit{TikTok} on \textit{Behavioral Intentions} ($+0.716$, [$0.147,,1.285$]), confidence intervals included $0$, indicating limited moderation by owner. See Appendix Table~\ref{app:tab:owner_effects:experience} for the full set of estimates.

%%%% Sections/Results/052_Experience.tex ends here %%%%

%%%% Sections/Results/053_OpinionShifts.tex starts here %%%%

\subsection{Perception Shifts: Pre- and Post-Perception Questionnaires}
\label{sec:opinion-shifts}

The pre- and post-perception questionnaires (Steps~2 and~6 in Section~\ref{sec:Procedure}) included Likert batteries measuring three perception constructs: (i) \emph{Confidence in Understanding} of the owner's privacy policies, (ii) \emph{Perceived Transparency} of the owner's practices, and (iii) \emph{Trust} in the policy owner. Both questionnaires also included single-item beliefs about (iv) the \textit{amount of data collected} according to the policy and (v) the \textit{number of third parties} receiving that data. These instruments capture baseline attitudes and any shifts after exposure to the assigned format and policy owner. In this section, we analyze these changes.

Because the change analyses use post–pre differences, we assessed internal consistency at both time points. Cronbach's $\alpha$ and McDonald's $\omega_{\text{total}}$ were acceptable to high for all the perception constructs, with higher values at post (Appendix Table~\ref{app:tab:perception_reliability} provides full details).

\setdiffscale{-0.4}{0.4}
\begin{table}[ht!]
\centering
\small
\setlength{\tabcolsep}{2pt}
\renewcommand{\arraystretch}{1.20}
\caption{Predicted perception shift (Post-Pre) by presentation format, averaged equally over policy owner. Each cell shows the predicted shift, with the difference from \textit{Scrollytelling} in parentheses. Parentheses are background-shaded by the magnitude of the difference on a diverging blue-white-red scale (white = 0; blue = larger shift, red = smaller shift).}
\label{tab:shift:pred:merged:color:compact}

\begin{tabular}{L{2cm} C{1.9cm} C{1.9cm} C{1.9cm}}
\toprule
\textbf{Format} & \textbf{Confidence in Understanding} &
                  \textbf{Perceived Transparency} &
                  \textbf{Trust} \\
\midrule
\textit{Scrollytelling} & 1.188 & 0.412 & -0.053 \\
\textit{Text} & 0.844 \diffcell{-0.344} & 0.366 \diffcell{-0.046} & 0.032 \diffcell{+0.085} \\
\textit{Nutrition Label} & 0.994 \diffcell{-0.194} & 0.517 \diffcell{+0.105} & -0.006 \diffcell{+0.047} \\
\textit{Nutr. Label + Text} & 0.979 \diffcell{-0.209} & 0.349 \diffcell{-0.063} & 0.214 \diffcell{+0.266} \\
\textit{Interactive Vis} & 1.297 \diffcell{+0.108} & 0.478 \diffcell{+0.066} & 0.330 \diffcell{+0.383} \\
\bottomrule
\end{tabular}
\Description{Predicted pre–post shift scores by format for three perception constructs. Scrollytelling serves as the baseline. Interactive Vis shows the largest predicted increase in Trust (+0.383) and Confidence (+0.108). Differences are small and non-significant but included for descriptive comparison. Each cell shows the predicted shift and the difference from Scrollytelling, with background color denoting magnitude.}
\end{table}

\subsubsection{Shifts in Perception Constructs}

For each participant and perception construct, we computed a difference score $\Delta Y = Y_{\text{post}} - Y_{\text{pre}}$ (1--5 scale) and estimated OLS models with HC3 standard errors including \textit{Format}, \textit{Owner}, and their interaction. Given limited per-cell sample sizes, the \textit{Format}$\times$\textit{Owner} interaction was treated as exploratory.

The coefficients resulting from these analyses are point differences relative to the \textit{Scrollytelling}/\textit{OpenAI} baseline and directly reflect shifts on the original rating scale (i.e., a positive $\Delta$ value implies a larger increase than the baseline).

Table~\ref{tab:shift:format_main_effects} summarizes the models' results. \keyfinding{Across constructs, none of the \textit{Format} contrasts yielded statistically reliable changes}; all 95\% CIs included 0. For \emph{Confidence in Understanding}, the largest point estimate was a small positive shift for \textit{Interactive Vis} ($\Delta=+0.073$, 95\% CI [$-0.39,\,0.53$]). \textit{Text}, \textit{Nutrition Label}, and \textit{Nutrition Label + Text} trended negative but were imprecise. For \emph{Perceived Transparency}, most estimates clustered near zero with a modest, non-significant positive trend for \textit{Interactive Vis} ($+0.183$). For \textit{Trust}, \textit{Interactive Vis} ($+0.159$) and \textit{Nutrition Label + Text} ($+0.157$) trended upward, \textit{Text} was near zero ($+0.018$), and \textit{Nutrition Label} trended negative ($-0.181$); again, all CIs covered 0. Owner effects and \textit{Format}$\times$\textit{Owner} interactions showed the same pattern: small estimates with 95\% CIs including 0. For exploratory purposes, Appendix Table~\ref{tab:appendix:owner_interactions:opinion_shifts} lists all \textit{Owner} and interaction estimates.

Model-based predictions averaged over policy owner (Table~\ref{tab:shift:pred:merged:color:compact}) align with the contrasts reported above. Relative to \textit{Scrollytelling}, the largest predicted increase occurs for \textit{Interactive Vis} in the \textit{Trust} construct ($+0.383$), followed by a more modest increment in \textit{Confidence in Understanding} ($+0.108$). For \textit{Perceived Transparency}, the largest increase was for \textit{Nutrition Label} ($+0.105$). The largest predicted decreases were for \textit{Text} in \textit{Confidence in Understanding} ($-0.344$) and for \textit{Nutrition Label + Text} in \textit{Perceived Transparency} ($-0.063$). For \textit{Trust}, all formats exceeded the baseline. These patterns, however, are descriptive; in all cases, the fixed-effects contrasts' 95\% CIs include 0.

\subsubsection{Shifts in Beliefs about Data and Third‐parties}

Beliefs about the \textit{amount of data collected} and the \textit{number of third parties} were measured with labeled four-option items: \emph{too little/few}, \emph{about right}, \emph{too much/many}, and \emph{I don't have enough information to say}. We treat responses as nominal and retain the uncertainty option as its own category.

We examined pre-post change with three approaches. First, Bowker's test of symmetry on the $4\times4$ pre-post table assessed directional transitions; second, a Stuart-Maxwell test assessed changes in the marginal distributions; third, a multinomial logit modeled post responses as a function of pre category, \textit{Format}, \textit{Owner}, and their interaction (HC3 CIs), using \emph{I don't have enough information} as the reference for pre and post to emphasize movement away from uncertainty.

\paragraph{Amount of Data Collected}

Bowker's test indicated strong, structured movement from pre to post ($\chi^2 = 223.23$, $df=6$, $p<.001$), whereas the Stuart–Maxwell test found no change in the overall composition of opinions (test statistic $\approx 0$, $df=3$, $p=1$). In short, there was substantial individual movement but no net shift in the margins. \keyfinding{Transitions out of uncertainty dominated} (Table~\ref{tab:transitions:amount}): only 13 participants remained \emph{uncertain}; most moved to \emph{too much} ($n{=}167$) or \emph{about right} ($n{=}49$). The multinomial model (Pseudo $R^2{=}0.114$, $p<.001$) points the same way: \keyfinding{pre-survey uncertainty strongly predicts a post-survey commitment, especially toward \emph{too much}}. After conditioning on pre-response, experimental factors contributed little; interaction terms were imprecise, suggesting initial beliefs were the main driver of post responses (see also Figure~\ref{figure:flows:amountDataCollected} in the Appendix, which depicts condition-wise flows).

\begin{table}[h!]
\centering
\small
\caption{Pre$\,\to\,$Post transitions for \emph{Amount of Data Collected}. Rows = Pre; Columns = Post.}
\label{tab:transitions:amount}
\begin{tabular}{lcccc}
\toprule
& Too little & About right & Too much & Uncertain \\
\midrule
Too little & 0 & 1 & 0 & 1 \\
About right & 0 & 39 & 40 & 2 \\
Too much & 1 & 11 & 126 & 2 \\
Uncertain & 2 & 49 & 167 & 13 \\
\bottomrule
\end{tabular}
\Description{4×4 contingency table showing transitions in beliefs about the amount of data collected, comparing pre- and post-survey responses. Rows are pre-survey responses (Too little, About right, Too much, Uncertain); columns are post-survey responses. Most participants move from Uncertain to Too much or About right. Very few remain in the Uncertain category.}
\end{table}

\paragraph{Number of Third Parties Receiving Data}

The same pattern held. Bowker's test showed pronounced directional movement ($\chi^2=233.10$, $df=6$, $p\approx1.7{\times}10^{-47}$), while the Stuart–Maxwell test indicated stable margins (test statistic $\approx 0$, $df=3$, $p=1$). Again, many participants left \emph{uncertain}: only 30 stayed there, versus 181 moving to \emph{too many} and 59 to \emph{about right} (Table~\ref{tab:transitions:thirdparties}; see also Figure~\ref{figure:flows:numberOfThirdParties} in the Appendix). The multinomial model pointed in the same direction, but convergence was fragile (non-finite solutions under Newton; BFGS with interactions required), so we emphasize the transition counts and flows rather than coefficient-level inferences.

\begin{table}[h!]
\centering
\small
\setlength{\tabcolsep}{6pt}
\renewcommand{\arraystretch}{1.20}
\caption{Pre$\,\to\,$Post transitions for \emph{Number of Third Parties Receiving Data}. Rows = Pre; Columns = Post.}
\label{tab:transitions:thirdparties}
\begin{tabular}{lcccc}
\toprule
 & Too few & About right & Too many & Uncertain \\
\midrule
Too few & 0  & 0  & 1 & 0 \\
About right & 0  & 30 & 15 & 1 \\
Too many & 2  & 25 & 105 & 3 \\
Uncertain & 2 & 59 & 181 & 30  \\
\bottomrule
\end{tabular}
\Description{4×4 contingency table showing transitions in beliefs about the number of third parties receiving data. As in Table 9, the table highlights strong movement from Uncertain to Too many or About right. Only a small number remain uncertain post-exposure.}
\end{table}

%%%% Sections/Results/053_OpinionShifts.tex ends here %%%%

%%%% Sections/Results/054_Feedback.tex starts here %%%%

\subsection{Participant Perspectives on Presentation Formats and Policy Owners}
\label{sec:results:feedback}

We qualitatively analyzed open-ended comments from two prompts: (i) a required post-format exposure \emph{feedback} item in the Experience questionnaire (min.\ 50 characters), and (ii) an optional \emph{final thoughts} item in the post-perception questionnaire. In total, we analyzed 668 comments: 454 feedback entries and 214 final thoughts. 

Because prompts were brief and required comments were often minimal (e.g., \quotes{nothing to add}), we do not claim formal themes. We instead report recurring topics across formats and owners, noting consistencies and tensions while acknowledging evidential limits.

\subsubsection{Perceptions of the Presentation Formats}

Comments on the different formats revealed a clear dichotomy between those who found the new approaches (\textit{Nutrition Labels}, \textit{Scrollytelling}, and the \textit{Interactive Vis}) to be an improvement over traditional text and those who felt they were still too complex or even deceptive. 

\paragraph{\textbf{Traditional Text}}

This format was consistently criticized for being \quotes{\textit{overly wordy and difficult to pick out the relevant information}} (P105, \textit{OpenAI}) and \quotes{\textit{very long-winded and taxing to read in full}} (P083, \textit{TikTok}). Many participants felt that its length and jargon were designed to deter them from reading. The comments from some participants were exemplary: \quotes{\textit{Trying to go through the policy document and isolate what's important is a nightmare}} (P058, \textit{OpenAI}); \quotes{\textit{These things are unfortunately always so long winded it is almost impossible to fully concentrate on the key points, far too much data to take in}} (P206, \textit{TikTok}); \quotes{\textit{If this popped up, I would ignore it as I normally do}} (P094, \textit{OpenAI}).

% One participant suggested it would be \quotes{\textit{better if using colour or use more examples to explain things}} (P059, \textit{OpenAI}).

% It was so clear and easy to find the relevant information to address my concerns. 

\paragraph{\textbf{Nutrition Label Variants}}

The \textit{Nutrition Label} (NL) and \textit{Nutrition Label + Text} (NL+T) formats were often described as faster to scan and easier to digest: \quotes{\textit{it was so easy to understand and read fast.}} (P016, NL+T, \textit{OpenAI}); \quotes{\textit{I find this format much easier to understand than the traditional format. It's very clear and helpful.}} (P036, NL, \textit{TikTok}); \quotes{\textit{Reading through pages of text is tedious and time consuming and so many people don't do it because of that, but this format made it so simple and engaging to find what you want without getting bored first.}} (P076, NL, \textit{OpenAI}). Some anticipated higher future use: \quotes{\textit{This would probably mean that I'd read more privacy policies in the future if they were in this format.}} (P023, NL, \textit{OpenAI}).  

% ; \quotes{\textit{A drop-down would be more accessible}} (P330, \textit{OpenAI})

At the same time, two friction points recurred. First, \textit{explanatory depth}: respondents wanted clearer definitions and more context (P004, P074), and some felt the grid remained \quotes{\textit{over complicated}} (P147). Second, \textit{interaction cost}: hover-to-reveal details was widely disliked and sometimes interpreted as concealing information: \quotes{\textit{It was inconvenient to hover over each box to see what data may or may not be collected.}} (P095, NL+T, \textit{OpenAI}); \quotes{\textit{Felt like it was trying to hide things}} (P309, \textit{TikTok}). Some also found these formats cognitively demanding: \quotes{\textit{Better than endless text, but still mentally demanding}} (P093, \textit{OpenAI}); \quotes{\textit{Too much information to digest}} (P026, \textit{TikTok}). Overall sentiment framed nutrition labels as a \quotes{\textit{step in the right direction}} but still imperfect and \quotes{\textit{a little ambiguous}} (P044, NL, \textit{OpenAI}; P410, NL+T, \textit{TikTok}).

\paragraph{\textbf{Scrollytelling}} This format was frequently described as more engaging and approachable than conventional ones. Participants often highlighted its clarity and readability: \quotes{\textit{Much easier to read in this format and highlights the key points in a much more user friendly way.}} (P134, \textit{TikTok}), and appreciated how effectively it highlighted key information: \quotes{\textit{It is well structured and emphasizes the most important aspects. Also, it has nice navigation and helper which is an advantage.}} (P395, \textit{OpenAI}); \quotes{\textit{It definitely helps in terms of drawing attention to key areas of the policy and directly attributes sections to how they impact on the individual.}} (P426, \textit{TikTok}). Others praised the visuals and how it breaks down the policy content: \quotes{\textit{Great to have visuals over large bulks of text which becomes boring to read.}} (P123, \textit{OpenAI}); \quotes{\textit{I loved the colour coded layout to explain different data aspects and who they're shared with; and the various symbols used to make the information stand out and take up less wording.}} (P208, \textit{OpenAI}); \quotes{\textit{This was certainly more user friendly as it broke down the categories of information collected in a readable way.}} (P380, \textit{TikTok}). Some even reported higher engagement: \quotes{\textit{I've tried to read through Privacy Policy's before in a standard text format. Using this system I was happy to engage and read all aspects and it even got me asking further questions. This format is definitely the best I've come across and I've spent more time on this than any other privacy policy before.}} (P070, \textit{OpenAI}).  

% ; \quotes{\textit{The format was more `Cannot see the wood for the trees'--too overwhelming to encourage engagement.}} (P203, \textit{TikTok})

Nevertheless, some participants found some mechanics tedious or distracting: \quotes{\textit{Too many transitions, I wished for it to be more streamlined.}} (P118, \textit{OpenAI}); \quotes{\textit{Very difficult to scroll through, seemed deliberately long winded and obtuse.}} (P008, \textit{TikTok}). Others reported the length and density of content still overwhelmed them: \quotes{\textit{It was still quite long and not the easiest to digest.}} (P124, \textit{OpenAI}); \quotes{\textit{Still too much info and not easy to understand. Why do these policies need to be so long winded?}} (P197, \textit{TikTok}). 
A recurring theme was that while scrollytelling could make the policy more digestible, it did not eliminate the inherent burden of privacy policies themselves. Several participants emphasized that no design could truly fix the problem of length and legal jargon: \quotes{\textit{All privacy policies are boring to read as they must contain so much content to cover themselves legally. I do not think there is a way to make them less boring.}} (P088, \textit{OpenAI}); \quotes{\textit{Format is not the issue, the context is. Policies are quite lengthy and difficult and boring to read through.}} (P400, \textit{TikTok}); \quotes{\textit{What a challenge... trying to make a Privacy Policy interesting. You helped a bit but it is what it is.}} (P204, \textit{OpenAI}).

\paragraph{\textbf{Interactive Vis}}

Reactions to this format were mixed to positive. Participants highlighted its clarity: \quotes{\textit{It put the policy in terms of what it actually does}} (P062, \textit{OpenAI}); \quotes{\textit{It seemed less designed to be hidden. The reading also felt less onerous than I expected.}} (P120, \textit{OpenAI}); \quotes{\textit{The format is very engaging. It requires attention and I find it to be very effective.}} (P438, \textit{OpenAI}). Others stressed usefulness for targeted information seeking: \quotes{\textit{I found it easier to find key facts and so if there was a particular worry that I had about my data being shared I would be able to look into it straightaway for more details.}} (P065, \textit{OpenAI}); \quotes{\textit{By being able to navigate to specific areas of the policy (such as `Financial Data' or `Location'), you are able to go straight to the information you are seeking or require clarification on.}} (P271, \textit{OpenAI}); \quotes{\textit{I could jump to specific parts easily. It made me more likely to look at the privacy policy rather than just skip through it.}} (P205, \textit{TikTok}). Some participants, however, noted unfamiliarity: \quotes{\textit{Confusing at first because I hadn’t seen this format}} (P084, \textit{OpenAI}); \quotes{\textit{Once I understood it, the task was easier}} (P162, \textit{TikTok}). As with other formats, some felt the volume of information remained daunting (P077, P024), though several still viewed it as an improvement over text (P243, P322, P342, P366).

\subsubsection{Perceptions of the Policy Owner}

Participants' comments also revealed distinct perceptions of the policy owners, often tied to existing public opinion and the perceived transparency of the policies. For OpenAI, a common theme was a growing sense of awareness and concern about data collection: \quotes{\textit{It has surprised me how much data Open AI collect, its much more than I realized. I also didn't know how they can share it and what its used for... Its a bit of a double edged sword as now I know that has in turn worried me more.}} (P070, \textit{Scrollytelling}, \textit{OpenAI}); \quotes{\textit{When reading this format I was extremely shocked at the amount of information that OpenAI has about a person.}} (NL, P093, \textit{OpenAI}).

For TikTok, the perception was often one of inherent distrust: \quotes{\textit{I frankly don't trust TikTok. I am not confident that it actually does what it says it does.}} (P003, NL+T, \textit{TikTok}); \quotes{\textit{the amount of information concealed behind each box made it feel as if TikTok were deliberately trying to hide things from me.}} (NL, P078, \textit{TikTok}); \quotes{\textit{innovative format but deceptive company.}} (P020, NL+T, \textit{TikTok}). In a few cases, the format was rated positively while the owner verdict remained negative: \quotes{\textit{very good format in terms of understaning the policies. At the same time, very good format in terms on coming to conclusion I do not want to use tiktok any longer.}} (P109, \textit{Scrollytelling}, \textit{TikTok}).

%%%% Sections/Results/054_Feedback.tex ends here %%%%

%%%% Sections/060-Results.tex ends here %%%%

%%%% 070-Discussion.tex starts here %%%%

\section{Discussion}
\label{sec:discussion}

In this section, we first summarize and interpret our results. We then situate \emph{scrollytelling} within the broader landscape of privacy policy communication formats, and reflect on potential deployment models in real-world settings. We close with a discussion of limitations and directions for future research.

\subsection{Summary and Interpretation of Findings}

Across formats, participants answered most comprehension items correctly, with only modest separation between conditions. The \textit{Nutrition Label} achieved the highest model-based accuracy (78.1\%), yielding a small but reliable advantage over \textit{Scrollytelling} (75.5\%). Other formats were statistically indistinguishable from \textit{Scrollytelling} on comprehension accuracy. The \textit{Nutrition Label} also produced the highest predicted comprehension confidence, whereas the remaining formats clustered near the \textit{Scrollytelling} baseline. These results hold when averaging over policy owner and item type.

Our completion-time analyses add nuance to this picture. Participants in the \textit{Scrollytelling} and \textit{Text} conditions completed the comprehension questionnaire fastest (median $\approx$ 240s), while \textit{Interactive Vis} required substantially more time (median $\approx$ 328 seconds) and showed greater variability (Figure~\ref{fig:CompletionTinesComprehension}). The nutrition label variants fell in between, descriptively slower than \textit{Scrollytelling} and \textit{Text} but not significantly different after correction. Post-hoc tests confirmed that only \textit{Interactive Vis} was reliably slower than both \textit{Text} and \textit{Scrollytelling}; all other contrasts were non-significant. Taken together, \textit{Scrollytelling} matched \textit{Text} on both accuracy and completion time, while \textit{Interactive Vis}, which lacked a coherent narrative structure, imposed a clear time cost without improving comprehension.

% , consistent with our confirmatory focus on \textit{Format} main effects. 

Subjective experience differed more clearly: \textit{Scrollytelling} delivered higher \textit{Enjoyment}, \textit{Engagement}, and \textit{Perceived Clarity} than \textit{Text} and, generally, strong format adoption intentions. Two frictions qualify these gains. First, participants frequently mentioned the effort of continuous scrolling (Section~\ref{sec:results:feedback}). Second, hidden or hover-to-reveal details in the nutrition label variants were perceived as extra work. Both frictions align with classic trade-offs in narrative and interactive visualization: sequencing and interaction can foster understanding, but they cost time and attention. 

The small shifts in \textit{Perceived Transparency}, \textit{Confidence in Understanding}, and \textit{Trust} likely reflect two constraints. First, these constructs are coarse attitudes with high prior inertia; shifting them may require stronger or repeated exposures than a single, brief exploration. Second, our models averaged over owners and item types. Thus, any subtle interaction (e.g., narrative helping most when policies contain many conditional clauses or exceptions) would be washed out by design. These small pre--post shifts in all perception constructs may also suggest that exposure helped uncertain participants consolidate views more than broadly changing their minds. 

% narrative scaffolding being especially helpful when policies contain many conditional clauses or exceptions

% We also observed substantial differences across policy \textit{owners}, consistent across formats  as reported by the Comprehension (Section~\ref{sec:results:comprehension}) and Confidence (Section~\ref{sec:results:experience}) models. These effects underscore that the policy's content and data practices significantly constrain what users can infer---regardless of how that content is presented. In this light, format alone is no silver bullet: gains from layout or narrative structure remain limited by the specificity, clarity, and stability of the underlying policy text.

We also observed large, consistent differences by policy owner that cut across formats. Interpretive comprehension items were systematically harder for OpenAI than for TikTok, a pattern visible in the owner-stratified accuracy estimates and model coefficients reported in Section~\ref{sec:results:comprehension} and mirrored by the self-reported Confidence models of Section~\ref{sec:results:experience}. These effects suggest that what users can infer may be bounded by the specificity, clarity, and stability of each owner's policy text and practices, not just by presentation. In this light, format alone is no silver bullet: narrative scaffolds, tabular arrangements, and visual summaries can reduce navigational and interpretive costs, but the upper bound on comprehension and confidence may be set by the underlying policy content. Two design corollaries follow. First, when the owner's policy uses conditional or ambiguous phrasing, formats should surface that uncertainty in place (not behind hover or page jumps) to prevent false certainty. Second, format evaluations should report owner-stratified results and include at least one policy with more precise language and one with more conditional language, so that format effects are not misread as content effects.

\subsection{Scrollytelling Within the Policy Format Landscape}

Prior work shows that long-form policy text imposes high cognitive and opportunity costs, often underperforming on usability and comprehension~\cite{McDonald2008Cost,Obar2016Biggest,Reidenberg2015Disagreeable}. In contrast, privacy nutrition labels consistently help users locate key facts more efficiently and report greater satisfaction~\cite{Kelley2009Nutrition,Kelley2010Standardizing,Schaub2017Designing,ReinhardtInteractivePP}. Meanwhile, narrative or comic-style formats tend to improve engagement and recall~\cite{Kay2010Textured,Knijnenburg2016Comics,Suh2022PrivacyToon}, but their effects on comprehension are mixed. Our findings place scrollytelling in the productive middle of this design space: it matches or exceeds other formats on comprehension, improves experience over plain text, and remains competitive on completion time with \textit{Text} and nutrition labels while clearly outperforming the interactive visualization in that regard.

Scrollytelling sits at the intersection of two design logics. From narrative visualization, it borrows pacing, ordering, and landmarks to reduce cold-start costs and orient readers~\cite{Segel2010NarrativeVisualization,Hullman2011Rhetoric,Tjarnhage2023Scrollytelling}. From accountable simplification, it preserves traceability to the source, enabling readers to verify visual claims. Our tool encoded these ideas as design principles: anchor simplifications in the source (\textbf{DP1}), unfold complexity step-by-step (\textbf{DP2}), scaffold orientation (\textbf{DP3}), visualize uncertainty (\textbf{DP4}), ensure bidirectional traceability (\textbf{DP5}), and defer heavy interaction until a scaffold exists (\textbf{DP6}). This combination likely explains why scrollytelling improved engagement and clarity without sacrificing accuracy or increasing completion time: orientation and verification reduced interpretive load, even as the narrative added some interaction cost.

These same principles may help explain the performance of the \emph{Interactive Vis}, which we introduced to isolate the narrative's contribution by omitting the narrative build-up, letting us assess the value of scaffolding and sequencing. Comprehension accuracy was statistically similar to both scrollytelling and plain text (Section~\ref{sec:results:comprehension}), and experience scores were roughly comparable---slightly higher behavioral intentions, but a modest uptick in cognitive load. Our completion-time analysis (Section~\ref{sec:results:comprehension-time}), however, showed that participants spent substantially longer answering the comprehension questions with the \textit{Interactive Vis} than those in the  \textit{Text} or \textit{Scrollytelling} conditions, without commensurate gains in accuracy. Two mechanisms likely account for this. First, narrative sequencing may serve as \emph{pre-training}, introducing key actors, data types, and relationships in a fixed order that helps readers build a mental model~\cite{Segel2010NarrativeVisualization,Hullman2011Rhetoric,Mayer2009MultimediaLearning}. Second, without that guidance, users face a higher interaction burden: they must decide where to begin, what to open, and how to explore, which increases cognitive load and takes time---especially when details are hidden or distributed. Prior work and our feedback suggest such unstructured exploration can suffer from split attention and uncertainty about coverage. The takeaway is not that the visualization is weak, but that scrollytelling adds value through orientation and pacing. For fast lookup, a static summary may suffice; for context and reasoning, lightweight narrative scaffolding appears to help.

Finally, the contrast between the two nutrition-label variants clarifies why only one surpassed the \textit{Scrollytelling} format on comprehension accuracy and confidence. \textit{Nutrition Label} outperformed, while \textit{Nutrition Label + Text} did not, even though the latter offered more information via a prominent link to the full policy. Our logs showed that readers used that link heavily: overall, 90\% followed it at least once. This supports two inferences. First, the compact grid in the \textit{Nutrition Label} condition creates a minimal fast path that reduces decisions and shortens time-to-answer. Adding a salient jump to the full policy introduces branching and context switches that can blunt the scanning advantage, even if our timing data show only small, non-significant differences between the two label variants. Second, both nutrition label variants carried interaction costs, and participants criticized hover-to-reveal for hiding details. When a full-page link is also present, the combined overhead rises. The design implication is to keep core lookup self-contained and to keep provenance in place. If access to the source text is required, anchored previews or side-by-side excerpts may be preferable over page-level navigation.

\subsection{Towards the Adoption of Scrollytelling for Privacy Policies}
\label{sec:TowardsAdoption}

Our results highlight a clear gap between what proves effective in controlled settings and what is typically deployed in real-world contexts. While text policies remain the standard on most websites, the text condition in our study consistently underperformed across most subjective measures (engagement, enjoyment, clarity, perceived adoption, and cognitive load) compared to \textbf{all} alternative formats. In this light, scrollytelling stands out as a compelling alternative to conventional privacy policy presentations. However, moving from a research prototype to real-world deployment requires addressing who would author these narratives, whether they can scale, and how to sustain user engagement over time.

As described in Section~\ref{sec:Implementation}, our tool generates narratives from two structured inputs grounded in the policy text. This design ensures that narrative elements are traceable to specific clauses. However, that traceability comes at an authoring cost involving three manual steps: identifying data-source facets, defining categories for data and actors, and extracting a data collection and sharing graph. While established NLP tools (e.g.,~\cite{Cui2023PoliGraph, Harkous2018Polisis}) or newer LLM-based approaches (e.g.,~\cite{Xie2025LLMBased}) could help automate parts of this process, human oversight remains essential to maintain narrative clarity and prevent misrepresentation. Semi-automated authoring workflows, in which humans refine machine-generated suggestions, may offer a practical compromise. This direction builds on established precedents. For instance, the W3C's Platform for Privacy Preferences (P3P) defined a machine-readable vocabulary to let user agents automatically retrieve and interpret websites' privacy practices~\cite{CranorP3P2008, Cranor2002P3P}. Its limited adoption, however, also illustrates the challenges of adopting standardized privacy vocabularies.

Once the required structured inputs are in place, the rest of the pipeline runs without further curation, generating a narrative ready for user-driven navigation. Separating authoring from rendering enables a range of deployment models. In one model, policy owners could author and maintain the structured inputs themselves, publishing the scrollytelling narrative alongside the official policy to ensure alignment with the full text. In an alternative model, independent third parties (e.g., researchers, nonprofits, consumer advocacy groups) could take on the authoring role. This could help make disclosures more accessible, especially for services unlikely to adopt the format themselves. However, shifting authorship outside policy owners raises questions about authority, long-term maintenance, and legal accountability.

To support broader adoption across both models, structured inputs will likely need standardized, publicly available schemas that specify structural elements such as actors, data types, and data practices. Such schemas would also need to encode explicit choices about narrative focus. The narrative of our tool centers on data sources, collection, and sharing, but privacy policies address many other topics that could also benefit from visual explanations. One of our workshop participants, for example, envisioned a narrative that also foregrounded how data is processed, safeguarded, transferred, and how users can exercise their rights (Figure~\ref{figure:workshop1:sketch04}). A standard schema for scrollytelling-based privacy policies would need to accommodate this breadth, specifying not only the entities and actions involved but also how policy fragments are presented and linked within the narrative. Standardization would, in turn, enable interoperability: the same inputs could be rendered by different tools, and a given narrative could be adapted to distinct audiences or use cases. For instance, a child-oriented version of our narrative might replace the visualization of Section~\ref{sect:InteractiveVisualization} with a more ludic element that prioritizes conceptual clarity over precision. Developing such schemas and their underlying taxonomies is therefore a necessary step toward making scrollytelling for privacy policies scalable, adaptable, and inclusive.

% To support broader adoption across both models, structured inputs will likely need to follow standardized, publicly available schemas that define structural elements, such as actors and data types, as well as data practices. Currently, there are no widely accepted schemas of this type. Our current tool encodes only one slice of a policy: data sources, collection, and sharing. In practice, privacy policies also describe how data is processed, safeguarded, transferred, and how users can exercise their rights. One of our workshop participants, for example, envisioned a narrative that foregrounded exactly these aspects (Figure~\ref{figure:workshop1:sketch04}). A standard schema would need to encode this diversity, specifying not only the entities and actions involved but also how policy fragments are presented and linked in the narrative. Standardization would also enable interoperability: the same inputs could be rendered by different tools, and the same narrative could be tailored to distinct audiences or use cases. For example, a version aimed at children might replace the interactive visualization of Section~\ref{sect:InteractiveVisualization} with a more playful element that prioritizes conceptual clarity over precision. Developing such schemas and their underlying taxonomies is therefore a necessary step toward making scrollytelling for privacy policies scalable, adaptable, and inclusive.

A further consideration for real-world adoption is habituation. As with any format, engagement may decline with repeated exposure. While scrollytelling does not eliminate this risk, its dynamic structure may offer partial mitigation. Unlike static formats (e.g., nutrition labels), scrollytelling allows for lightweight variation in layout, pacing, and sequencing even when the underlying data remains constant. This variability could help delay habituation, particularly when narratives are adapted to different audiences, devices, or contexts. Deployment strategies may also play a role. For example, scrollytelling narratives could be embedded selectively (e.g., shown only during onboarding or when data practices change) rather than as the default presentation for all users at all times. Scalability is closely linked to this issue. By separating authoring from rendering and relying on structured inputs, our approach could support reusable narrative engines with tailored surface elements. Open-sourcing the tool and input schemas in the future would also enable broader experimentation and help counter habituation through variation and personalization.

\subsection{Limitations and Open Questions}

Our study evaluated privacy policy formats under conditions designed to balance ecological realism and experimental control. While this yields actionable insights, several limitations and open questions remain. Our use of a between-subjects design minimized cross-condition contamination, particularly for perception measures, but it also reduced power to detect subtle differences and prevented within-participant contrasts. A natural extension would be a within-subjects design in which participants experience two formats for the same policy. Such a design would enable direct comparisons of usability and clarity, reveal transfer and retention effects, and test whether narrative sequencing provides measurable advantages when alternatives are visible side by side.

We also used real privacy policies from widely known platforms (OpenAI and TikTok), which allowed us to capture pre-post perception shifts in a realistic setting. At the same time, this choice introduces confounds: some answers may have reflected prior knowledge or stable attitudes rather than comprehension during the study task. Future work could explore this tradeoff explicitly, for example, by contrasting real and fictional policies or measuring how prior familiarity modulates comprehension and trust.

Furthermore, our study does not test whether scrollytelling leads to \textit{better} privacy decisions (e.g., different settings choices, app adoption, or data-sharing behavior). Addressing this would require tasks where participants face concrete trade-offs and where \quotes{better} or \quotes{more accurate} decisions are defined relative to their stated preferences, privacy concerns, and risk tolerance, which was beyond the scope of this work. Instead, we focused on the necessary conditions for any decision-making impact: whether a narrative, visual format can match or surpass the benefits of traditional textual notices and existing alternatives. We view our findings as a necessary, albeit insufficient, step toward addressing the broader question of how privacy notices influence real-world decisions. Future work could couple our format with consequential decision tasks and follow-up measures of choices to evaluate behavioral impact.

Another limitation lies in how we assessed interaction costs. We relied primarily on subjective scales and qualitative feedback, and while we analyzed time-on-task measures, we did not assess detailed navigation traces. A more comprehensive understanding of how narrative scaffolding shapes effort would benefit from measures such as scroll depth, dwell time, and interaction sequence analysis.

Finally, our study focused on English-speaking participants on desktop web. This leaves open how scrollytelling performs on mobile platforms, which lack hover interactions and impose different pacing and screen real estate constraints, as well as in localized versions across languages and privacy cultures.

These limitations open questions about how scrollytelling for privacy policies can be adapted and deployed. Our results suggest that its strength lies in providing orientation and traceability, raising key design questions: Can narrative length adjust to user expertise, offering shorter versions for return visits? Could static labels support an optional handoff for added context? Might mobile flows use taps or swipes instead of scrolling? And, beyond one-off exposures, what value might emerge if scrollytelling were embedded into repeated interactions, such as onboarding, consent prompts, or updates? All these questions are exciting avenues for future research.

%%%% Sections/070-Discussion.tex ends here %%%%

%%%% 080-Conclusion.tex starts here %%%%

\section{Conclusion}

In this paper, we investigated the feasibility of \textit{Scrollytelling} as a format for privacy policy communication. Drawing on principles of narrative scaffolding and accountable simplification, we developed a prototype that interleaves policy text with animated, traceable visual explanations. The tool served as a concrete instantiation to test whether narrative pacing and structured visual framing can help users better navigate and comprehend privacy policies.

In a between-subjects, online study with 454 participants, we compared scrollytelling against four alternatives: full-text, two nutrition-label variants, and a standalone interactive visualization. Comprehension scores were high across all formats. The \textit{Nutrition Label} format yielded slightly higher accuracy, but scrollytelling improved subjective experience, with gains in engagement, perceived clarity, and lower cognitive load. The same visualization, when shown without narrative scaffolding, performed slightly worse---suggesting that orientation, framing, and sequencing are critical to how users process policy content. Our results also reveal large differences by policy owner, underscoring that design cannot fully compensate for vague or opaque language.

Taken together, our findings position scrollytelling as a promising format for privacy policy presentation, one that preserves comprehension while enhancing usability, supporting traceability, and easing cognitive effort. Beyond our specific implementation, our study contributes empirical evidence that narrative formats can play a productive role in privacy communication, bridging the gap between usability and transparency.

%%%% Sections/080-Conclusion.tex ends here %%%%

%%%% 090-Acknowledgements.tex starts here %%%%

\begin{acks}

This work was supported by INCIBE's strategic SPRINT (\textit{Seguridad y Privacidad en Sistemas con Inteligencia Artificial}) C063/23 project with funds from the EU-NextGenerationEU through the Spanish government's \textit{Plan de Recuperaci\'{o}n, Transformaci\'{o}n y Resiliencia}, by the Spanish Government project PID2023-151536OB-I00, and by the Generalitat Valenciana's project PROMETEO CIPROM/2023/23.

\end{acks}

%%%% Sections/090-Acknowledgements.tex ends here %%%%

\bibliographystyle{ACM-Reference-Format}

%%%% bibliography starts here %%%%

%%% -*-BibTeX-*-
%%% Do NOT edit. File created by BibTeX with style
%%% ACM-Reference-Format-Journals [18-Jan-2012].

%\bibliography{000-bibliography}%% Commented by merge tool

\begin{thebibliography}{76}

%%% ====================================================================
%%% NOTE TO THE USER: you can override these defaults by providing
%%% customized versions of any of these macros before the \bibliography
%%% command.  Each of them MUST provide its own final punctuation,
%%% except for \shownote{} and \showURL{}.  The latter two
%%% do not use final punctuation, in order to avoid confusing it with
%%% the Web address.
%%%
%%% To suppress output of a particular field, define its macro to expand
%%% to an empty string, or better, \unskip, like this:
%%%
%%% \newcommand{\showURL}[1]{\unskip}   % LaTeX syntax
%%%
%%% \def \showURL #1{\unskip}           % plain TeX syntax
%%%
%%% ====================================================================

\ifx \showCODEN    \undefined \def \showCODEN     #1{\unskip}     \fi
\ifx \showISBNx    \undefined \def \showISBNx     #1{\unskip}     \fi
\ifx \showISBNxiii \undefined \def \showISBNxiii  #1{\unskip}     \fi
\ifx \showISSN     \undefined \def \showISSN      #1{\unskip}     \fi
\ifx \showLCCN     \undefined \def \showLCCN      #1{\unskip}     \fi
\ifx \shownote     \undefined \def \shownote      #1{#1}          \fi
\ifx \showarticletitle \undefined \def \showarticletitle #1{#1}   \fi
\ifx \showURL      \undefined \def \showURL       {\relax}        \fi
% The following commands are used for tagged output and should be
% invisible to TeX
\providecommand\bibfield[2]{#2}
\providecommand\bibinfo[2]{#2}
\providecommand\natexlab[1]{#1}
\providecommand\showeprint[2][]{arXiv:#2}

\bibitem[Abu-Salma et~al\mbox{.}(2025)]%
        {abu-salma2025grand}
\bibfield{author}{\bibinfo{person}{Ruba Abu-Salma}, \bibinfo{person}{Pauline
  Anthonysamy}, \bibinfo{person}{Zinaida Benenson}, \bibinfo{person}{Benjamin
  Berens}, \bibinfo{person}{Kovila P.~L. Coopamootoo}, \bibinfo{person}{Andreas
  Gutmann}, \bibinfo{person}{Adam Jenkins}, \bibinfo{person}{Sameer Patil},
  \bibinfo{person}{Sören Preibusch}, \bibinfo{person}{Florian Schaub},
  \bibinfo{person}{William Seymour}, \bibinfo{person}{Jose Such},
  \bibinfo{person}{Mohammad Tahaei}, \bibinfo{person}{Aybars Tuncdogan},
  \bibinfo{person}{Max~Van Kleek}, {and} \bibinfo{person}{Daricia Wilkinson}.}
  \bibinfo{year}{2025}\natexlab{}.
\newblock \showarticletitle{Grand Challenges in Human-Centered Privacy}.
\newblock \bibinfo{journal}{\emph{IEEE Security \& Privacy}}
  \bibinfo{volume}{23}, \bibinfo{number}{4} (\bibinfo{year}{2025}),
  \bibinfo{pages}{103--110}.
\newblock
\href{https://doi.org/10.1109/MSEC.2025.3566451}{doi:\nolinkurl{10.1109/MSEC.2025.3566451}}


\bibitem[Acquisti et~al\mbox{.}(2015)]%
        {acquisti2015privacy}
\bibfield{author}{\bibinfo{person}{Alessandro Acquisti}, \bibinfo{person}{Laura
  Brandimarte}, {and} \bibinfo{person}{George Loewenstein}.}
  \bibinfo{year}{2015}\natexlab{}.
\newblock \showarticletitle{Privacy and human behavior in the age of
  information}.
\newblock \bibinfo{journal}{\emph{Science}} \bibinfo{volume}{347},
  \bibinfo{number}{6221} (\bibinfo{year}{2015}), \bibinfo{pages}{509--514}.
\newblock
\href{https://doi.org/10.1126/science.aaa1465}{doi:\nolinkurl{10.1126/science.aaa1465}}


\bibitem[Adhikari et~al\mbox{.}(2023)]%
        {adhikari2023evolution}
\bibfield{author}{\bibinfo{person}{Andrick Adhikari}, \bibinfo{person}{Sanchari
  Das}, {and} \bibinfo{person}{Rinku Dewri}.} \bibinfo{year}{2023}\natexlab{}.
\newblock \showarticletitle{Evolution of composition, readability, and
  structure of privacy policies over two decades}.
\newblock \bibinfo{journal}{\emph{Proceedings on Privacy Enhancing
  Technologies}} (\bibinfo{year}{2023}).
\newblock
\href{https://doi.org/10.56553/popets-2023-0074}{doi:\nolinkurl{10.56553/popets-2023-0074}}


\bibitem[Adhikari et~al\mbox{.}(2025)]%
        {Adhikari2023PolicyPulse}
\bibfield{author}{\bibinfo{person}{Andrick Adhikari}, \bibinfo{person}{Sanchari
  Das}, {and} \bibinfo{person}{Rinku Dewri}.} \bibinfo{year}{2025}\natexlab{}.
\newblock \showarticletitle{{PolicyPulse}: Precision Semantic Role Extraction
  for Enhanced Privacy Policy Comprehension}. In
  \bibinfo{booktitle}{\emph{Network and Distributed System Security Symposium
  (NDSS)}}.
\newblock
\href{https://doi.org/10.14722/ndss.2025.240916}{doi:\nolinkurl{10.14722/ndss.2025.240916}}


\bibitem[Anaraky et~al\mbox{.}(2019)]%
        {Anaraky2019TestingAComic}
\bibfield{author}{\bibinfo{person}{Reza~Ghaiumy Anaraky},
  \bibinfo{person}{David Cherry}, \bibinfo{person}{Marie Jarrell}, {and}
  \bibinfo{person}{Bart~P. Knijnenburg}.} \bibinfo{year}{2019}\natexlab{}.
\newblock \showarticletitle{Testing a Comic-Based Privacy Policy}. In
  \bibinfo{booktitle}{\emph{SOUPS 2019 Posters}}.
\newblock
\urldef\tempurl%
\url{https://www.usenix.org/sites/default/files/soups2019posters-anaraky.pdf}
\showURL{%
\tempurl}


\bibitem[{Bloomberg}(2015)]%
        {Bloomberg2015USAutoSales}
\bibfield{author}{\bibinfo{person}{{Bloomberg}}.}
  \bibinfo{year}{2015}\natexlab{}.
\newblock \bibinfo{booktitle}{\emph{US Auto Sales}}.
\newblock Bloomberg.
\newblock
\urldef\tempurl%
\url{https://www.bloomberg.com/graphics/2015-auto-sales}
\showURL{%
\tempurl}
\newblock
\shownote{Bloomberg Graphics}.


\bibitem[Bui et~al\mbox{.}(2021)]%
        {Bui2021PIExtract}
\bibfield{author}{\bibinfo{person}{Duc Bui}, \bibinfo{person}{Kang~G. Shin},
  \bibinfo{person}{Jong{-}Min Choi}, {and} \bibinfo{person}{Junbum Shin}.}
  \bibinfo{year}{2021}\natexlab{}.
\newblock \showarticletitle{Automated Extraction and Presentation of Data
  Practices in Privacy Policies}.
\newblock \bibinfo{journal}{\emph{Proceedings on Privacy Enhancing Technologies
  (PoPETs)}} \bibinfo{volume}{2021}, \bibinfo{number}{2}
  (\bibinfo{year}{2021}), \bibinfo{pages}{88--110}.
\newblock
\href{https://doi.org/10.2478/popets-2021-0019}{doi:\nolinkurl{10.2478/popets-2021-0019}}


\bibitem[Chmielewski and Kucker(2020)]%
        {Chmielewski2020}
\bibfield{author}{\bibinfo{person}{Michael Chmielewski} {and}
  \bibinfo{person}{Sarah~A. Kucker}.} \bibinfo{year}{2020}\natexlab{}.
\newblock \showarticletitle{An MTurk Crisis? Shifts in Data Quality and the
  Impact on Study Results}.
\newblock \bibinfo{journal}{\emph{Social Psychological and Personality
  Science}} \bibinfo{volume}{11}, \bibinfo{number}{4} (\bibinfo{year}{2020}),
  \bibinfo{pages}{464--473}.
\newblock
\href{https://doi.org/10.1177/1948550619875149}{doi:\nolinkurl{10.1177/1948550619875149}}


\bibitem[Chow({[n.\,d.]})]%
        {ChowBestCollegeMajors}
\bibfield{author}{\bibinfo{person}{Cuthbert Chow}.}
  \bibinfo{year}{[n.\,d.]}\natexlab{}.
\newblock \bibinfo{booktitle}{\emph{Best College Majors}}.
\newblock
\urldef\tempurl%
\url{https://cuthchow.github.io/college-majors-visualisation/}
\showURL{%
\tempurl}


\bibitem[Cranor et~al\mbox{.}(2002)]%
        {Cranor2002P3P}
\bibfield{author}{\bibinfo{person}{Lorrie Cranor}, \bibinfo{person}{Marc
  Langheinrich}, \bibinfo{person}{Massimo Marchiori}, \bibinfo{person}{Martin
  Presler-Marshall}, {and} \bibinfo{person}{Joseph Reagle}.}
  \bibinfo{year}{2002}\natexlab{}.
\newblock \bibinfo{title}{The Platform for Privacy Preferences 1.0 (P3P1.0)
  Specification}.
\newblock
\urldef\tempurl%
\url{https://www.w3.org/TR/P3P/}
\showURL{%
\tempurl}
\newblock
\shownote{W3C Recommendation, 16 April 2002 (obsoleted 30 August 2018)}.


\bibitem[Cranor(2012)]%
        {Cranor2012Necessary}
\bibfield{author}{\bibinfo{person}{Lorrie~Faith Cranor}.}
  \bibinfo{year}{2012}\natexlab{}.
\newblock \showarticletitle{Necessary But Not Sufficient: Standardized
  Mechanisms for Privacy Notice and Choice}.
\newblock \bibinfo{journal}{\emph{Journal on Telecommunications and High
  Technology Law}}  \bibinfo{volume}{10} (\bibinfo{year}{2012}),
  \bibinfo{pages}{273--304}.
\newblock
\urldef\tempurl%
\url{https://scholar.law.colorado.edu/ctlj/vol10/iss2/7/}
\showURL{%
\tempurl}


\bibitem[Cranor et~al\mbox{.}(2008)]%
        {CranorP3P2008}
\bibfield{author}{\bibinfo{person}{Lorrie~Faith Cranor}, \bibinfo{person}{Serge
  Egelman}, \bibinfo{person}{Steve Sheng}, \bibinfo{person}{Aleecia~M.
  McDonald}, {and} \bibinfo{person}{Abdur Chowdhury}.}
  \bibinfo{year}{2008}\natexlab{}.
\newblock \showarticletitle{deployment on websites}.
\newblock \bibinfo{journal}{\emph{Electron. Commer. Rec. Appl.}}
  \bibinfo{volume}{7}, \bibinfo{number}{3} (\bibinfo{date}{Nov.}
  \bibinfo{year}{2008}), \bibinfo{pages}{274–293}.
\newblock
\showISSN{1567-4223}
\href{https://doi.org/10.1016/j.elerap.2008.04.003}{doi:\nolinkurl{10.1016/j.elerap.2008.04.003}}


\bibitem[Cui et~al\mbox{.}(2023)]%
        {Cui2023PoliGraph}
\bibfield{author}{\bibinfo{person}{Hao Cui}, \bibinfo{person}{Rahmadi
  Trimananda}, \bibinfo{person}{Athina Markopoulou}, {and}
  \bibinfo{person}{Scott Jordan}.} \bibinfo{year}{2023}\natexlab{}.
\newblock \showarticletitle{PoliGraph: Automated Privacy Policy Analysis using
  Knowledge Graphs}. In \bibinfo{booktitle}{\emph{32nd USENIX Security
  Symposium}}.
\newblock
\urldef\tempurl%
\url{https://www.usenix.org/system/files/usenixsecurity23-cui.pdf}
\showURL{%
\tempurl}


\bibitem[Dincelli and Chengalur-Smith(2019)]%
        {Dincelli2019ChooseYourHackingAdventure}
\bibfield{author}{\bibinfo{person}{Ersin Dincelli} {and}
  \bibinfo{person}{InduShobha Chengalur-Smith}.}
  \bibinfo{year}{2019}\natexlab{}.
\newblock \showarticletitle{Choose your own hacking adventure: Contextualized
  storytelling to enhance security education and training}. In
  \bibinfo{booktitle}{\emph{Proceedings of the 15th Symposium on Usable Privacy
  and Security (SOUPS)}}.
\newblock


\bibitem[Eiband et~al\mbox{.}(2018)]%
        {Eiband2018TransparencyDesign}
\bibfield{author}{\bibinfo{person}{Malin Eiband}, \bibinfo{person}{Hanna
  Schneider}, \bibinfo{person}{Mark Bilandzic}, \bibinfo{person}{Julian
  Fazekas-Con}, \bibinfo{person}{Mareike Haug}, {and} \bibinfo{person}{Heinrich
  Hussmann}.} \bibinfo{year}{2018}\natexlab{}.
\newblock \showarticletitle{Bringing Transparency Design into Practice}. In
  \bibinfo{booktitle}{\emph{Proceedings of the 23rd International Conference on
  Intelligent User Interfaces}} (Tokyo, Japan) \emph{(\bibinfo{series}{IUI
  '18})}. \bibinfo{publisher}{Association for Computing Machinery},
  \bibinfo{address}{New York, NY, USA}, \bibinfo{pages}{211–223}.
\newblock
\showISBNx{9781450349451}
\href{https://doi.org/10.1145/3172944.3172961}{doi:\nolinkurl{10.1145/3172944.3172961}}


\bibitem[Emami-Naeini et~al\mbox{.}(2020)]%
        {EmamiNaeini2020IoTLabels}
\bibfield{author}{\bibinfo{person}{Pardis Emami-Naeini},
  \bibinfo{person}{Yuvraj Agarwal}, \bibinfo{person}{Lorrie Faith~Cranor},
  {and} \bibinfo{person}{Hanan Hibshi}.} \bibinfo{year}{2020}\natexlab{}.
\newblock \showarticletitle{Ask the Experts: What Should Be on an IoT Privacy
  and Security Label?}. In \bibinfo{booktitle}{\emph{2020 IEEE Symposium on
  Security and Privacy (SP)}}. \bibinfo{pages}{447--464}.
\newblock
\href{https://doi.org/10.1109/SP40000.2020.00043}{doi:\nolinkurl{10.1109/SP40000.2020.00043}}


\bibitem[Feng et~al\mbox{.}(2021)]%
        {feng_design_2021}
\bibfield{author}{\bibinfo{person}{Yuanyuan Feng}, \bibinfo{person}{Yaxing
  Yao}, {and} \bibinfo{person}{Norman Sadeh}.} \bibinfo{year}{2021}\natexlab{}.
\newblock \showarticletitle{A Design Space for Privacy Choices: Towards
  Meaningful Privacy Control in the Internet of Things}. In
  \bibinfo{booktitle}{\emph{Proceedings of the 2021 {CHI} Conference on Human
  Factors in Computing Systems}} (Yokohama Japan, 2021-05-06).
  \bibinfo{publisher}{{ACM}}, \bibinfo{pages}{1--16}.
\newblock
\showISBNx{978-1-4503-8096-6}
\href{https://doi.org/10.1145/3411764.3445148}{doi:\nolinkurl{10.1145/3411764.3445148}}


\bibitem[Flagg et~al\mbox{.}(2014)]%
        {FlaggCraigBruno2014_CaliforniasGettingFracked}
\bibfield{author}{\bibinfo{person}{Anna Flagg}, \bibinfo{person}{Sarah Craig},
  {and} \bibinfo{person}{Antonia Bruno}.} \bibinfo{year}{2014}\natexlab{}.
\newblock \bibinfo{booktitle}{\emph{California’s Getting Fracked}}.
\newblock Faces of Fracking.
\newblock
\urldef\tempurl%
\url{http://www.facesoffracking.org/data-visualization}
\showURL{%
\tempurl}


\bibitem[{Guo, et al.}(2020)]%
        {Guo2020Polisee}
\bibfield{author}{\bibinfo{person}{{Guo, et al.}}}
  \bibinfo{year}{2020}\natexlab{}.
\newblock \showarticletitle{PoliSee: Visualizing Privacy Policies}. In
  \bibinfo{booktitle}{\emph{Proceedings of the 19th Workshop on Privacy in the
  Electronic Society (WPES '20)}}. \bibinfo{publisher}{ACM}.
\newblock
\href{https://doi.org/10.1145/3411497.3420221}{doi:\nolinkurl{10.1145/3411497.3420221}}


\bibitem[Habib et~al\mbox{.}(2021)]%
        {Habib2021Icons}
\bibfield{author}{\bibinfo{person}{Hana Habib}, \bibinfo{person}{Yixin Zou},
  \bibinfo{person}{Yaxing Yao}, \bibinfo{person}{Alessandro Acquisti},
  \bibinfo{person}{Lorrie~Faith Cranor}, \bibinfo{person}{Joel~R. Reidenberg},
  \bibinfo{person}{Norman Sadeh}, {and} \bibinfo{person}{Florian Schaub}.}
  \bibinfo{year}{2021}\natexlab{}.
\newblock \showarticletitle{Toggles, Dollar Signs, and Triangles: How to
  (In)Effectively Convey Privacy Choices with Icons and Link Texts}. In
  \bibinfo{booktitle}{\emph{CHI Conference on Human Factors in Computing
  Systems (CHI '21)}}.
\newblock
\href{https://doi.org/10.1145/3411764.3445387}{doi:\nolinkurl{10.1145/3411764.3445387}}


\bibitem[Harkous et~al\mbox{.}(2018)]%
        {Harkous2018Polisis}
\bibfield{author}{\bibinfo{person}{Hamza Harkous}, \bibinfo{person}{Kassem
  Fawaz}, \bibinfo{person}{R{\'e}mi Lebret}, \bibinfo{person}{Florian Schaub},
  \bibinfo{person}{Kang~G. Shin}, {and} \bibinfo{person}{Karl Aberer}.}
  \bibinfo{year}{2018}\natexlab{}.
\newblock \showarticletitle{Polisis: Automated Analysis and Presentation of
  Privacy Policies Using Deep Learning}. In \bibinfo{booktitle}{\emph{27th
  USENIX Security Symposium (USENIX Security 18)}}. \bibinfo{publisher}{USENIX
  Association}, \bibinfo{address}{Baltimore, MD}, \bibinfo{pages}{531--548}.
\newblock
\showISBNx{978-1-939133-04-5}
\urldef\tempurl%
\url{https://www.usenix.org/conference/usenixsecurity18/presentation/harkous}
\showURL{%
\tempurl}


\bibitem[Heer and Robertson(2007)]%
        {Heer2007AnimatedTransitions}
\bibfield{author}{\bibinfo{person}{Jeffrey Heer} {and}
  \bibinfo{person}{George~G. Robertson}.} \bibinfo{year}{2007}\natexlab{}.
\newblock \showarticletitle{Animated Transitions in Statistical Data Graphics}.
\newblock \bibinfo{journal}{\emph{IEEE Transactions on Visualization and
  Computer Graphics}} \bibinfo{volume}{13}, \bibinfo{number}{6}
  (\bibinfo{year}{2007}), \bibinfo{pages}{1240--1247}.
\newblock
\href{https://doi.org/10.1109/TVCG.2007.70539}{doi:\nolinkurl{10.1109/TVCG.2007.70539}}


\bibitem[Hullman and Diakopoulos(2011)]%
        {Hullman2011Rhetoric}
\bibfield{author}{\bibinfo{person}{Jessica Hullman} {and} \bibinfo{person}{Nick
  Diakopoulos}.} \bibinfo{year}{2011}\natexlab{}.
\newblock \showarticletitle{Visualization Rhetoric: Framing Effects in
  Narrative Visualization}.
\newblock \bibinfo{journal}{\emph{IEEE Transactions on Visualization and
  Computer Graphics}} \bibinfo{volume}{17}, \bibinfo{number}{12}
  (\bibinfo{year}{2011}), \bibinfo{pages}{2231--2240}.
\newblock
\href{https://doi.org/10.1109/TVCG.2011.255}{doi:\nolinkurl{10.1109/TVCG.2011.255}}


\bibitem[Jensen and Potts(2004)]%
        {Jensen2004}
\bibfield{author}{\bibinfo{person}{Carlos Jensen} {and} \bibinfo{person}{Colin
  Potts}.} \bibinfo{year}{2004}\natexlab{}.
\newblock \showarticletitle{Privacy Policies as Decision-Making Tools: An
  Evaluation of Online Privacy Notices}. In
  \bibinfo{booktitle}{\emph{Proceedings of the SIGCHI Conference on Human
  Factors in Computing Systems}} (Vienna, Austria) \emph{(\bibinfo{series}{CHI
  '04})}. \bibinfo{publisher}{Association for Computing Machinery},
  \bibinfo{address}{New York, NY, USA}, \bibinfo{pages}{471–478}.
\newblock
\showISBNx{1581137028}
\href{https://doi.org/10.1145/985692.985752}{doi:\nolinkurl{10.1145/985692.985752}}


\bibitem[Kay and Terry(2010)]%
        {Kay2010Textured}
\bibfield{author}{\bibinfo{person}{Matthew Kay} {and} \bibinfo{person}{Michael
  Terry}.} \bibinfo{year}{2010}\natexlab{}.
\newblock \showarticletitle{Textured Agreements: Re-envisioning Electronic
  Consent}. In \bibinfo{booktitle}{\emph{Proceedings of the Symposium on Usable
  Privacy and Security (SOUPS '10)}}. \bibinfo{address}{Redmond, WA, USA}.
\newblock
\urldef\tempurl%
\url{https://www.mjskay.com/papers/soups_2010_textured.pdf}
\showURL{%
\tempurl}


\bibitem[Kelley et~al\mbox{.}(2009)]%
        {Kelley2009Nutrition}
\bibfield{author}{\bibinfo{person}{Patrick~Gage Kelley},
  \bibinfo{person}{Joanna Bresee}, \bibinfo{person}{Lorrie~Faith Cranor}, {and}
  \bibinfo{person}{Robert~W. Reeder}.} \bibinfo{year}{2009}\natexlab{}.
\newblock \showarticletitle{A "nutrition label" for privacy}. In
  \bibinfo{booktitle}{\emph{Proceedings of the 5th Symposium on Usable Privacy
  and Security}} (Mountain View, California, USA) \emph{(\bibinfo{series}{SOUPS
  '09})}. \bibinfo{publisher}{Association for Computing Machinery},
  \bibinfo{address}{New York, NY, USA}, Article \bibinfo{articleno}{4},
  \bibinfo{numpages}{12}~pages.
\newblock
\showISBNx{9781605587363}
\href{https://doi.org/10.1145/1572532.1572538}{doi:\nolinkurl{10.1145/1572532.1572538}}


\bibitem[Kelley et~al\mbox{.}(2010)]%
        {Kelley2010Standardizing}
\bibfield{author}{\bibinfo{person}{Patrick~Gage Kelley},
  \bibinfo{person}{Lucian Cesca}, \bibinfo{person}{Joanna Bresee}, {and}
  \bibinfo{person}{Lorrie~Faith Cranor}.} \bibinfo{year}{2010}\natexlab{}.
\newblock \showarticletitle{Standardizing privacy notices: an online study of
  the nutrition label approach}. In \bibinfo{booktitle}{\emph{Proceedings of
  the SIGCHI Conference on Human Factors in Computing Systems}} (Atlanta,
  Georgia, USA) \emph{(\bibinfo{series}{CHI '10})}.
  \bibinfo{publisher}{Association for Computing Machinery},
  \bibinfo{address}{New York, NY, USA}, \bibinfo{pages}{1573–1582}.
\newblock
\showISBNx{9781605589299}
\href{https://doi.org/10.1145/1753326.1753561}{doi:\nolinkurl{10.1145/1753326.1753561}}


\bibitem[Kelley et~al\mbox{.}(2013)]%
        {KelleyAppDecision}
\bibfield{author}{\bibinfo{person}{Patrick~Gage Kelley},
  \bibinfo{person}{Lorrie~Faith Cranor}, {and} \bibinfo{person}{Norman Sadeh}.}
  \bibinfo{year}{2013}\natexlab{}.
\newblock \showarticletitle{Privacy as part of the app decision-making
  process}. In \bibinfo{booktitle}{\emph{Proceedings of the SIGCHI Conference
  on Human Factors in Computing Systems}} (Paris, France)
  \emph{(\bibinfo{series}{CHI '13})}. \bibinfo{publisher}{Association for
  Computing Machinery}, \bibinfo{address}{New York, NY, USA},
  \bibinfo{pages}{3393–3402}.
\newblock
\showISBNx{9781450318990}
\href{https://doi.org/10.1145/2470654.2466466}{doi:\nolinkurl{10.1145/2470654.2466466}}


\bibitem[Knijnenburg and Cherry(2016)]%
        {Knijnenburg2016Comics}
\bibfield{author}{\bibinfo{person}{Bart Knijnenburg} {and}
  \bibinfo{person}{David Cherry}.} \bibinfo{year}{2016}\natexlab{}.
\newblock \showarticletitle{Comics as a medium for privacy notices}. In
  \bibinfo{booktitle}{\emph{Twelfth Symposium on Usable Privacy and Security
  (SOUPS 2016)}}.
\newblock


\bibitem[Kumar et~al\mbox{.}(2018)]%
        {Kumar2018CoDesign}
\bibfield{author}{\bibinfo{person}{Priya Kumar}, \bibinfo{person}{Jessica
  Vitak}, \bibinfo{person}{Marshini Chetty}, \bibinfo{person}{Tamara~L. Clegg},
  \bibinfo{person}{Jonathan Yang}, \bibinfo{person}{Brenna McNally}, {and}
  \bibinfo{person}{Elizabeth Bonsignore}.} \bibinfo{year}{2018}\natexlab{}.
\newblock \showarticletitle{Co-designing online privacy-related games and
  stories with children}. In \bibinfo{booktitle}{\emph{Proceedings of the 17th
  ACM Conference on Interaction Design and Children}} (Trondheim, Norway)
  \emph{(\bibinfo{series}{IDC '18})}. \bibinfo{publisher}{Association for
  Computing Machinery}, \bibinfo{address}{New York, NY, USA},
  \bibinfo{pages}{67–79}.
\newblock
\showISBNx{9781450351522}
\href{https://doi.org/10.1145/3202185.3202735}{doi:\nolinkurl{10.1145/3202185.3202735}}


\bibitem[Lee et~al\mbox{.}(2015)]%
        {Lee2015MoreThanTelling}
\bibfield{author}{\bibinfo{person}{Bongshin Lee}, \bibinfo{person}{Nathalie
  Henry~Riche Kim}, \bibinfo{person}{Petra Isenberg}, {and}
  \bibinfo{person}{Sheelagh Carpendale}.} \bibinfo{year}{2015}\natexlab{}.
\newblock \showarticletitle{More Than Telling a Story: Transforming Data into
  Visually Shared Stories}. In \bibinfo{booktitle}{\emph{Proceedings of the
  2015 Eurographics Conference on Visualization (EuroVis)}}.
  \bibinfo{pages}{491--500}.
\newblock
\href{https://doi.org/10.1109/MCG.2015.99}{doi:\nolinkurl{10.1109/MCG.2015.99}}


\bibitem[Leppink et~al\mbox{.}(2013)]%
        {leppink2013cls}
\bibfield{author}{\bibinfo{person}{Jimmie Leppink}, \bibinfo{person}{Fred
  Paas}, \bibinfo{person}{Cees P.~M. Van~der Vleuten}, \bibinfo{person}{Tamara
  Van~Gog}, {and} \bibinfo{person}{Jeroen J.~G. Van~Merri{\"e}nboer}.}
  \bibinfo{year}{2013}\natexlab{}.
\newblock \showarticletitle{Development of an instrument for measuring
  different types of cognitive load}.
\newblock \bibinfo{journal}{\emph{Behavior Research Methods}}
  \bibinfo{volume}{45}, \bibinfo{number}{4} (\bibinfo{year}{2013}),
  \bibinfo{pages}{1058--1072}.
\newblock
\href{https://doi.org/10.3758/s13428-013-0334-1}{doi:\nolinkurl{10.3758/s13428-013-0334-1}}


\bibitem[Liu(2024)]%
        {Liu2024EffectsOfSegmentation}
\bibfield{author}{\bibinfo{person}{Dongyang Liu}.}
  \bibinfo{year}{2024}\natexlab{}.
\newblock \showarticletitle{The effects of segmentation on cognitive load,
  vocabulary learning and retention, and reading comprehension in a multimedia
  learning environment}.
\newblock \bibinfo{journal}{\emph{BMC Psychology}} \bibinfo{volume}{12},
  \bibinfo{number}{1} (\bibinfo{year}{2024}), \bibinfo{pages}{4}.
\newblock
\showISSN{2050-7283}
\href{https://doi.org/10.1186/s40359-023-01489-5}{doi:\nolinkurl{10.1186/s40359-023-01489-5}}


\bibitem[Mathur et~al\mbox{.}(2019)]%
        {Mathur2019DarkPatterns}
\bibfield{author}{\bibinfo{person}{Arunesh Mathur}, \bibinfo{person}{Gunes
  Acar}, \bibinfo{person}{Michael~J. Friedman}, \bibinfo{person}{Eli
  Lucherini}, \bibinfo{person}{Jonathan Mayer}, \bibinfo{person}{Marshini
  Chetty}, {and} \bibinfo{person}{Arvind Narayanan}.}
  \bibinfo{year}{2019}\natexlab{}.
\newblock \showarticletitle{Dark Patterns at Scale: Findings from a Crawl of
  11K Shopping Websites}.
\newblock \bibinfo{journal}{\emph{Proc. ACM Hum.-Comput. Interact.}}
  \bibinfo{volume}{3}, \bibinfo{number}{CSCW}, Article \bibinfo{articleno}{81}
  (\bibinfo{date}{Nov.} \bibinfo{year}{2019}), \bibinfo{numpages}{32}~pages.
\newblock
\href{https://doi.org/10.1145/3359183}{doi:\nolinkurl{10.1145/3359183}}


\bibitem[Mayer(2009)]%
        {Mayer2009MultimediaLearning}
\bibfield{author}{\bibinfo{person}{Richard~E. Mayer}.}
  \bibinfo{year}{2009}\natexlab{}.
\newblock \bibinfo{booktitle}{\emph{Multimedia Learning} (\bibinfo{edition}{2}
  ed.)}.
\newblock \bibinfo{publisher}{Cambridge University Press},
  \bibinfo{address}{New York, NY}.
\newblock
\href{https://doi.org/10.1017/CBO9780511811678}{doi:\nolinkurl{10.1017/CBO9780511811678}}


\bibitem[Mazza et~al\mbox{.}(2023)]%
        {Mazza2023Pictograms}
\bibfield{author}{\bibinfo{person}{Larissa~Ugaya Mazza},
  \bibinfo{person}{Laura~Xavier Fadrique}, \bibinfo{person}{Amethyst Kuang},
  \bibinfo{person}{Tania Donovska}, \bibinfo{person}{H{\'e}l{\`e}ne
  Vaillancourt}, \bibinfo{person}{Jennifer Teague},
  \bibinfo{person}{Victoria~A. Hailey}, \bibinfo{person}{Stephen Michell},
  {and} \bibinfo{person}{Plinio~P. Morita}.} \bibinfo{year}{2023}\natexlab{}.
\newblock \showarticletitle{Exploring the Use of Pictograms in Privacy
  Agreements to Facilitate Communication Between Users and Data Collecting
  Entities: Randomized Controlled Trial}.
\newblock \bibinfo{journal}{\emph{JMIR Human Factors}}  \bibinfo{volume}{10}
  (\bibinfo{year}{2023}), \bibinfo{pages}{e34855}.
\newblock
\href{https://doi.org/10.2196/34855}{doi:\nolinkurl{10.2196/34855}}


\bibitem[McAuley et~al\mbox{.}(1989)]%
        {mcauley1989imi}
\bibfield{author}{\bibinfo{person}{Edward McAuley}, \bibinfo{person}{Terry~E.
  Duncan}, {and} \bibinfo{person}{Vance~V. Tammen}.}
  \bibinfo{year}{1989}\natexlab{}.
\newblock \showarticletitle{Psychometric properties of the Intrinsic Motivation
  Inventory in a competitive sport setting: A confirmatory factor analysis}.
\newblock \bibinfo{journal}{\emph{Research Quarterly for Exercise and Sport}}
  \bibinfo{volume}{60}, \bibinfo{number}{1} (\bibinfo{year}{1989}),
  \bibinfo{pages}{48--58}.
\newblock
\href{https://doi.org/10.1080/02701367.1989.10607413}{doi:\nolinkurl{10.1080/02701367.1989.10607413}}


\bibitem[McDonald and Cranor(2008)]%
        {McDonald2008Cost}
\bibfield{author}{\bibinfo{person}{Aleecia~M McDonald} {and}
  \bibinfo{person}{Lorrie~Faith Cranor}.} \bibinfo{year}{2008}\natexlab{}.
\newblock \showarticletitle{The Cost of Reading Privacy Policies}.
\newblock \bibinfo{journal}{\emph{I/S: A Journal of Law and Policy for the
  Information Society}} \bibinfo{volume}{4}, \bibinfo{number}{3}
  (\bibinfo{year}{2008}), \bibinfo{pages}{543--568}.
\newblock
\urldef\tempurl%
\url{https://heinonline.org/HOL/LandingPage?handle=hein.journals/isjlpsoc4&div=27&id=&page=}
\showURL{%
\tempurl}


\bibitem[McKnight et~al\mbox{.}(2002)]%
        {mcknight2002trust}
\bibfield{author}{\bibinfo{person}{D.~Harrison McKnight},
  \bibinfo{person}{Vivek Choudhury}, {and} \bibinfo{person}{Charles Kacmar}.}
  \bibinfo{year}{2002}\natexlab{}.
\newblock \showarticletitle{Developing and Validating Trust Measures for
  e-Commerce: An Integrative Typology}.
\newblock \bibinfo{journal}{\emph{Information Systems Research}}
  \bibinfo{volume}{13}, \bibinfo{number}{3} (\bibinfo{year}{2002}),
  \bibinfo{pages}{334--359}.
\newblock
\href{https://doi.org/10.1287/isre.13.3.334.81}{doi:\nolinkurl{10.1287/isre.13.3.334.81}}


\bibitem[M\'{e}ndez and Mendoza(2021)]%
        {Mendez2021Scrollytelling}
\bibfield{author}{\bibinfo{person}{Gonzalo~Gabriel M\'{e}ndez} {and}
  \bibinfo{person}{Patricio Mendoza}.} \bibinfo{year}{2021}\natexlab{}.
\newblock \showarticletitle{Using Scrollytelling to Explain Voting Power in
  Ecuador}. In \bibinfo{booktitle}{\emph{Proceedings of the 14th International
  Symposium on Visual Information Communication and Interaction}} (Potsdam,
  Germany) \emph{(\bibinfo{series}{VINCI '21})}.
  \bibinfo{publisher}{Association for Computing Machinery},
  \bibinfo{address}{New York, NY, USA}, Article \bibinfo{articleno}{19},
  \bibinfo{numpages}{2}~pages.
\newblock
\showISBNx{9781450386470}
\href{https://doi.org/10.1145/3481549.3481572}{doi:\nolinkurl{10.1145/3481549.3481572}}


\bibitem[Nielsen(2006)]%
        {Nielsen2006ProgressiveDisclosure}
\bibfield{author}{\bibinfo{person}{Jakob Nielsen}.}
  \bibinfo{year}{2006}\natexlab{}.
\newblock \bibinfo{title}{Progressive Disclosure}.
\newblock \bibinfo{howpublished}{Nielsen Norman Group}.
\newblock
\urldef\tempurl%
\url{https://www.nngroup.com/articles/progressive-disclosure}
\showURL{%
\tempurl}


\bibitem[Obar and Oeldorf-Hirsch(2016)]%
        {Obar2016Biggest}
\bibfield{author}{\bibinfo{person}{Jonathan~A. Obar} {and}
  \bibinfo{person}{Anne Oeldorf-Hirsch}.} \bibinfo{year}{2016}\natexlab{}.
\newblock \showarticletitle{The Biggest Lie on the Internet: Ignoring the
  Privacy Policies and Terms of Service Policies of Social Networking
  Services}.
\newblock \bibinfo{journal}{\emph{Information, Communication \& Society}}
  \bibinfo{volume}{19}, \bibinfo{number}{3} (\bibinfo{year}{2016}),
  \bibinfo{pages}{1--20}.
\newblock
\href{https://doi.org/10.1080/1369118X.2016.1203451}{doi:\nolinkurl{10.1080/1369118X.2016.1203451}}


\bibitem[O'Brien et~al\mbox{.}(2018)]%
        {obrien2018ues}
\bibfield{author}{\bibinfo{person}{Heather~L. O'Brien}, \bibinfo{person}{Paul
  Cairns}, {and} \bibinfo{person}{Mark Hall}.} \bibinfo{year}{2018}\natexlab{}.
\newblock \showarticletitle{A practical approach to measuring user engagement
  with the refined user engagement scale (UES) and new UES short form}.
\newblock \bibinfo{journal}{\emph{International Journal of Human-Computer
  Studies}}  \bibinfo{volume}{112} (\bibinfo{year}{2018}),
  \bibinfo{pages}{28--39}.
\newblock
\href{https://doi.org/10.1016/j.ijhcs.2018.01.004}{doi:\nolinkurl{10.1016/j.ijhcs.2018.01.004}}


\bibitem[Paivio(1991)]%
        {Paivio1991DualCoding}
\bibfield{author}{\bibinfo{person}{Allan Paivio}.}
  \bibinfo{year}{1991}\natexlab{}.
\newblock \showarticletitle{Dual Coding Theory: Retrospect and Current Status}.
\newblock \bibinfo{journal}{\emph{Canadian Journal of Psychology}}
  \bibinfo{volume}{45}, \bibinfo{number}{3} (\bibinfo{year}{1991}),
  \bibinfo{pages}{255--287}.
\newblock
\href{https://doi.org/10.1037/h0084295}{doi:\nolinkurl{10.1037/h0084295}}


\bibitem[Palan and Schitter(2018)]%
        {Palan2018}
\bibfield{author}{\bibinfo{person}{Stefan Palan} {and}
  \bibinfo{person}{Christian Schitter}.} \bibinfo{year}{2018}\natexlab{}.
\newblock \showarticletitle{Prolific.ac—A subject pool for online
  experiments}.
\newblock \bibinfo{journal}{\emph{Journal of Behavioral and Experimental
  Finance}}  \bibinfo{volume}{17} (\bibinfo{year}{2018}),
  \bibinfo{pages}{22--27}.
\newblock
\href{https://doi.org/10.1016/j.jbef.2017.12.004}{doi:\nolinkurl{10.1016/j.jbef.2017.12.004}}


\bibitem[Peer et~al\mbox{.}(2017)]%
        {Peer2017}
\bibfield{author}{\bibinfo{person}{Eyal Peer}, \bibinfo{person}{Laura
  Brandimarte}, \bibinfo{person}{Sonam Samat}, {and}
  \bibinfo{person}{Alessandro Acquisti}.} \bibinfo{year}{2017}\natexlab{}.
\newblock \showarticletitle{Beyond the Turk: Alternative platforms for
  crowdsourcing behavioral research}.
\newblock \bibinfo{journal}{\emph{Journal of Experimental Social Psychology}}
  \bibinfo{volume}{70} (\bibinfo{year}{2017}), \bibinfo{pages}{153--163}.
\newblock
\href{https://doi.org/10.1016/j.jesp.2017.01.006}{doi:\nolinkurl{10.1016/j.jesp.2017.01.006}}


\bibitem[Perez et~al\mbox{.}(2018)]%
        {Perez2018Review}
\bibfield{author}{\bibinfo{person}{Alfredo~J. Perez}, \bibinfo{person}{Sherali
  Zeadally}, {and} \bibinfo{person}{Jonathan Cochran}.}
  \bibinfo{year}{2018}\natexlab{}.
\newblock \showarticletitle{A review and an empirical analysis of privacy
  policy and notices for consumer Internet of things}.
\newblock \bibinfo{journal}{\emph{SECURITY AND PRIVACY}} \bibinfo{volume}{1},
  \bibinfo{number}{3} (\bibinfo{year}{2018}), \bibinfo{pages}{e15}.
\newblock
\href{https://doi.org/10.1002/spy2.15}{doi:\nolinkurl{10.1002/spy2.15}}


\bibitem[Quayyum(2025)]%
        {Dincelli2025CoDesigningCybersecurity}
\bibfield{author}{\bibinfo{person}{Farzana Quayyum}.}
  \bibinfo{year}{2025}\natexlab{}.
\newblock \showarticletitle{Co-designing cybersecurity-related stories with
  children: Perceptions on cybersecurity risks and parental involvement}.
\newblock \bibinfo{journal}{\emph{Entertainment Computing}}
  \bibinfo{volume}{52} (\bibinfo{year}{2025}), \bibinfo{pages}{100753}.
\newblock
\showISSN{1875-9521}
\href{https://doi.org/10.1016/j.entcom.2024.100753}{doi:\nolinkurl{10.1016/j.entcom.2024.100753}}


\bibitem[Reidenberg et~al\mbox{.}(2015)]%
        {Reidenberg2015Disagreeable}
\bibfield{author}{\bibinfo{person}{Joel~R. Reidenberg}, \bibinfo{person}{Travis
  Breaux}, \bibinfo{person}{Lorrie~Faith Cranor}, \bibinfo{person}{Brian
  French}, \bibinfo{person}{Amanda Grannis}, \bibinfo{person}{James~T. Graves},
  \bibinfo{person}{Fei Liu}, \bibinfo{person}{Aleecia~M. McDonald},
  \bibinfo{person}{Thomas~B. Norton}, \bibinfo{person}{Rohan Ramanath},
  \bibinfo{person}{N.~Cameron Russell}, \bibinfo{person}{Florian Schaub},
  \bibinfo{person}{Jaspreet Singh}, {and} \bibinfo{person}{Nan Wang}.}
  \bibinfo{year}{2015}\natexlab{}.
\newblock \showarticletitle{Disagreeable Privacy Policies: Mismatches Between
  Meaning and Users’ Understanding}.
\newblock \bibinfo{journal}{\emph{Berkeley Technology Law Journal}}
  \bibinfo{volume}{30}, \bibinfo{number}{1} (\bibinfo{year}{2015}),
  \bibinfo{pages}{39--88}.
\newblock
\href{https://doi.org/10.2139/ssrn.2418297}{doi:\nolinkurl{10.2139/ssrn.2418297}}


\bibitem[Reinhardt et~al\mbox{.}(2021)]%
        {ReinhardtInteractivePP}
\bibfield{author}{\bibinfo{person}{Daniel Reinhardt}, \bibinfo{person}{Johannes
  Borchard}, {and} \bibinfo{person}{J\"{o}rn Hurtienne}.}
  \bibinfo{year}{2021}\natexlab{}.
\newblock \showarticletitle{Visual Interactive Privacy Policy: The Better
  Choice?}. In \bibinfo{booktitle}{\emph{Proceedings of the 2021 CHI Conference
  on Human Factors in Computing Systems}} (Yokohama, Japan)
  \emph{(\bibinfo{series}{CHI '21})}. \bibinfo{publisher}{Association for
  Computing Machinery}, \bibinfo{address}{New York, NY, USA}, Article
  \bibinfo{articleno}{66}, \bibinfo{numpages}{12}~pages.
\newblock
\showISBNx{9781450380966}
\href{https://doi.org/10.1145/3411764.3445465}{doi:\nolinkurl{10.1145/3411764.3445465}}


\bibitem[Rey et~al\mbox{.}(2019)]%
        {Rey2019Metaanalysis}
\bibfield{author}{\bibinfo{person}{Günter~Daniel Rey}, \bibinfo{person}{Maik
  Beege}, \bibinfo{person}{Steve Nebel}, \bibinfo{person}{Maria Wirzberger},
  \bibinfo{person}{Tobias~H. Schmitt}, {and} \bibinfo{person}{Sascha
  Schneider}.} \bibinfo{year}{2019}\natexlab{}.
\newblock \showarticletitle{A Meta-analysis of the Segmenting Effect}.
\newblock \bibinfo{journal}{\emph{Educational Psychology Review}}
  \bibinfo{volume}{31}, \bibinfo{number}{2} (\bibinfo{year}{2019}),
  \bibinfo{pages}{389--419}.
\newblock
\showISSN{1573-336X}
\href{https://doi.org/10.1007/s10648-018-9456-4}{doi:\nolinkurl{10.1007/s10648-018-9456-4}}


\bibitem[Schaub et~al\mbox{.}(2015)]%
        {Schaub2015DesignSpace}
\bibfield{author}{\bibinfo{person}{Florian Schaub}, \bibinfo{person}{Rebecca
  Balebako}, \bibinfo{person}{Adam~L. Durity}, {and}
  \bibinfo{person}{Lorrie~Faith Cranor}.} \bibinfo{year}{2015}\natexlab{}.
\newblock \showarticletitle{A Design Space for Effective Privacy Notices}. In
  \bibinfo{booktitle}{\emph{Proceedings of the Eleventh Symposium on Usable
  Privacy and Security (SOUPS 2015)}}. \bibinfo{publisher}{USENIX Association},
  \bibinfo{address}{Ottawa, Canada}.
\newblock
\urldef\tempurl%
\url{https://www.usenix.org/system/files/conference/soups2015/soups15-paper-schaub.pdf}
\showURL{%
\tempurl}


\bibitem[Schaub et~al\mbox{.}(2017)]%
        {Schaub2017Designing}
\bibfield{author}{\bibinfo{person}{Florian Schaub}, \bibinfo{person}{Rebecca
  Balebako}, \bibinfo{person}{Adam~L. Durity}, {and}
  \bibinfo{person}{Lorrie~Faith Cranor}.} \bibinfo{year}{2017}\natexlab{}.
\newblock \showarticletitle{Designing Effective Privacy Notices and Controls}.
\newblock \bibinfo{journal}{\emph{IEEE Internet Computing}}
  \bibinfo{volume}{21}, \bibinfo{number}{3} (\bibinfo{year}{2017}),
  \bibinfo{pages}{70--77}.
\newblock
\href{https://doi.org/10.1109/MIC.2017.75}{doi:\nolinkurl{10.1109/MIC.2017.75}}


\bibitem[Schnackenberg et~al\mbox{.}(2021)]%
        {schnackenberg2021transparency}
\bibfield{author}{\bibinfo{person}{Andrew~K. Schnackenberg},
  \bibinfo{person}{Edward~C. Tomlinson}, {and} \bibinfo{person}{Corinne Coen}.}
  \bibinfo{year}{2021}\natexlab{}.
\newblock \showarticletitle{The dimensional structure of transparency: A
  construct validation of transparency as disclosure, clarity, and accuracy in
  organizations}.
\newblock \bibinfo{journal}{\emph{Human Relations}} \bibinfo{volume}{74},
  \bibinfo{number}{10} (\bibinfo{year}{2021}), \bibinfo{pages}{1628--1660}.
\newblock
\href{https://doi.org/10.1177/0018726720933317}{doi:\nolinkurl{10.1177/0018726720933317}}


\bibitem[Segel and Heer(2010)]%
        {Segel2010NarrativeVisualization}
\bibfield{author}{\bibinfo{person}{Edward Segel} {and} \bibinfo{person}{Jeffrey
  Heer}.} \bibinfo{year}{2010}\natexlab{}.
\newblock \showarticletitle{Narrative Visualization: Telling Stories with
  Data}.
\newblock \bibinfo{journal}{\emph{IEEE Transactions on Visualization and
  Computer Graphics}} \bibinfo{volume}{16}, \bibinfo{number}{6}
  (\bibinfo{year}{2010}), \bibinfo{pages}{1139--1148}.
\newblock
\href{https://doi.org/10.1109/TVCG.2010.179}{doi:\nolinkurl{10.1109/TVCG.2010.179}}


\bibitem[Sengers et~al\mbox{.}(2005)]%
        {Sengers2005ReflectiveDesign}
\bibfield{author}{\bibinfo{person}{Phoebe Sengers}, \bibinfo{person}{Kirsten
  Boehner}, \bibinfo{person}{Shay David}, {and}
  \bibinfo{person}{Joseph~'Jofish' Kaye}.} \bibinfo{year}{2005}\natexlab{}.
\newblock \showarticletitle{Reflective design}. In
  \bibinfo{booktitle}{\emph{Proceedings of the 4th Decennial Conference on
  Critical Computing: Between Sense and Sensibility}} (Aarhus, Denmark)
  \emph{(\bibinfo{series}{CC '05})}. \bibinfo{publisher}{Association for
  Computing Machinery}, \bibinfo{address}{New York, NY, USA},
  \bibinfo{pages}{49–58}.
\newblock
\showISBNx{1595932038}
\href{https://doi.org/10.1145/1094562.1094569}{doi:\nolinkurl{10.1145/1094562.1094569}}


\bibitem[Shneiderman(1996)]%
        {Shneiderman1996Eyes}
\bibfield{author}{\bibinfo{person}{Ben Shneiderman}.}
  \bibinfo{year}{1996}\natexlab{}.
\newblock \showarticletitle{The Eyes Have It: A Task by Data Type Taxonomy for
  Information Visualizations}. In \bibinfo{booktitle}{\emph{Proceedings of the
  IEEE Symposium on Visual Languages}}. \bibinfo{pages}{336--343}.
\newblock
\href{https://doi.org/10.1109/VL.1996.545307}{doi:\nolinkurl{10.1109/VL.1996.545307}}


\bibitem[Solove(2013)]%
        {Solove2013Nothing}
\bibfield{author}{\bibinfo{person}{Daniel~J. Solove}.}
  \bibinfo{year}{2013}\natexlab{}.
\newblock \showarticletitle{Privacy Self-Management and the Consent Dilemma}.
\newblock \bibinfo{journal}{\emph{Harvard Law Review}} \bibinfo{volume}{126},
  \bibinfo{number}{7} (\bibinfo{year}{2013}), \bibinfo{pages}{1880--1903}.
\newblock
\urldef\tempurl%
\url{https://harvardlawreview.org/2013/05/privacy-self-management-and-the-consent-dilemma/}
\showURL{%
\tempurl}


\bibitem[Soumelidou and Tsohou(2020)]%
        {Soumelidou2019Visualization}
\bibfield{author}{\bibinfo{person}{Athanasia Soumelidou} {and}
  \bibinfo{person}{Maria Tsohou}.} \bibinfo{year}{2020}\natexlab{}.
\newblock \showarticletitle{Data privacy policies for social media users: A
  novel visualization approach on Instagram's privacy policy}.
\newblock \bibinfo{journal}{\emph{Information Technology \& People}}
  \bibinfo{volume}{33}, \bibinfo{number}{2} (\bibinfo{year}{2020}),
  \bibinfo{pages}{502--534}.
\newblock
\href{https://doi.org/10.1108/ITP-08-2017-0241}{doi:\nolinkurl{10.1108/ITP-08-2017-0241}}


\bibitem[Spanjers et~al\mbox{.}(2012)]%
        {spanjers2012explaining}
\bibfield{author}{\bibinfo{person}{Ingrid~AE Spanjers}, \bibinfo{person}{Tamara
  Van~Gog}, \bibinfo{person}{Pieter Wouters}, {and} \bibinfo{person}{Jeroen~JG
  Van~Merri{\"e}nboer}.} \bibinfo{year}{2012}\natexlab{}.
\newblock \showarticletitle{Explaining the segmentation effect in learning from
  animations: The role of pausing and temporal cueing}.
\newblock \bibinfo{journal}{\emph{Computers \& Education}}
  \bibinfo{volume}{59}, \bibinfo{number}{2} (\bibinfo{year}{2012}),
  \bibinfo{pages}{274--280}.
\newblock
\href{https://doi.org/10.1016/j.compedu.2011.12.024}{doi:\nolinkurl{10.1016/j.compedu.2011.12.024}}


\bibitem[Springer and Whittaker(2018)]%
        {Springer2018ProgressiveDisclosure}
\bibfield{author}{\bibinfo{person}{Aaron Springer} {and} \bibinfo{person}{Steve
  Whittaker}.} \bibinfo{year}{2018}\natexlab{}.
\newblock \showarticletitle{Progressive Disclosure: Designing for Effective
  Transparency}.
\newblock \bibinfo{journal}{\emph{arXiv preprint arXiv:1811.02164}}
  (\bibinfo{year}{2018}).
\newblock
\urldef\tempurl%
\url{https://arxiv.org/abs/1811.02164}
\showURL{%
\tempurl}


\bibitem[Staff(2011)]%
        {Staff_2011}
\bibfield{author}{\bibinfo{person}{FTC Staff}.}
  \bibinfo{year}{2011}\natexlab{}.
\newblock \showarticletitle{Protecting Consumer Privacy in an Era of Rapid
  Change–A Proposed Framework for Businesses and Policymakers}.
\newblock \bibinfo{journal}{\emph{Journal of Privacy and Confidentiality}}
  \bibinfo{volume}{3}, \bibinfo{number}{1} (\bibinfo{date}{Jun.}
  \bibinfo{year}{2011}).
\newblock
\href{https://doi.org/10.29012/jpc.v3i1.596}{doi:\nolinkurl{10.29012/jpc.v3i1.596}}


\bibitem[Suh et~al\mbox{.}(2022)]%
        {Suh2022PrivacyToon}
\bibfield{author}{\bibinfo{person}{Sangho Suh}, \bibinfo{person}{Sydney
  Lamorea}, \bibinfo{person}{Edith Law}, {and} \bibinfo{person}{Leah
  Zhang{-}Kennedy}.} \bibinfo{year}{2022}\natexlab{}.
\newblock \showarticletitle{PrivacyToon: Concept-driven Storytelling with
  Creativity Support for Privacy Concepts}. In
  \bibinfo{booktitle}{\emph{Proceedings of the Designing Interactive Systems
  Conference (DIS '22)}}. \bibinfo{publisher}{ACM}.
\newblock
\href{https://doi.org/10.1145/3532106.3533557}{doi:\nolinkurl{10.1145/3532106.3533557}}


\bibitem[Tabassum et~al\mbox{.}(2018)]%
        {Tabassum2018ComicPolicy}
\bibfield{author}{\bibinfo{person}{Madiha Tabassum},
  \bibinfo{person}{Abdulmajeed Alqhatani}, \bibinfo{person}{Marran Aldossari},
  {and} \bibinfo{person}{Heather Richter~Lipford}.}
  \bibinfo{year}{2018}\natexlab{}.
\newblock \showarticletitle{Increasing User Attention with a Comic-based
  Policy}. In \bibinfo{booktitle}{\emph{Proceedings of the 2018 CHI Conference
  on Human Factors in Computing Systems (CHI '18)}}. \bibinfo{publisher}{ACM},
  Article \bibinfo{articleno}{200}, \bibinfo{numpages}{6}~pages.
\newblock
\href{https://doi.org/10.1145/3173574.3173774}{doi:\nolinkurl{10.1145/3173574.3173774}}


\bibitem[Tj{\"a}rnhage et~al\mbox{.}(2023)]%
        {Tjarnhage2023Scrollytelling}
\bibfield{author}{\bibinfo{person}{Anja Tj{\"a}rnhage}, \bibinfo{person}{Ulrik
  S{\"o}derstr{\"o}m}, \bibinfo{person}{Ole Norberg}, \bibinfo{person}{Mattias
  Andersson}, {and} \bibinfo{person}{Thomas Mejtoft}.}
  \bibinfo{year}{2023}\natexlab{}.
\newblock \showarticletitle{The Impact of Scrollytelling on the Reading
  Experience of Long-Form Journalism}. In \bibinfo{booktitle}{\emph{Proceedings
  of the European Conference on Cognitive Ergonomics (ECCE 2023)}}.
  \bibinfo{publisher}{ACM}.
\newblock
\href{https://doi.org/10.1145/3605655.3605683}{doi:\nolinkurl{10.1145/3605655.3605683}}


\bibitem[Venkatesh et~al\mbox{.}(2003)]%
        {venkatesh2003utaut}
\bibfield{author}{\bibinfo{person}{Viswanath Venkatesh},
  \bibinfo{person}{Michael~G. Morris}, \bibinfo{person}{Gordon~B. Davis}, {and}
  \bibinfo{person}{Fred~D. Davis}.} \bibinfo{year}{2003}\natexlab{}.
\newblock \showarticletitle{User Acceptance of Information Technology: Toward a
  Unified View}.
\newblock \bibinfo{journal}{\emph{MIS Quarterly}} \bibinfo{volume}{27},
  \bibinfo{number}{3} (\bibinfo{year}{2003}), \bibinfo{pages}{425--478}.
\newblock
\href{https://doi.org/10.2307/30036540}{doi:\nolinkurl{10.2307/30036540}}


\bibitem[Venkatesh et~al\mbox{.}(2012)]%
        {venkatesh2012utaut2}
\bibfield{author}{\bibinfo{person}{Viswanath Venkatesh}, \bibinfo{person}{James
  Y.~L. Thong}, {and} \bibinfo{person}{Xin Xu}.}
  \bibinfo{year}{2012}\natexlab{}.
\newblock \showarticletitle{Consumer Acceptance and Use of Information
  Technology: Extending the Unified Theory of Acceptance and Use of
  Technology}.
\newblock \bibinfo{journal}{\emph{MIS Quarterly}} \bibinfo{volume}{36},
  \bibinfo{number}{1} (\bibinfo{year}{2012}), \bibinfo{pages}{157--178}.
\newblock
\href{https://doi.org/10.2307/41410412}{doi:\nolinkurl{10.2307/41410412}}


\bibitem[Wagner(2023)]%
        {Wagner2023}
\bibfield{author}{\bibinfo{person}{Isabel Wagner}.}
  \bibinfo{year}{2023}\natexlab{}.
\newblock \showarticletitle{Privacy Policies across the Ages: Content of
  Privacy Policies 1996–2021}.
\newblock \bibinfo{journal}{\emph{ACM Trans. Priv. Secur.}}
  \bibinfo{volume}{26}, \bibinfo{number}{3}, Article \bibinfo{articleno}{32}
  (\bibinfo{date}{May} \bibinfo{year}{2023}), \bibinfo{numpages}{32}~pages.
\newblock
\showISSN{2471-2566}
\href{https://doi.org/10.1145/3590152}{doi:\nolinkurl{10.1145/3590152}}


\bibitem[Watson et~al\mbox{.}(2021)]%
        {Watson2021InvestigationComic}
\bibfield{author}{\bibinfo{person}{Katie Watson}, \bibinfo{person}{Mike Just},
  {and} \bibinfo{person}{Tessa Berg}.} \bibinfo{year}{2021}\natexlab{}.
\newblock \showarticletitle{An Investigation of Comic-Based Permission
  Requests}. In \bibinfo{booktitle}{\emph{Secure IT Systems}},
  \bibfield{editor}{\bibinfo{person}{Mikael Asplund} {and}
  \bibinfo{person}{Simin Nadjm-Tehrani}} (Eds.). \bibinfo{publisher}{Springer
  International Publishing}, \bibinfo{address}{Cham},
  \bibinfo{pages}{246--261}.
\newblock
\showISBNx{978-3-030-70852-8}


\bibitem[Watson et~al\mbox{.}(2023)]%
        {Watson2023ComicToPermission}
\bibfield{author}{\bibinfo{person}{Katie Watson}, \bibinfo{person}{Mike Just},
  {and} \bibinfo{person}{Tessa Berg}.} \bibinfo{year}{2023}\natexlab{}.
\newblock \showarticletitle{A comic-based approach to permission request
  communication}.
\newblock \bibinfo{journal}{\emph{Comput. Secur.}} \bibinfo{volume}{124},
  \bibinfo{number}{C} (\bibinfo{date}{Jan.} \bibinfo{year}{2023}),
  \bibinfo{numpages}{13}~pages.
\newblock
\showISSN{0167-4048}
\href{https://doi.org/10.1016/j.cose.2022.102942}{doi:\nolinkurl{10.1016/j.cose.2022.102942}}


\bibitem[Winkler and Zeadally(2016)]%
        {Winkler2016}
\bibfield{author}{\bibinfo{person}{Stephanie Winkler} {and}
  \bibinfo{person}{Sherali Zeadally}.} \bibinfo{year}{2016}\natexlab{}.
\newblock \showarticletitle{Privacy Policy Analysis of Popular Web Platforms}.
\newblock \bibinfo{journal}{\emph{IEEE Technology and Society Magazine}}
  \bibinfo{volume}{35}, \bibinfo{number}{2} (\bibinfo{year}{2016}),
  \bibinfo{pages}{75--85}.
\newblock
\href{https://doi.org/10.1109/MTS.2016.2554419}{doi:\nolinkurl{10.1109/MTS.2016.2554419}}


\bibitem[Wixom and Todd(2005)]%
        {wixom2005integration}
\bibfield{author}{\bibinfo{person}{Barbara~H. Wixom} {and}
  \bibinfo{person}{Peter~A. Todd}.} \bibinfo{year}{2005}\natexlab{}.
\newblock \showarticletitle{A theoretical integration of user satisfaction and
  technology acceptance}.
\newblock \bibinfo{journal}{\emph{Information Systems Research}}
  \bibinfo{volume}{16}, \bibinfo{number}{1} (\bibinfo{year}{2005}),
  \bibinfo{pages}{85--102}.
\newblock
\href{https://doi.org/10.1287/isre.1050.0042}{doi:\nolinkurl{10.1287/isre.1050.0042}}


\bibitem[Xie et~al\mbox{.}(2025)]%
        {Xie2025LLMBased}
\bibfield{author}{\bibinfo{person}{Qinge Xie}, \bibinfo{person}{Karthik
  Ramakrishnan}, {and} \bibinfo{person}{Frank Li}.}
  \bibinfo{year}{2025}\natexlab{}.
\newblock \showarticletitle{Evaluating privacy policies under modern privacy
  laws at scale: an LLM-based automated approach}. In
  \bibinfo{booktitle}{\emph{Proceedings of the 34th USENIX Conference on
  Security Symposium}} (Seattle, WA, USA) \emph{(\bibinfo{series}{SEC '25})}.
  \bibinfo{publisher}{USENIX Association}, \bibinfo{address}{USA}, Article
  \bibinfo{articleno}{298}, \bibinfo{numpages}{20}~pages.
\newblock
\showISBNx{978-1-939133-52-6}


\bibitem[Zaeem and Barber(2020)]%
        {Zaeem2020}
\bibfield{author}{\bibinfo{person}{Razieh~Nokhbeh Zaeem} {and}
  \bibinfo{person}{K.~Suzanne Barber}.} \bibinfo{year}{2020}\natexlab{}.
\newblock \showarticletitle{The Effect of the GDPR on Privacy Policies: Recent
  Progress and Future Promise}.
\newblock \bibinfo{journal}{\emph{ACM Trans. Manage. Inf. Syst.}}
  \bibinfo{volume}{12}, \bibinfo{number}{1}, Article \bibinfo{articleno}{2}
  (\bibinfo{date}{Dec.} \bibinfo{year}{2020}), \bibinfo{numpages}{20}~pages.
\newblock
\showISSN{2158-656X}
\href{https://doi.org/10.1145/3389685}{doi:\nolinkurl{10.1145/3389685}}


\bibitem[Zhang-Kennedy et~al\mbox{.}(2017a)]%
        {ZHANGKENNEDY201710}
\bibfield{author}{\bibinfo{person}{Leah Zhang-Kennedy}, \bibinfo{person}{Yomna
  Abdelaziz}, {and} \bibinfo{person}{Sonia Chiasson}.}
  \bibinfo{year}{2017}\natexlab{a}.
\newblock \showarticletitle{Cyberheroes: The design and evaluation of an
  interactive ebook to educate children about online privacy}.
\newblock \bibinfo{journal}{\emph{International Journal of Child-Computer
  Interaction}}  \bibinfo{volume}{13} (\bibinfo{year}{2017}),
  \bibinfo{pages}{10--18}.
\newblock
\showISSN{2212-8689}
\href{https://doi.org/10.1016/j.ijcci.2017.05.001}{doi:\nolinkurl{10.1016/j.ijcci.2017.05.001}}


\bibitem[Zhang-Kennedy et~al\mbox{.}(2017b)]%
        {ZhangKennedy2017ChildrenComics}
\bibfield{author}{\bibinfo{person}{Leah Zhang-Kennedy},
  \bibinfo{person}{Khadija Baig}, {and} \bibinfo{person}{Sonia Chiasson}.}
  \bibinfo{year}{2017}\natexlab{b}.
\newblock \showarticletitle{Engaging children about online privacy through
  storytelling in an interactive comic}. In
  \bibinfo{booktitle}{\emph{Electronic Visualisation and the Arts (EVA 2017)}}.
  BCS Learning \& Development.
\newblock
\href{https://doi.org/10.14236/ewic/HCI2017.45}{doi:\nolinkurl{10.14236/ewic/HCI2017.45}}


\end{thebibliography}

\appendix

%%%% Study.tex starts here %%%%

\section{Sample Size Determination}
\label{app:samplesize}

We ran Monte Carlo simulations under the planned $5{\times}2$ between-subjects design (balanced allocation in simulation) to calibrate the per-cell sample size for confirmatory \textit{Format} main effects (two-sided $\alpha=.05$). The simulations reproduced the fixed-effect structure of the planned models and their variance structure: (i) item-level Binomial GLM for \emph{accuracy} with logit link, treatment coding for \textit{Format}, \textit{Owner}, and \textit{Item type}, and cluster-robust (by participant) Wald tests for omnibus \textit{Format} contrasts (approximating the mixed-effects analysis used in the main text); (ii) item-level OLS for \emph{confidence} with HC3 standard errors and omnibus \textit{Format} contrasts (approximating the linear mixed-effects analysis); (iii) per-participant OLS with HC3 for the six \emph{experience} scales (omnibus \textit{Format} tests); and (iv) per-participant OLS with HC3 on pre--post differences ($\Delta$) for the perception constructs (omnibus \textit{Format} tests).

For \textit{Accuracy}, we assumed baseline correctness $p_0=0.65$ in the \textit{Scrollytelling}/\textit{OpenAI}/\textit{factual} reference and modeled \textit{Format} effects as odds-ratio shifts. For continuous outcomes (\textit{Confidence}, \textit{Experience}), we defined a medium effect as minimally important ($f{=}0.25$). For $\Delta$, we set pre--post correlation $\rho{=}0.50$ and tuned means to yield $f{=}0.25$ on the change scale.

We simulated grids of per-cell sizes and selected the smallest $n$ achieving \emph{power $\ge .95$} for the confirmatory families (the \textit{Format} main effect on \emph{accuracy} and $\Delta$), yielding a target of \emph{$n\approx45$ per \textit{Format}$\times$\textit{Owner} cell} (analyzable $N\approx450$). Other families (e.g., \emph{confidence}, nominal symmetry) met this threshold at lower $n$ and inherit $n=45$. Exploratory analyses (e.g., multinomial transitions) did not inform $n$ and are reported descriptively.

Post-screening cells ranged 41---52 participants (Figure~\ref{fig:participantsPerCondition}). Simulations assumed balance; the main-text estimators (Bayesian GLMM for accuracy; linear mixed-effects model for confidence; HC3-robust OLS for experience and $\Delta$) provide valid owner-averaged contrasts under such mild imbalance. Accordingly, we retain the planned confirmatory scope (Format main effects) and treat \textit{Format}$\times$\textit{Owner} interactions as exploratory.

\section{Pilot Testing and Data Quality Measures}
\label{app:pilots}

% Some would close the privacy policy and try to answer the items from memory. So we allowed reopening of the policy both when answering the comprehension, experience, and post-perception questionnaire. 

% We actually introduce the waiting time for the pilot 2 Also, initially, the waiting time was too long 

Before launching the study on Prolific, we conducted two in-person pilot sessions with invited individuals who tested the experimental materials and provided qualitative feedback through post-session interviews. These sessions enabled us to identify and address language clarity issues, as well as evaluate the perceived length of the questionnaires. As a result, we refined question wording and eliminated redundant items from specific constructs to reduce respondent burden.

We then conducted three rounds of online pilot studies on Prolific. In each pilot, participants were also provided a dedicated space to give qualitative feedback on the experimental materials. In the first pilot (N = 10 per condition), we focused on the overall study duration. Analysis revealed that a few participants completed the study in as little as four minutes---far shorter than the 15--20 minutes observed in the in-person pilots---suggesting low engagement. While the small sample precluded definitive conclusions about data quality, these results highlighted the need for more precise instrumentation and controls.

In the second online pilot (10 participants per condition), we implemented several improvements: the interface was instrumented to capture detailed response timestamps for each item, enabling granular analysis of time-on-task; and both instructions and question wording were further refined based on participant feedback from the online pilot 1. The number of comprehension items was reduced from ten to eight to address concerns about study length. The data of this pilot showed moderate improvement in response durations. Although we did not observe strong evidence of straight-lining, there were still indications of random answering, as revealed by high response entropy across certain questionnaires.

For the third online pilot (20 participants per condition), we introduced additional mechanisms to ensure data quality. The compensation policy was clarified, and participants were explicitly warned---both in the Prolific study description and within the experimental interface---that inattentive or non-engaged responses, including failing attention checks or submitting implausibly fast completions, would result in exclusion and forfeiture of compensation. Minimum completion times were enforced for each section, and real-time warnings were added for submissions that were too rapid, requiring explicit participant acknowledgement. These cumulative changes resulted in engagement and response patterns that closely matched those of the in-person pilots.

Finally, with the data from the third pilot, we assessed the internal consistency of each questionnaire construct using Cronbach's alpha. While most constructs demonstrated satisfactory reliability, one construct had an alpha below 0.7; accordingly, we revised two of its items to improve coherence before the final study.

%%%% Sections/Appendices/Study.tex ends here %%%%

%%%% Results.tex starts here %%%%

\section{Supplementary Results}
\label{app:results}

This section reports supplementary analyses and diagnostic summaries that complement the main results.

\subsection{Comprehension}
\label{app:Comprehension}
\aptLtoX{\begin{table}[t]
\centering
\small
\caption{Observed accuracy by experimental condition. Shows the proportion of correct responses for each combination Format$\times$Owner$\times$Question type.}
\label{tab:accuracy:accuracy_by_condition}
\begin{tabular}{lllcc}
\toprule
Format & Owner & Question Type & $n$ & Accuracy \\
\midrule
\textit{Scrollytelling} & OpenAI & Factual & 164 & \cellcolor[HTML]{A8DECF}{0.622} \\
\textit{Scrollytelling} & OpenAI & Interpretative & 164 & \cellcolor[HTML]{D1EEE6}{0.537} \\
\textit{Scrollytelling} & TikTok & Factual & 176 & \cellcolor[HTML]{42B797}{0.818} \\
\textit{Scrollytelling} & TikTok & Interpretative & 176 & \cellcolor[HTML]{0FA47B}{0.915} \\

\textit{Text} & OpenAI & Factual & 200 & \cellcolor[HTML]{99D8C7}{0.650} \\
\textit{Text} & OpenAI & Interpretative & 200 & \cellcolor[HTML]{BDE6DB}{0.575} \\
\textit{Text} & TikTok & Factual & 172 & \cellcolor[HTML]{2EB08C}{0.855} \\
\textit{Text} & TikTok & Interpretative & 172 & \cellcolor[HTML]{14A67E}{0.907} \\

\textit{Nutrition Label} & OpenAI & Factual & 168 & \cellcolor[HTML]{8AD2BE}{0.679} \\
\textit{Nutrition Label} & OpenAI & Interpretative & 168 & \cellcolor[HTML]{94D6C4}{0.655} \\
\textit{Nutrition Label} & TikTok & Factual & 188 & \cellcolor[HTML]{33B18F}{0.846} \\
\textit{Nutrition Label} & TikTok & Interpretative & 188 & \cellcolor[HTML]{2EB08C}{0.856} \\

\textit{Nutr. Label + Text} & OpenAI & Factual & 208 & \cellcolor[HTML]{ADE0D2}{0.606} \\
\textit{Nutr. Label + Text} & OpenAI & Interpretative & 208 & \cellcolor[HTML]{D1EEE6}{0.538} \\
\textit{Nutr. Label + Text} & TikTok & Factual & 176 & \cellcolor[HTML]{2EB08C}{0.864} \\
\textit{Nutr. Label + Text} & TikTok & Interpretative & 176 & \cellcolor[HTML]{1FAA84}{0.886} \\

\textit{Interactive Vis} & OpenAI & Factual & 164 & \cellcolor[HTML]{B2E2D5}{0.604} \\
\textit{Interactive Vis} & OpenAI & Interpretative & 164 & \cellcolor[HTML]{DBF1EB}{0.518} \\
\textit{Interactive Vis} & TikTok & Factual & 200 & \cellcolor[HTML]{47B99A}{0.810} \\
\textit{Interactive Vis} & TikTok & Interpretative & 200 & \cellcolor[HTML]{0FA47B}{0.915} \\
\bottomrule
\end{tabular}
\Description{This table presents observed accuracy rates for comprehension questions across all experimental conditions, organized by format, policy owner, and question type (factual or interpretive). Each row corresponds to a unique combination of these variables, and reports the sample size (n) and the proportion of correct responses (Accuracy). The table includes 20 rows, covering five presentation formats (Scrollytelling, Text, Nutrition Label, Nutrition Label + Text, and Interactive Visualization), two policy owners (OpenAI and TikTok), and both factual and interpretive question types. Accuracy scores range from 0.518 to 0.915. Higher accuracy values are most often associated with TikTok policies, particularly for interpretive questions. The values are visually color-coded using a green gradient, with darker shades indicating higher accuracy, which helps reveal consistent patterns across conditions.}
\end{table}}{\begin{table}[t]
\centering
\small
\caption{Observed accuracy by experimental condition. Shows the proportion of correct responses for each combination Format$\times$Owner$\times$Question type.}
\label{tab:accuracy:accuracy_by_condition}
\begin{tabular}{lllcc}
\toprule
Format & Owner & Question Type & $n$ & Accuracy \\
\midrule
\textit{Scrollytelling} & OpenAI & Factual & 164 & \gradientcell[myGreen]{0.622} \\
\textit{Scrollytelling} & OpenAI & Interpretative & 164 & \gradientcell[myGreen]{0.537} \\
\textit{Scrollytelling} & TikTok & Factual & 176 & \gradientcell[myGreen]{0.818} \\
\textit{Scrollytelling} & TikTok & Interpretative & 176 & \gradientcell[myGreen]{0.915} \\

\addlinespace

\textit{Text} & OpenAI & Factual & 200 & \gradientcell[myGreen]{0.650} \\
\textit{Text} & OpenAI & Interpretative & 200 & \gradientcell[myGreen]{0.575} \\
\textit{Text} & TikTok & Factual & 172 & \gradientcell[myGreen]{0.855} \\
\textit{Text} & TikTok & Interpretative & 172 & \gradientcell[myGreen]{0.907} \\

\addlinespace

\textit{Nutrition Label} & OpenAI & Factual & 168 & \gradientcell[myGreen]{0.679} \\
\textit{Nutrition Label} & OpenAI & Interpretative & 168 & \gradientcell[myGreen]{0.655} \\
\textit{Nutrition Label} & TikTok & Factual & 188 & \gradientcell[myGreen]{0.846} \\
\textit{Nutrition Label} & TikTok & Interpretative & 188 & \gradientcell[myGreen]{0.856} \\

 \addlinespace
 
\textit{Nutr. Label + Text} & OpenAI & Factual & 208 & \gradientcell[myGreen]{0.606} \\
\textit{Nutr. Label + Text} & OpenAI & Interpretative & 208 & \gradientcell[myGreen]{0.538} \\
\textit{Nutr. Label + Text} & TikTok & Factual & 176 & \gradientcell[myGreen]{0.864} \\
\textit{Nutr. Label + Text} & TikTok & Interpretative & 176 & \gradientcell[myGreen]{0.886} \\

\addlinespace

\textit{Interactive Vis} & OpenAI & Factual & 164 & \gradientcell[myGreen]{0.604} \\
\textit{Interactive Vis} & OpenAI & Interpretative & 164 & \gradientcell[myGreen]{0.518} \\
\textit{Interactive Vis} & TikTok & Factual & 200 & \gradientcell[myGreen]{0.810} \\
\textit{Interactive Vis} & TikTok & Interpretative & 200 & \gradientcell[myGreen]{0.915} \\
\bottomrule
\end{tabular}
\Description{This table presents observed accuracy rates for comprehension questions across all experimental conditions, organized by format, policy owner, and question type (factual or interpretive). Each row corresponds to a unique combination of these variables, and reports the sample size (n) and the proportion of correct responses (Accuracy). The table includes 20 rows, covering five presentation formats (Scrollytelling, Text, Nutrition Label, Nutrition Label + Text, and Interactive Visualization), two policy owners (OpenAI and TikTok), and both factual and interpretive question types. Accuracy scores range from 0.518 to 0.915. Higher accuracy values are most often associated with TikTok policies, particularly for interpretive questions. The values are visually color-coded using a green gradient, with darker shades indicating higher accuracy, which helps reveal consistent patterns across conditions.}
\end{table}}

Table~\ref{tab:accuracy:accuracy_by_condition} reports the \emph{raw, unadjusted} proportion of correct responses in every \textit{Format}$\times$\textit{Owner}$\times$\textit{Question type} cell, together with the number of item-level trials ($n$). Here, ``Accuracy'' is the fraction correct on a 0-1 scale; counts reflect item responses (participants $\times$ items within a question type), not the number of participants. This table is provided to show the scale of variation across conditions and to make the heterogeneity in the data transparent.

Because cell sizes are unequal and items differ in baseline difficulty (with items crossed with participants and nested within owner), these descriptive proportions should not be used for hypothesis testing. All statistical conclusions in the main text come from the mixed-effects models, which estimate condition effects while accounting for stable between-person differences and item$\times$owner difficulty, and which yield intervals and marginal predictions that are comparable across conditions.

For orientation, the spreads visible in Table~\ref{tab:accuracy:accuracy_by_condition} align with the model-based results. For example, within \textit{scrollytelling}, interpretative accuracy is lower for \textit{OpenAI} than for \textit{TikTok} (0.537 vs.\ 0.915), with a similar pattern for \textit{vis} (0.518 vs.\ 0.915). These wide gaps are consistent with the fixed-effect contrasts and predicted-by-format means in the main text (Tables~\ref{tab:accuracy:vb_fixed_effects} and~\ref{tab:accuracy:predicted_accuracy_by_format}).

\begin{table}[ht!]
\centering
\small
\caption{Internal consistency for the Experience constructs. We report Cronbach's $\alpha$ and McDonald's $\omega_{\text{total}}$ (one-factor solution on item correlations). Reverse-worded items were reverse-scored before analysis.}
\label{app:tab:xp_reliability}
\begin{tabular}{lccc}
\toprule
Construct & $\alpha$ & $\omega_{\text{total}}$ \\
\midrule
Behavioral Intentions & .74 & .85 \\
Cognitive Load        & .78 & .87 \\
Engagement           & .85 & .91 \\
Enjoyment            & .80 & .88 \\
Format Adoption       & .84 & .91 \\
Perceived Clarity     & .86 & .91 \\
\bottomrule
\end{tabular}
\end{table}

% ---- width knobs (tune as needed) ----
\newlength{\wEffA}\setlength{\wEffA}{0.37\linewidth}  % Effect column
\newlength{\wTxtA}\setlength{\wTxtA}{0.33\linewidth}  % Δ text column
\newlength{\wCIA}\setlength{\wCIA}{0.22\linewidth}    % CI image column

\begin{table*}[ht!]
% \begin{table*}
\centering
\small
\setlength{\tabcolsep}{1pt}
\renewcommand{\arraystretch}{1.20}

\caption{Owner main effects (TikTok vs.\ OpenAI) and format $\times$ owner interactions (OLS with HC3 SEs). Entries are point differences relative to the \textit{scrollytelling}/\textit{OpenAI} baseline. Right subcolumns plot 95\% CIs on a common difference axis (vertical line at 0). Filled markers indicate CIs excluding 0.}
\label{app:tab:owner_effects:experience}

% ---------- Panel A ----------
\begin{tabular}{L{\wEffA} R{\wTxtA} C{\wCIA} R{\wTxtA} C{\wCIA} R{\wTxtA} C{\wCIA}}
\toprule
& \multicolumn{2}{c}{\textbf{Behavioral Intentions}} &
  \multicolumn{2}{c}{\textbf{Cognitive Load}} &
  \multicolumn{2}{c}{\textbf{Engagement}} \\
\cmidrule(lr){2-3}\cmidrule(lr){4-5}\cmidrule(lr){6-7}
\textbf{Effect} & $\Delta$ [95\% CI] & CI & $\Delta$ [95\% CI] & CI & $\Delta$ [95\% CI] & CI \\
\midrule

TikTok vs.\ OpenAI
& $-0.674$ [$-1.113,\,-0.234$]
& \multirow{5}{*}{\includegraphics[height=2cm,keepaspectratio]{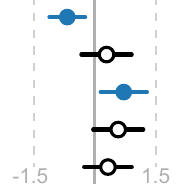}}
& $+0.144$ [$-0.319,\,+0.608$]
& \multirow{5}{*}{\includegraphics[height=2cm,keepaspectratio]{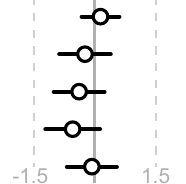}}
& $-0.362$ [$-0.830,\,+0.106$]
& \multirow{5}{*}{\includegraphics[height=2cm,keepaspectratio]{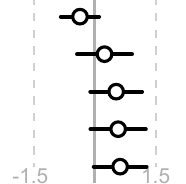}}
\\

\textit{Text} $\times$ TikTok
& $+0.292$ [$-0.327,\,+0.911$] &
& $-0.238$ [$-0.871,\,+0.395$] &
& $+0.242$ [$-0.431,\,+0.916$] &
\\

\textit{Nutr.\ Label + Text} $\times$ TikTok
& $+0.716$ [$+0.147,\,+1.285$] &
& $-0.382$ [$-1.006,\,+0.242$] &
& $+0.529$ [$-0.106,\,+1.164$] &
\\

\textit{Nutrition Label} $\times$ TikTok
& $+0.579$ [$-0.037,\,+1.196$] &
& $-0.541$ [$-1.211,\,+0.130$] &
& $+0.577$ [$-0.101,\,+1.255$] &
\\

\textit{Interactive Vis} $\times$ TikTok
& $+0.327$ [$-0.258,\,+0.913$] &
& $-0.068$ [$-0.681,\,+0.545$] &
& $+0.626$ [$-0.021,\,+1.274$] &
\\
\end{tabular}

\vspace{0.15em}

% ---------- Panel B ----------
\begin{tabular}{L{\wEffA} R{\wTxtA} C{\wCIA} R{\wTxtA} C{\wCIA} R{\wTxtA} C{\wCIA}}
\toprule
& \multicolumn{2}{c}{\textbf{Enjoyment}} &
  \multicolumn{2}{c}{\textbf{Format Adoption}} &
  \multicolumn{2}{c}{\textbf{Perceived Clarity}} \\
\cmidrule(lr){2-3}\cmidrule(lr){4-5}\cmidrule(lr){6-7}
\textbf{Effect} & $\Delta$ [95\% CI] & CI & $\Delta$ [95\% CI] & CI & $\Delta$ [95\% CI] & CI \\
\midrule

TikTok vs.\ OpenAI
& $-0.360$ [$-0.862,\,+0.142$]
& \multirow{5}{*}{\includegraphics[height=2cm,keepaspectratio]{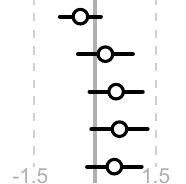}}
& $-0.353$ [$-0.861,\,+0.155$]
& \multirow{5}{*}{\includegraphics[height=2cm,keepaspectratio]{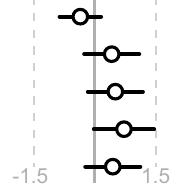}}
& $-0.329$ [$-0.792,\,+0.133$]
& \multirow{5}{*}{\includegraphics[height=2cm,keepaspectratio]{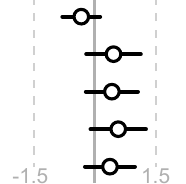}}
\\

\textit{Text} $\times$ TikTok
& $+0.256$ [$-0.417,\,+0.929$] &
& $+0.419$ [$-0.256,\,+1.095$] &
& $+0.464$ [$-0.205,\,+1.133$] &
\\

\textit{Nutr.\ Label + Text} $\times$ TikTok
& $+0.519$ [$-0.133,\,+1.172$] &
& $+0.509$ [$-0.166,\,+1.184$] &
& $+0.424$ [$-0.215,\,+1.063$] &
\\

\textit{Nutrition Label} $\times$ TikTok
& $+0.602$ [$-0.088,\,+1.293$] &
& $+0.721$ [$-0.019,\,+1.462$] &
& $+0.580$ [$-0.100,\,+1.260$] &
\\

\textit{Interactive Vis} $\times$ TikTok
& $+0.476$ [$-0.192,\,+1.145$] &
& $+0.446$ [$-0.229,\,+1.121$] &
& $+0.376$ [$-0.236,\,+0.989$] &
\\

\bottomrule
\end{tabular}

\Description{This table reports OLS (HC3) estimates for the main effect of owner (TikTok vs.\ OpenAI) and Format $\times$ TikTok interactions, relative to the \textit{Scrollytelling--OpenAI} baseline. Each cell reports a point difference and 95\% CI; adjacent CI columns visualize the same intervals on a common axis (0 marked). Filled markers indicate CIs excluding 0.}
\end{table*}

\subsection{Subjective Experience}
\label{app:Experience}

\subsubsection{Internal Consistency}
\label{app:InternalConsistency}

Table~\ref{app:tab:xp_reliability} reports the internal consistency of the six constructs used in the Experience questionnaire.

\subsubsection{Owner Main Effects (TikTok vs.\ OpenAI) and Format $\times$ Owner Interactions on Opinion Shifts}

Only one effect is statistically different from zero (Table~\ref{app:tab:owner_effects:experience}): the owner main effect for \textit{Behavioral Intentions}. Within the scrollytelling baseline, participants assigned to TikTok report lower willingness to adopt the format than those assigned to OpenAI by about $0.67$ points on the $1$--$5$ scale (95\% CI $[-1.11,-0.23]$; filled marker). All other owner effects are imprecise: estimates lean negative for \textit{Engagement}, \textit{Enjoyment}, \textit{Format Adoption}, and \textit{Perceived Clarity}, and slightly positive for \textit{Cognitive Load}, but their CIs include $0$. 

The \textit{Nutrition Label + Text} $\times$ TikTok interaction for \textit{Behavioral Intentions} is positive and excludes $0$ ($+0.72$, CI $[0.15,1.29]$), indicating that the TikTok--OpenAI gap seen under scrollytelling is largely neutralized in that format (owner $+$ interaction $\approx -0.67 + 0.72 \approx +0.04$). Interactions for the other constructs are directionally suggestive but inconclusive (CIs span $0$): label variants tend to reduce the TikTok--OpenAI difference in \textit{Cognitive Load} (negative interactions) and increase \textit{Engagement} for TikTok (positive interactions), while \textit{Text} and \textit{Interactive Vis} show smaller, non-decisive shifts. In short, a TikTok penalty appears only for \textit{Behavioral Intentions} under scrollytelling and is mitigated when paired with a nutrition label plus full text; elsewhere, owner differences and format-by-owner modulations are small and uncertain.

\subsection{Perception Shifts}
\label{app:PerceptionShifts}

\subsubsection{Internal Consistency}
\label{app:perceptionshifts:InternalConsistency}

Table~\ref{app:tab:perception_reliability} shows the internal consistency of the three subjective constructs measured before (Pre) and after (Post) exposure to the assigned privacy policy format.

\begin{table}[hb!]
\centering
\small
\caption{Internal consistency of the perception constructs at pre and post. We report Cronbach’s $\alpha$ and McDonald’s $\omega_{\text{total}}$ (one-factor solution on item correlations); reverse-worded items were reverse-scored prior to analysis.}
\label{app:tab:perception_reliability}
\begin{tabular}{lcccc}
\toprule
& \multicolumn{2}{c}{Pre} & \multicolumn{2}{c}{Post} \\
\cmidrule(lr){2-3}\cmidrule(lr){4-5}
Construct & $\alpha$ & $\omega_{\text{total}}$ & $\alpha$ & $\omega_{\text{total}}$ \\
\midrule
Confidence in Understanding & .81 & .91 & .86 & .93 \\
Transparency & .68 & .86 & .80 & .91 \\
Trust & .67 & .86 & .85 & .93 \\
\bottomrule
\end{tabular}
\Description{This table presents internal consistency reliability estimates for three constructs measured before and after the intervention: Confidence in Understanding, Transparency, and Trust. For each construct, the table reports Cronbach's alpha ($\alpha$) and McDonald's omega total ($\omega_{\text{total}}$) based on item correlations. Pre-intervention values are shown on the left; post-intervention values on the right.}
\end{table}

\subsubsection{Owner and Format $\times$ Owner Effects for Confidence in Understanding, Perceived Transparency, and Trust}

Table~\ref{tab:appendix:owner_interactions:opinion_shifts} reports owner main effects (TikTok vs.\ OpenAI) and \textit{Format}$\times$\textit{Owner} interactions on opinion shifts for \textit{Confidence in Understanding}, \textit{Perceived Transparency}, and \textit{Trust}. Coefficients are point differences (\textit{Post}--\textit{Pre}) on the 1--5 scale relative to the \textit{Scrollytelling}/\textit{OpenAI} reference. A negative owner coefficient indicates a smaller gain (or larger loss) for TikTok than for OpenAI. Interaction rows show how the TikTok--OpenAI contrast changes for each format relative to the scrollytelling reference; to obtain the TikTok--OpenAI contrast \textit{within} a non-reference format, sum the owner coefficient and the corresponding interaction. Confidence intervals are 95\%; all include 0, so owner-specific differences and format-by-owner modulations of shifts are statistically inconclusive in this sample. Sparklines display the point estimate (hollow circle) and its interval (horizontal bar) around 0 (vertical line). Magnitudes are modest in raw units ($|\Delta|\leq 0.44$).

% ---- width knobs (tune as needed) ----
\newlength{\wEffB}\setlength{\wEffB}{0.50\linewidth}  % Effect column
\newlength{\wTxtB}\setlength{\wTxtB}{0.30\linewidth}  % Δ text columns
\newlength{\wCIB}\setlength{\wCIB}{0.21\linewidth}    % CI image columns

\begin{table*}[h!]
% \begin{table*}
\centering
\small
\setlength{\tabcolsep}{1pt}
\renewcommand{\arraystretch}{1.20}
\caption{Owner main effects (TikTok vs.\ OpenAI) and Format(F)$\times$Owner(O) (TikTok) interaction contrasts on opinion shifts, by construct. Entries are point differences (Post--Pre) on the 1--5 scale relative to the \textit{scrollytelling}/\textit{OpenAI} baseline. All 95\% CIs include 0.}
\label{tab:appendix:owner_interactions:opinion_shifts}

\begin{tabular}{L{\wEffB}
  R{\wTxtB} C{\wCIB}
  R{\wTxtB} C{\wCIB}
  R{\wTxtB} C{\wCIB}}
\toprule
& \multicolumn{2}{c}{\textbf{Confidence in Understanding}} &
  \multicolumn{2}{c}{\textbf{Perceived Transparency}} &
  \multicolumn{2}{c}{\textbf{Trust}} \\
\cmidrule(lr){2-3}\cmidrule(lr){4-5}\cmidrule(lr){6-7}
\textbf{Effect} & $\Delta$ [95\% CI] & CI & $\Delta$ [95\% CI] & CI & $\Delta$ [95\% CI] & CI \\
\midrule

Owner: TikTok
& $-0.084$ [$-0.55,\,+0.38$]
& \multirow{5}{*}{\includegraphics[height=2cm,keepaspectratio]{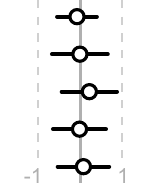}}
& $+0.112$ [$-0.29,\,+0.52$]
& \multirow{5}{*}{\includegraphics[height=2cm,keepaspectratio]{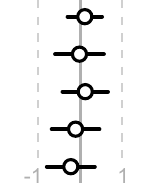}}
& $-0.338$ [$-0.71,\,+0.04$]
& \multirow{5}{*}{\includegraphics[height=2cm,keepaspectratio]{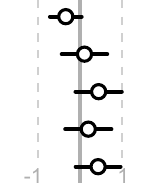}}
\\

F$\times$O: \textit{Text} $\times$ TikTok
& $-0.015$ [$-0.67,\,+0.64$] &
& $-0.014$ [$-0.59,\,+0.56$] &
& $+0.105$ [$-0.43,\,+0.64$] &
\\

F$\times$O: \textit{Nutrition Label} $\times$ TikTok
& $+0.209$ [$-0.44,\,+0.86$] &
& $+0.123$ [$-0.41,\,+0.66$] &
& $+0.440$ [$-0.11,\,+0.99$] &
\\

F$\times$O: \textit{Nutrition Label + Text} $\times$ TikTok
& $-0.024$ [$-0.65,\,+0.60$] &
& $-0.106$ [$-0.67,\,+0.46$] &
& $+0.195$ [$-0.35,\,+0.74$] &
\\

F$\times$O: \textit{Interactive Vis} $\times$ TikTok
& $+0.069$ [$-0.54,\,+0.68$] &
& $-0.219$ [$-0.79,\,+0.35$] &
& $+0.427$ [$-0.10,\,+0.96$] &
\\

\bottomrule
\end{tabular}
\end{table*}

% \clearpage

\subsubsection{Pre-post transitions in participants' beliefs about the \textit{amount of data collected}}

\begin{figure*}[hb!]
  \centering

  % Row 1
  \begin{subfigure}[t]{0.45\textwidth}
    \centering
    \includegraphics[width=0.85\linewidth]{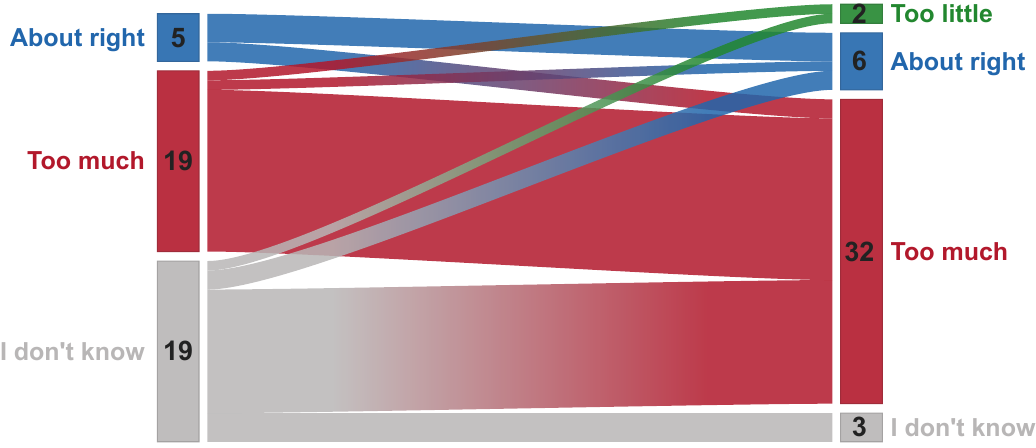}
    \caption{Text -- OpenAI}
    \label{figure:DataCollected:text:OpenAI}
  \end{subfigure}\hfill
  \begin{subfigure}[t]{0.45\textwidth}
    \centering
    \includegraphics[width=0.85\linewidth]{Sections/Appendices/Figures/Flows/flowAmountDataCollected-text-TikTok.pdf}
    \caption{Text -- TikTok}
   \label{figure:DataCollected:text:TikTok}
  \end{subfigure}

  \vspace{0.12in}

  % Row 2
  \begin{subfigure}[t]{0.45\textwidth}
    \centering
    \includegraphics[width=0.85\linewidth]{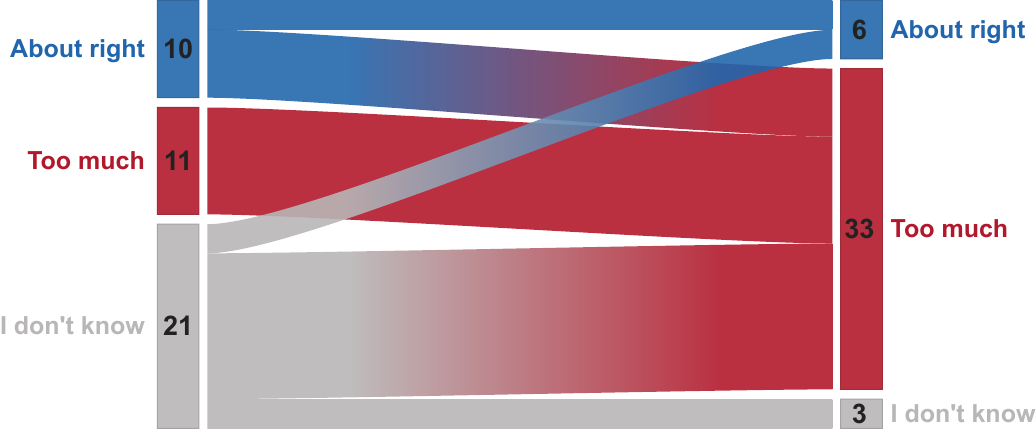}
    \caption{Nutrition Label -- OpenAI}
    \label{figure:DataCollected:NutritionOnly:OpenAI}
  \end{subfigure}\hfill
  \begin{subfigure}[t]{0.45\textwidth}
    \centering
    \includegraphics[width=0.85\linewidth]{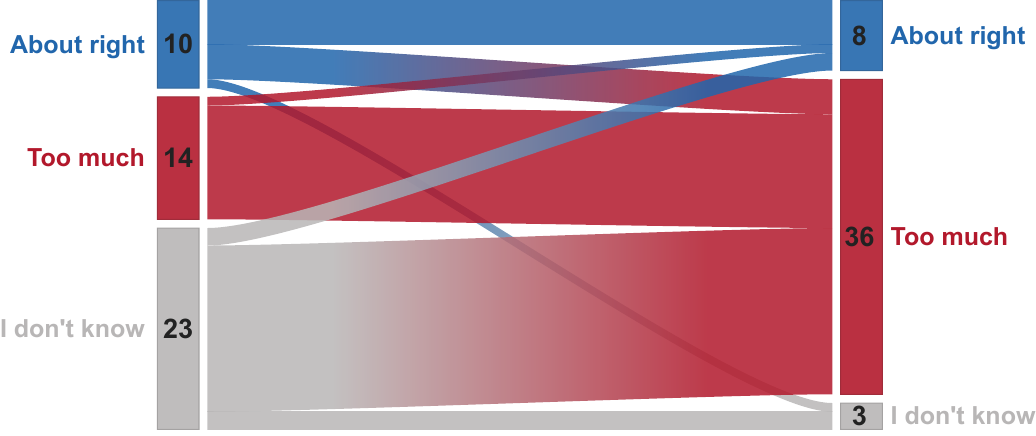}
    \caption{Nutrition Label -- TikTok}
    \label{figure:DataCollected:NutritionOnly:TikTok}
  \end{subfigure}

  \vspace{0.12in}

  % Row 3
  \begin{subfigure}[t]{0.45\textwidth}
    \centering
    \includegraphics[width=0.85\linewidth]{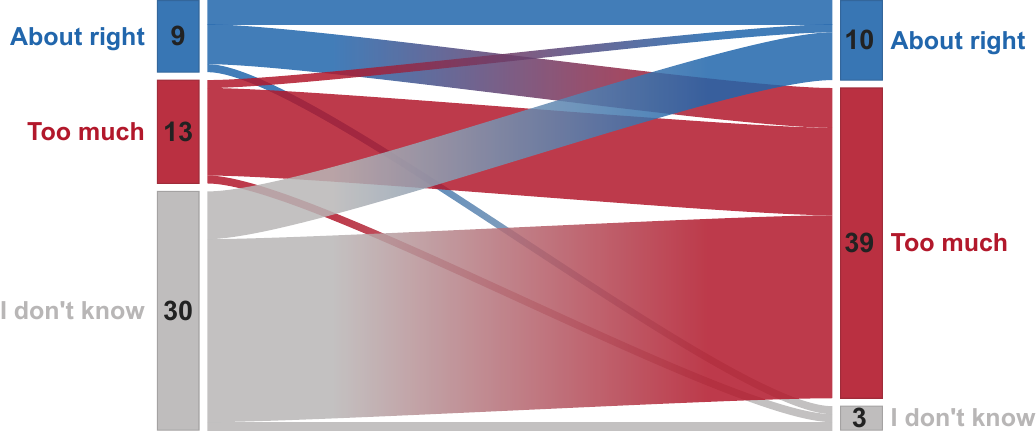}
    \caption{Nutrition Label + Text -- OpenAI}
    \label{figure:DataCollected:Nutrition:OpenAI}
  \end{subfigure}\hfill
  \begin{subfigure}[t]{0.45\textwidth}
    \centering
    \includegraphics[width=0.85\linewidth]{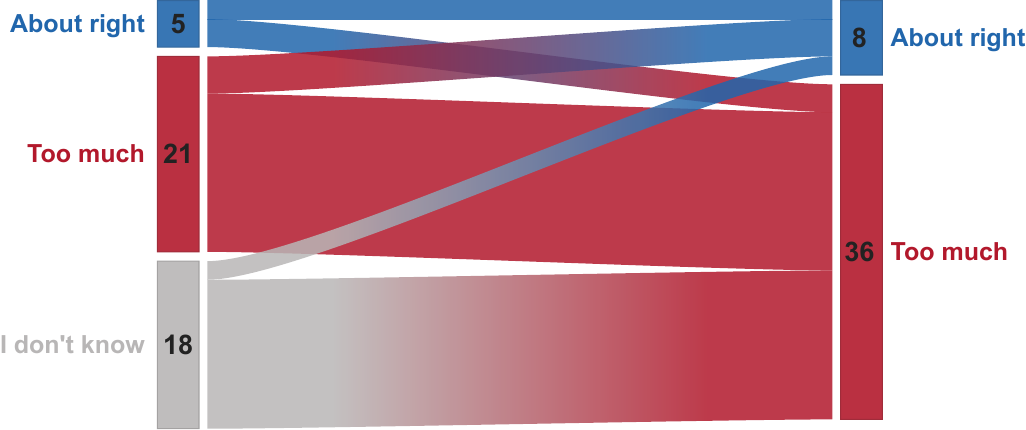}
    \caption{Nutrition Label + Text -- TikTok}
    \label{figure:DataCollected:Nutrition:TikTok}
  \end{subfigure}

  \vspace{0.12in}

  % Row 4
  \begin{subfigure}[t]{0.45\textwidth}
    \centering
    \includegraphics[width=0.85\linewidth]{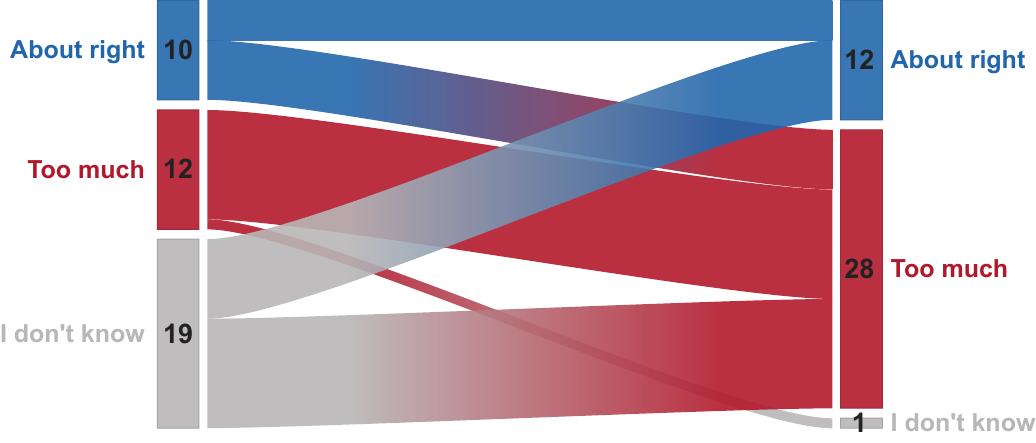}
    \caption{Scrollytelling -- OpenAI}
    \label{figure:DataCollected:scrollytelling:OpenAI}
  \end{subfigure}\hfill
  \begin{subfigure}[t]{0.45\textwidth}
    \centering
    \includegraphics[width=0.85\linewidth]{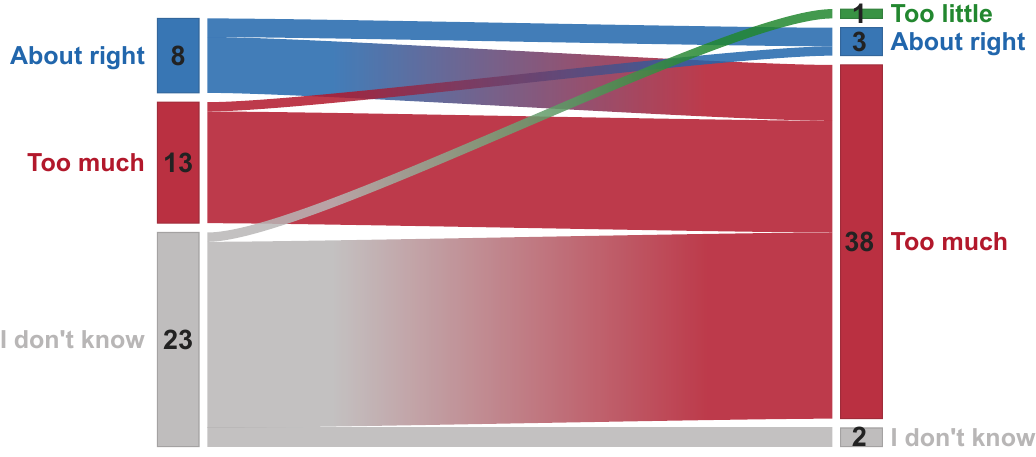}
    \caption{Scrollytelling -- TikTok}
    \label{figure:DataCollected:scrollytelling:TikTok}
  \end{subfigure}

  \vspace{0.12in}

  % Row 5
  \begin{subfigure}[t]{0.45\textwidth}
    \centering
    \includegraphics[width=0.85\linewidth]{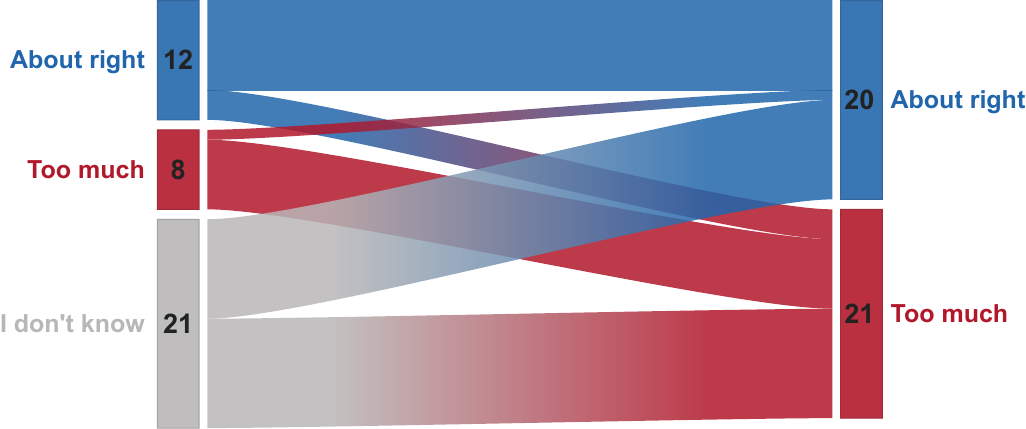}
    \caption{Interactive Vis -- OpenaAI}
    \label{figure:DataCollected:vis:OpenAI}
  \end{subfigure}\hfill
  \begin{subfigure}[t]{0.45\textwidth}
    \centering
    \includegraphics[width=0.85\linewidth]{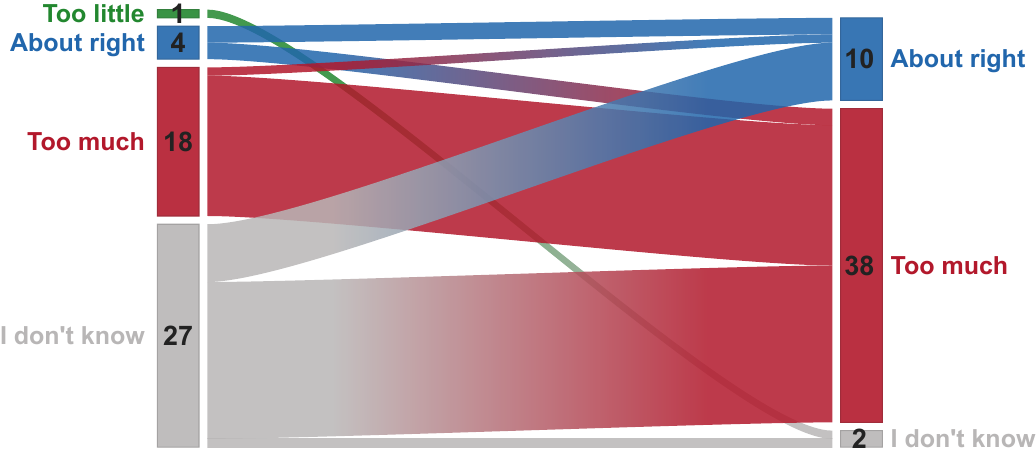}
    \caption{Interactive Vis -- TikTok}
    \label{figure:DataCollected:vis:TikTok}
  \end{subfigure}

  \caption{Pre-post transitions in participants' beliefs about the \textit{amount of data collected}, visualized as Sankey-style flows. Each panel corresponds to a format-owner condition. Nodes on the left indicate pre-format exposure responses; nodes on the right indicate post-exposure responses. Edge thickness reflects the number of participants shifting between categories.}

  \label{figure:flows:amountDataCollected}
  \Description{This figure shows ten Sankey-style flow diagrams illustrating pre–post transitions in participants' beliefs about the amount of data collected by a platform. Each panel represents a unique combination of format and policy owner (e.g., Text–OpenAI, Nutrition Label–TikTok), with nodes on the left indicating pre-exposure responses and nodes on the right indicating post-exposure responses. The response categories are color-coded: "Too much" (red), "About right" (blue), "Too little" (green), and "I don't know" (gray). Flow thickness corresponds to the number of participants shifting from one belief to another. Most transitions occur between "Too much" and "About right" or "I don't know." In nearly all panels, a substantial portion of participants began with "I don't know" (left gray node) and shifted toward either "Too much" or "About right." Transitions from "Too much" to "About right" are most visible in formats like Scrollytelling and Interactive Vis, especially under the TikTok condition. These diagrams highlight how different presentation formats and data owners influenced changes in perceived data collection quantity.}
\end{figure*}

Figure~\ref{figure:flows:amountDataCollected} provides a condition-level view of the same pre--post dynamics analyzed in the main text. Each panel shows, for one Format$\times$Owner condition, how participants moved between response categories (\textit{too much}, \textit{about right}, \textit{too little}, \textit{I don't know}). Across panels, the dominant pattern is movement out of \textit{I don't know} into committed judgments, with relatively little evidence of systematic net shifts that would differ by format or owner. The figure is intended as a diagnostic complement to the aggregate transition table and tests: it makes the heterogeneity of individual pathways visible while underscoring that condition-level differences are modest relative to the overall \quotes{exit from uncertainty} trend.

% \begin{figure*}

% \clearpage

\subsubsection{Pre-post transitions in participants' beliefs about the \textit{number of third parties receiving data}}

\begin{figure*}[hb!]
  \centering

  % Row 1
  \begin{subfigure}[t]{0.45\textwidth}
    \centering
    \includegraphics[width=0.85\linewidth]{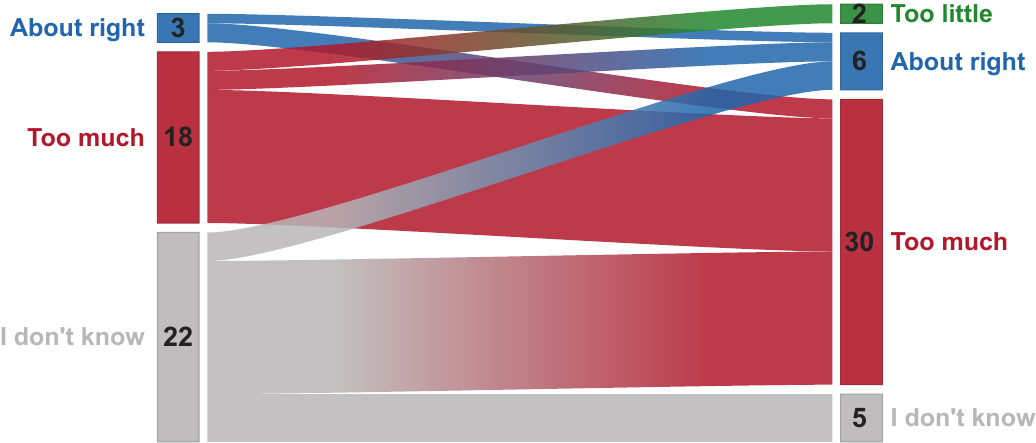}
    \caption{Text -- OpenAI}
    \label{figure:ThirdParties:text:OpenAI}
  \end{subfigure}\hfill
  \begin{subfigure}[t]{0.45\textwidth}
    \centering
    \includegraphics[width=0.85\linewidth]{Sections/Appendices/Figures/Flows/flowNumberThirdParties-text-TikTok.pdf}
    \caption{Text -- TikTok}
   \label{figure:ThirdParties:text:TikTok}
  \end{subfigure}

  \vspace{0.12in}

  % Row 2
  \begin{subfigure}[t]{0.45\textwidth}
    \centering
    \includegraphics[width=0.85\linewidth]{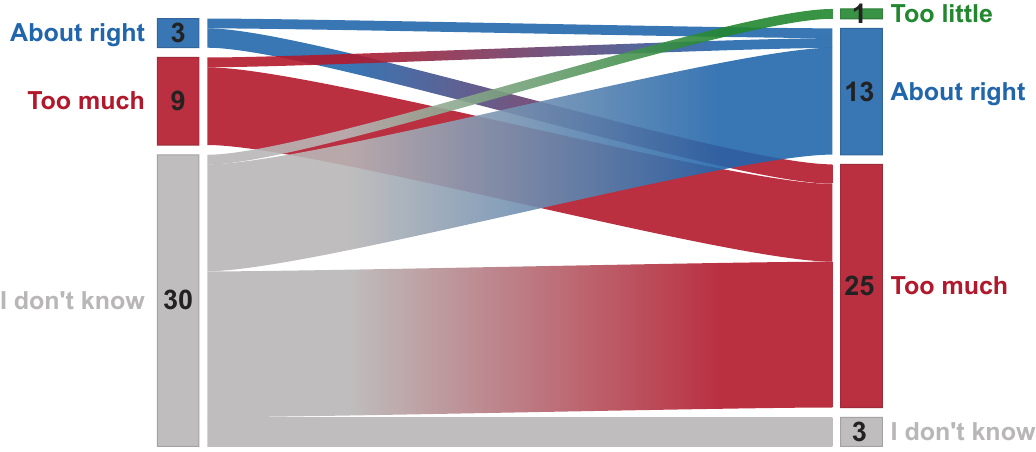}
    \caption{Nutrition Label -- OpenAI}
    \label{figure:ThirdParties:NutritionOnly:OpenAI}
  \end{subfigure}\hfill
  \begin{subfigure}[t]{0.45\textwidth}
    \centering
    \includegraphics[width=0.85\linewidth]{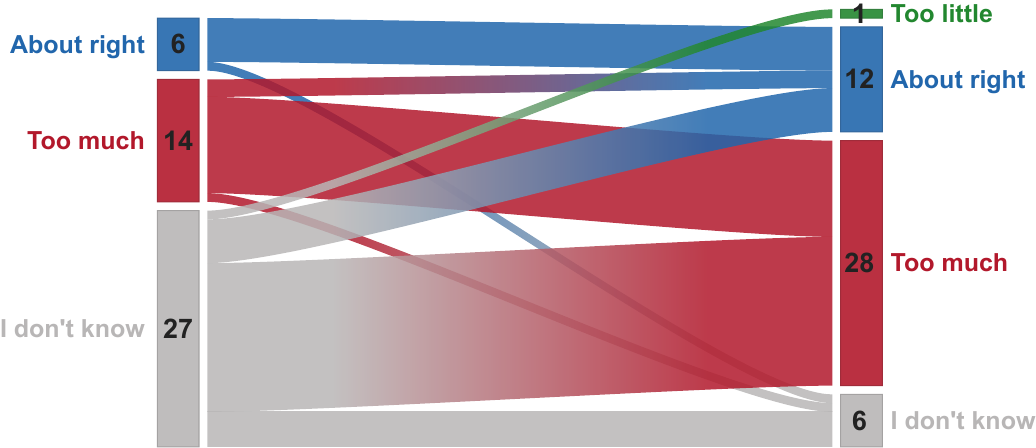}
    \caption{Nutrition Label -- TikTok}
    \label{figure:ThirdParties:NutritionOnly:TikTok}
  \end{subfigure}

  \vspace{0.12in}

  % Row 3
  \begin{subfigure}[t]{0.45\textwidth}
    \centering
    \includegraphics[width=0.85\linewidth]{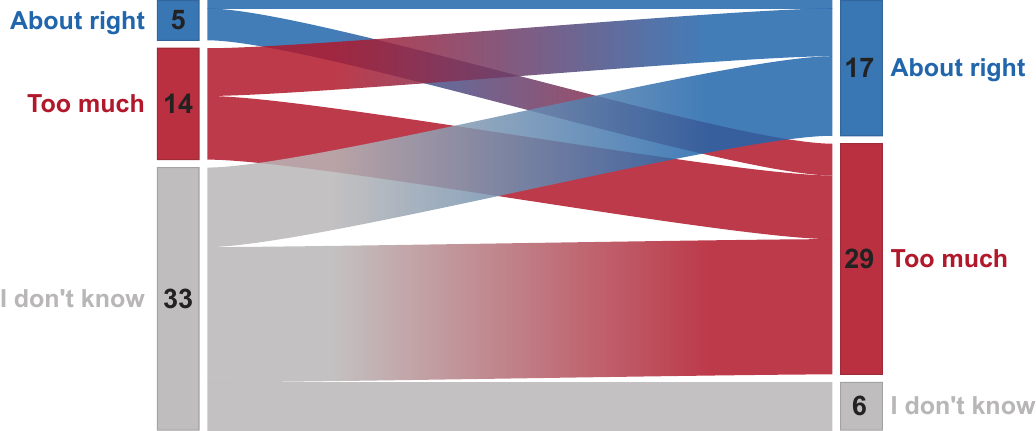}
    \caption{Nutrition Label + Text -- OpenAI}
    \label{figure:ThirdParties:Nutrition:OpenAI}
  \end{subfigure}\hfill
  \begin{subfigure}[t]{0.45\textwidth}
    \centering
    \includegraphics[width=0.85\linewidth]{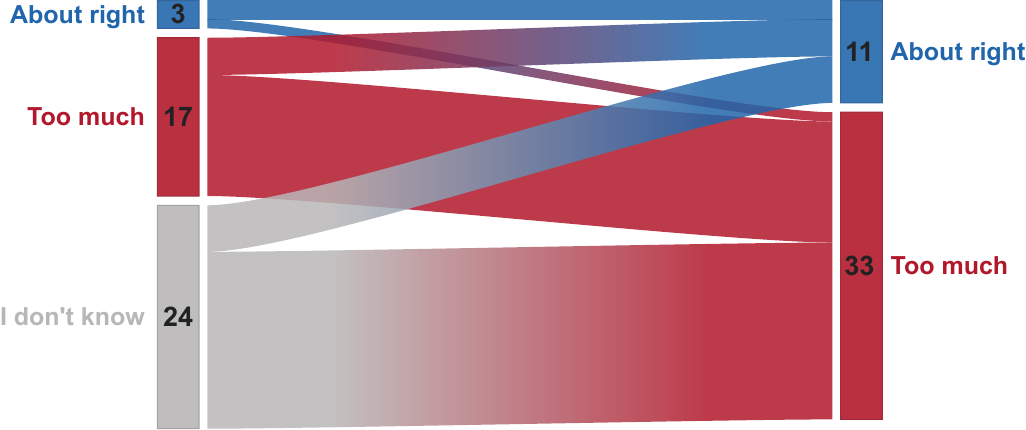}
    \caption{Nutrition Label + Text -- TikTok}
    \label{figure:ThirdParties:Nutrition:TikTok}
  \end{subfigure}

  \vspace{0.12in}

  % Row 4
  \begin{subfigure}[t]{0.45\textwidth}
    \centering
    \includegraphics[width=0.85\linewidth]{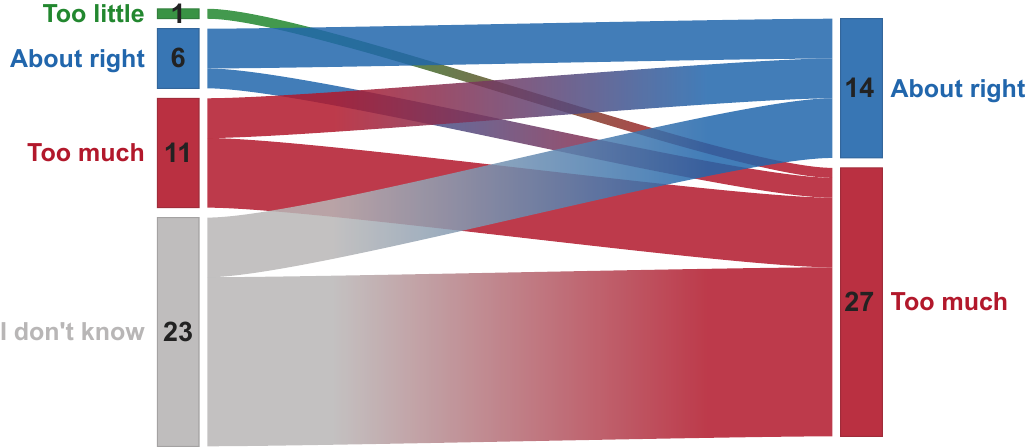}
    \caption{Scrollytelling -- OpenAI}
    \label{figure:ThirdParties:scrollytelling:OpenAI}
  \end{subfigure}\hfill
  \begin{subfigure}[t]{0.45\textwidth}
    \centering
    \includegraphics[width=0.85\linewidth]{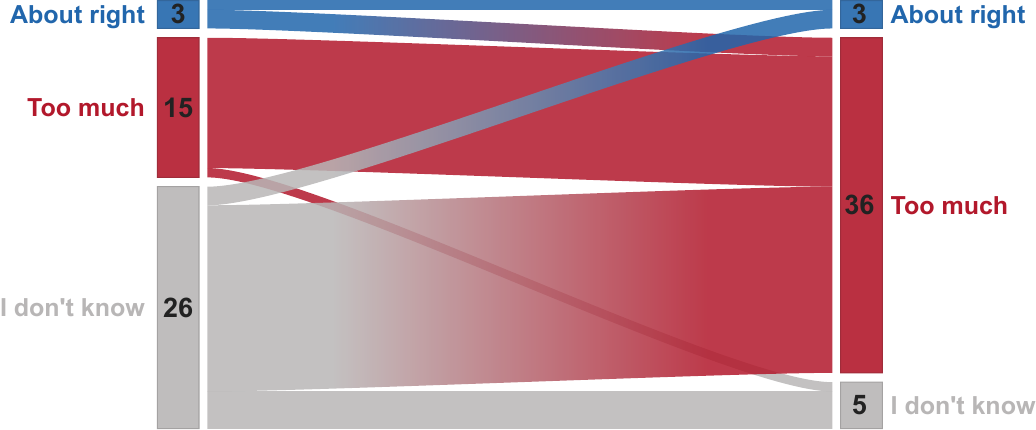}
    \caption{Scrollytelling -- TikTok}
    \label{figure:ThirdParties:scrollytelling:TikTok}
  \end{subfigure}

  \vspace{0.12in}

  % Row 5
  \begin{subfigure}[t]{0.45\textwidth}
    \centering
    \includegraphics[width=0.85\linewidth]{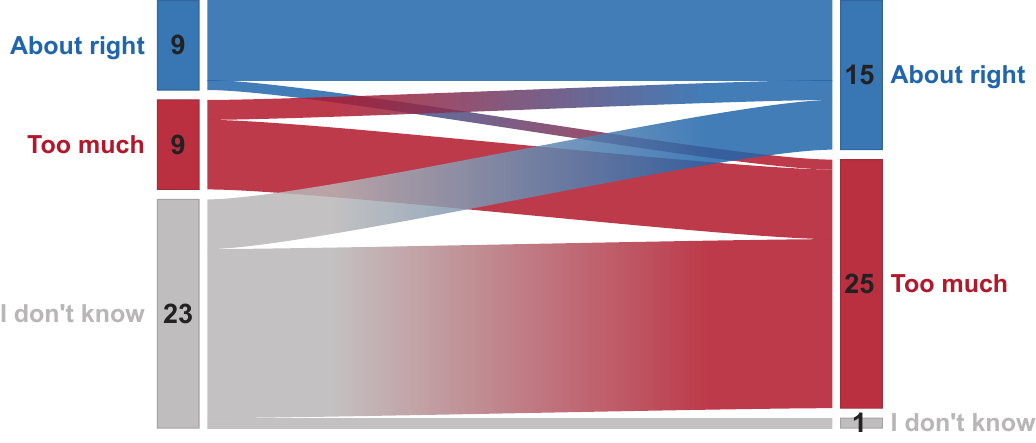}
    \caption{Interactive Vis -- OpenaAI}
    \label{figure:ThirdParties:vis:OpenAI}
  \end{subfigure}\hfill
  \begin{subfigure}[t]{0.45\textwidth}
    \centering
    \includegraphics[width=0.85\linewidth]{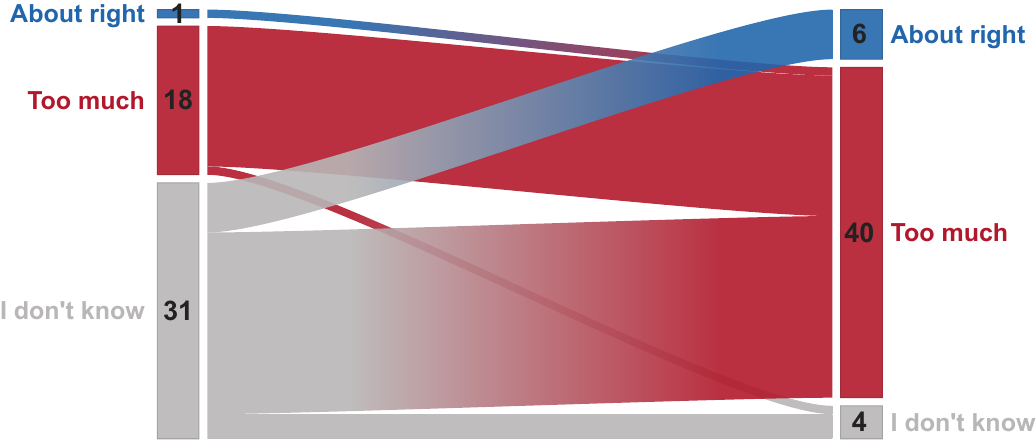}
    \caption{Interactive Vis -- TikTok}
    \label{figure:ThirdParties:vis:TikTok}
  \end{subfigure}

  \caption{Pre-post transitions in participants' beliefs about the \textit{number of third parties receiving data}, visualized as Sankey-style flows. Each panel corresponds to a format-owner condition. Nodes on the left indicate pre-format exposure responses; nodes on the right indicate post-exposure responses. Edge thickness reflects the number of participants shifting between categories.}

  \label{figure:flows:numberOfThirdParties}

  \Description{This figure presents ten Sankey-style diagrams that depict pre–post transitions in participants' beliefs about the number of third parties receiving data. Each panel corresponds to a different combination of format and policy owner (e.g., Text–OpenAI, Nutrition Label–TikTok). On the left side of each panel are pre-exposure response categories, and on the right are post-exposure responses. Categories are color-coded: "Too much" (red), "About right" (blue), "Too little" (green), and "I don't know" (gray). Flows between nodes represent how participants' beliefs shifted, with thicker edges indicating more participants. Across most conditions, many participants moved from "I don't know" to more definitive responses, particularly "Too much." This trend is especially pronounced in panels for Nutrition Label + Text and Interactive Vis formats under TikTok ownership. Only a few participants selected "Too little" before or after exposure. The flows suggest that participants often gained clarity or reinforced concern about third-party data sharing after engaging with the visualized policy formats.}
\end{figure*}

Figure~\ref{figure:flows:numberOfThirdParties} visualizes the pre--post transitions for perceived third-party sharing within each Format$\times$Owner condition. As in the aggregate transitions (Table~\ref{tab:transitions:thirdparties}), the dominant movement in every panel is an exit from \textit{I don't know} into committed judgments, most often toward \textit{too many}. Differences across formats and owners are comparatively small in magnitude relative to this shared pattern. Given the fragility of the multinomial model fit for this item, we treat the figure as corroborating descriptive evidence: it makes the underlying response pathways transparent and helps verify that the main result is driven by widespread \quotes{leaving uncertainty} rather than a condition-specific reallocation of the margins.

% \begin{figure*}

%%%% Sections/Appendices/Results.tex ends here %%%%

\balance{}

\end{document}